\documentclass[floats,floatfix,showpacs,amssymb,prd,twocolumn,superscriptaddress,nofootinbib,nolongbibliography,reprint]{revtex4-2}

\usepackage{amssymb,amsmath,verbatim,mathtools,needspace,enumitem,etoolbox,graphicx,physics,microtype,afterpage,bigints,gensymb,tabularx,multirow}

\usepackage[dvipsnames, usenames, table]{xcolor}
\definecolor{linkcolor}{rgb}{0.0,0.3,0.5}
\definecolor{dodgerblue}{HTML}{1E90FF}
\usepackage[unicode, colorlinks=true, linkcolor=linkcolor, citecolor=linkcolor, filecolor=linkcolor,urlcolor=linkcolor, pdfusetitle]{hyperref}
\usepackage[all]{hypcap}
\usepackage[T1]{fontenc}
\usepackage[utf8]{inputenc}

\usepackage{longtable}

\usepackage{orcidlink}

\renewcommand{\arraystretch}{1.4}

\makeatletter
\newcommand*{\balancecolsandclearpage}{\close@column@grid \cleardoublepage \twocolumngrid}
\makeatother

\def\apj{{Astrophys.~J.}}\def\apjl{{Astrophys.~J.~Lett.}}
\def\apjs{{Astrophys.~J.~Supp.}}

\def\aap{{Astron.~Astrophys.}}

\def\mnras{{Mon.~Not.~R.~Astron.~Soc.}}

\def\prd{{Phys.~Rev.~D}}

\def\prx{{Phys.~Rev.~X}}
\def\prl{{Phys.~Rev.~Lett.}}
\def\pasa{{PASA}}

\def\nat{{Nature}}

\def\physrep{{Phys.~Rep.}}

\def\lrr{{Living Rev. Relativ.}}
\def\cqg{{Class. Quantum Gravity}}
\def\natastro{{Nat. Astron.}}

\newcommand{\bham}{\affiliation{School of Physics and Astronomy \& Institute for Gravitational Wave Astronomy, \\ University of Birmingham, Birmingham, B15 2TT, United Kingdom}}
\newcommand{\vanderbilt}{\affiliation{Department of Physics and Astronomy, Vanderbilt University,  2301 Vanderbilt Place, Nashville, TN 37235, USA}}
\newcommand{\milan}{
    \affiliation{Dipartimento di Fisica ``G. Occhialini'', Universit\'a degli Studi di Milano-Bicocca, Piazza della Scienza 3, 20126 Milano, Italy}
    \affiliation{INFN, Sezione di Milano-Bicocca, Piazza della Scienza 3, 20126 Milano, Italy}
}

\renewcommand{\comment}[1]{}

\newcommand\orcid[1]{\href{https://orcid.org/#1}{$\!$\includegraphics[scale=0.006]{orcidicon.png} $\!\!$}}

\newcommand{\chieff}{\chi_\mathrm{eff}}
\newcommand{\chip}{\chi_\mathrm{p}}
\newcommand{\mmax}{m_\mathrm{max}}
\newcommand{\chimax}{\chi_\mathrm{max}}
\newcommand{\Mchirp}{M_\mathrm{c}}
\newcommand{\Msun}{M_\odot}
\newcommand{\vesc}{v_\mathrm{esc}}
\newcommand{\ppop}{p_\mathrm{pop}}
\newcommand{\Pdet}{P_\mathrm{det}}
\newcommand{\Npop}{N_\lambda}
\newcommand{\NBH}{N_\mathrm{BH}}
\newcommand{\Ncl}{N_\mathrm{cl}}

\newcommand{\Nobs}{N_\mathrm{obs}}
\newcommand{\Ndet}{N_\mathrm{det}}

\newcommand{\zmax}{z_\mathrm{max}}
\newcommand{\Vh}{V_\mathrm{h}}
\newcommand{\Vp}{V_\mathrm{p}}
\newcommand{\Nh}{N_\mathrm{h}}
\newcommand{\Np}{N_\mathrm{p}}
\renewcommand{\L}{\mathcal{L}}
\newcommand{\ssetminus}{{\setminus}}
\newcommand{\thetab}{\bar{\theta}}
\newcommand{\dH}{d_\mathrm{H}}

\newcommand{\moneg}{m_{1\mathrm{g}}}
\newcommand{\kmps}{\mathrm{km\,s^{-1}}}
\newcommand{\vkick}{v_\mathrm{kick}}
\newcommand{\PPD}{\mathrm{PPD}}

\begin{document}

\title{Deep learning and Bayesian inference of gravitational-wave populations:\\ Hierarchical black-hole mergers}

\author{Matthew Mould\,\orcidlink{0000-0001-5460-2910}}
\email{mmould@star.sr.bham.ac.uk}
\bham
\author{Davide Gerosa$\,$\orcidlink{0000-0002-0933-3579}}
 \milan \bham
\author{Stephen R. Taylor$\,$\orcidlink{0000-0001-8217-1599}}
\vanderbilt

\pacs{}

\date{\today}

\begin{abstract}
The catalog of gravitational-wave events is growing, and so are our hopes of constraining the underlying astrophysics of stellar-mass black-hole mergers by inferring the distributions of, e.g., masses and spins. While conventional analyses parametrize this population with simple phenomenological models, we propose an emulation-based approach that can compare astrophysical simulations against gravitational-wave data. We combine state-of-the-art deep-learning techniques with hierarchical Bayesian inference and exploit our approach to constrain the properties of repeated black-hole mergers from the gravitational-wave events in the most recent LIGO/Virgo catalog. Deep neural networks allow us to (i) construct a flexible single-channel population model that accurately emulates simple parametrized numerical simulations of hierarchical mergers, (ii) estimate selection effects, and (iii) recover the branching ratios of repeated-merger generations. Among our results, we find the following: The distribution of host-environment escape speeds favors values less than $100$~km\,s$^{-1}$ but is relatively flat, with around $37\%$ of first-generation mergers retained in their host environments; first-generation black holes are born with a maximum mass that is compatible with current estimates from pair-instability supernovae; there is multimodal substructure in both the mass and spin distributions, which, in our model, can be explained by repeated mergers; and binaries with a higher-generation component make up at least 14\% of the underlying population. Though these results are inferred through emulation of a simplified model, the deep-learning pipeline we present is readily applicable to realistic astrophysical simulations.
\end{abstract}

\maketitle

\section{Introduction}
\label{sec:intro}

The Advanced LIGO~\cite{2015CQGra..32g4001L} and Virgo~\cite{2015CQGra..32b4001A} gravitational-wave (GW) detectors are revealing the previously unseen landscape of compact binary coalescences. To date, nearly 100 GW signals from merging stellar-mass compact objects have been observed, the majority being black holes (BHs)~\cite{2019PhRvX...9c1040A,2021PhRvX..11b1053A,2021arXiv210801045T,2021arXiv211103606T,2021ApJ...922...76N,2021arXiv211206878N,2022arXiv220102252O}. Accurate estimation of the intrinsic properties of individual sources, such as component masses and spins, allows us to view the distribution of merging binary BHs as a whole. Crucially, the binary parameters inferred at merger are influenced by the formation history and astrophysical environment in which the progenitor systems were born; conversely, cumulative measurements of those source properties allow constraints to be placed at the population level, which can ultimately be compared to the predictions from likely binary formation scenarios.

Two examples include isolated stellar binary evolution~\cite{2014LRR....17....3P} and dynamical interactions in star clusters~\cite{2013LRR....16....4B}. While the former predicts a forbidden mass region for stellar remnants~\cite{2019ApJ...887...53F, 2021ApJ...912L..31W} and spins that favor small misalignments with the binary orbital angular momentum~\cite{2000ApJ...541..319K,2018PhRvD..98h4036G,2021PhRvD.103f3032S}, binaries formed in the latter channel may repeatedly interact and merge with other members of the cluster and thus be pushed to higher masses and isotropic spin orientations~\cite{2021NatAs...5..749G} (with GW-driven inspiral preserving the spin isotropy~\cite{2007ApJ...661L.147B,2015PhRvD..92f4016G}).

Given the catalog of GW detections, one can take two approaches to assess the underlying astrophysical population of binary BHs. In a simulation-based analysis, sources are synthesized ---accounting for as many detailed astrophysical processes as are known or are computationally feasible--- to form distributions of detectable merging binaries. By varying population-level input parameters controlling binary evolution (e.g., common envelope efficiency, and strength of supernova kicks), one can assess the degree of consistency with the observed events (for reviews see, e.g., Refs.~\cite{2022PhR...955....1M,2022LRR....25....1M,2021hgwa.bookE...4M}). However, such simulations are typically computationally intensive and large uncertainties remain on key parameters (see, e.g., Refs.~\cite{2018MNRAS.477.4685B,2022ApJ...925...69B}).

The second approach is to first construct a model of the astrophysical distribution of source parameters ---which is conditionally dependent on given population-level parameters controlling its shape (the ``hyperparameters'', e.g., mass cutoffs or spectral indices)--- and use the observed catalog to perform a hierarchical Bayesian inference that accounts for observational biases (e.g., that heavier sources are easier to detect). This statistical analysis is hierarchical in the sense that one uses previous Bayesian measurements of the binary BH source parameters to then measure said hyperparameters \cite{2019MNRAS.486.1086M,2022hgwa.bookE..45V}. The population model used could be as in the previous approach such that the distribution is known only at discrete values of the hyperparameters, but this would allow only for single posterior evaluations for relative comparisons (e.g., via Bayes factors) and leave some of the hyperparameters unconstrained (see, e.g., Refs.~\cite{2021ApJ...910..152Z,2021MNRAS.507.5224B} for examples of this approach).

On the other hand, a population model that can be continuously evaluated across the population-level parameter space can be used to make Bayesian measurements of the hyperparameters. This requirement typically necessitates simple, quick-to-evaluate parametric forms with statistical independence between source parameters (see, e.g., the models used in Refs.~\cite{2019ApJ...882L..24A,2021ApJ...913L...7A,2021arXiv211103634T}) to enable efficient hyperposterior sampling. The disadvantage of this approach is that it is inherently phenomenological with a discretionary selection of the employed functional forms. Recent work has sought to improve parametric population models by addressing potential correlations between mass and spin parameters~\cite{2021ApJ...922L...5C,2021arXiv211103634T,2022PhRvD.105l3024F} and assessing the suitability of spin parametrizations~\cite{2021PhRvD.104h3010R,2021ApJ...921L..15G,2022PhRvD.105b4076M} since accurate inference requires appropriate models~\cite{2022PASA...39...25R}. Along other lines, the flexibility of population analyses can be improved with semiparametric and nonparametric modeling techniques~\cite{2022ApJ...924..101E,2021CQGra..38o5007T,2017MNRAS.465.3254M,2022PhRvD.105l3014S,2022MNRAS.509.5454R}.

Previous studies have focused on combining the simulation-based and parametric approaches: a simulation emulator constructed with sufficient accuracy to rapidly synthesize predictive distributions over the hyperparameter space can be adopted in place of parametrized phenomenological models in the Bayesian inference pipeline. Such models leverage the advantages of efficient hyperposterior sampling and direct astrophysics-to-GW data comparison provided by each approach.

A first step in this direction within the context of GW population inference was taken by some of the authors in Ref.~\cite{2018PhRvD..98h3017T}. Compressed principal components of binned simulation data were emulated over low- (typically one- or two-) dimensional source- and population-level parameter spaces using Gaussian process regression (GPR). However, this emulation approach was shown to be unsuitable for extension to more complex higher-dimensional modeling scenarios due to poor predictive accuracy and infeasible computational requirements~\cite{2019PhRvD.100h3015W,2021arXiv211206707C}. These issues were tackled in Ref.~\cite{2020PhRvD.101l3005W} by employing deep-learning techniques to construct simulation-informed population models; in particular, the conditional density estimator takes the form of a flow-based generative neural network known as a normalizing flow~\cite{2017arXiv170507057P}. In general, neural networks are powerful tools that offer greater flexibility when employed as functional emulators. In this case, normalizing flows prompted population studies considering the scenarios of primordial BHs~\cite{2021PhRvD.103b3026W} and mixture models between isolated and dynamical evolution~\cite{2021PhRvD.103h3021W}.

In this work, we develop complementary deep-learning techniques that build on the advancements of Refs.~\cite{2018PhRvD..98h3017T,2020PhRvD.101l3005W} by pushing the emulated parameter space dimensionality and introducing new neural network applications. We employ fully connected deep neural networks (DNNs; also referred to as multilayer perceptrons) to act as the conditional density estimator of a population model and to capture the effect of GW detection biases on the population of observed binary BH events (see also Refs.~\cite{2020PhRvD.102j3020G,2022ApJ...927...76T,2020arXiv200710350W} for machine-learning approaches to estimating selection effects).

Motivated by evidence for large masses in the observed GW catalog, we apply these deep-learning techniques to binary BH populations containing hierarchical mergers, in which component BHs may be the remnants of (multiple) previous mergers~\cite{2021NatAs...5..749G}. These so-called ``higher-generation'' BHs may explain the outlier properties of events such as GW190412~\cite{2020PhRvD.102d3015A,2020PhRvL.125j1103G,2020ApJ...896L..10R,2020ApJ...898...99H,2020ApJ...897L...7S}, GW190521~\cite{2020PhRvL.125j1102A,2020ApJ...900L..13A,2020ApJ...903L...5R,2021ApJ...915L..35K,2020ApJ...902L..26F,2021MNRAS.505..339M,2020arXiv201009765S,2021ApJ...920..128A} (though see also Refs.~\cite{2020ApJ...904L..26F,2022ApJ...926...34E}, which find that these events may in fact be consistent with the population), and GW190814~\cite{2020ApJ...896L..44A,2021MNRAS.502.2049L,2021ApJ...908..194T,2021MNRAS.500.1817L}. The presence of hierarchical mergers in binary BH populations is crucially dependent on the escape speeds of dynamical host environments (e.g., young star clusters, globular clusters, and nuclear star clusters~\cite{2021MNRAS.505..339M}) and the magnitudes of gravitational recoils received due to the anisotropic emission of GWs~\cite{2007PhRvL..98w1101G,2007ApJ...659L...5C,2018PhRvD..97j4049G}.

Our DNN population model learns from simple simulations of clusterlike environments~\cite{2019PhRvD.100d1301G}, which account for the retention and ejection of merger remnants due to GW kicks. We model the \emph{joint} distribution of four source parameters ---two masses and two effective spins, which present identifiable features due to the influence of higher-generation BHs~\cite{2021PhRvD.104h4002B,2021ApJ...915...56G}--- and six population-level (hyper) parameters. These hyperparameters control the population properties of first-generation BHs born in stellar collapse, the binary pairing process, and host escape speeds. We also train a DNN to predict the fractional contributions of the population-dependent first-, mixed-, second-, and higher-generation binary BHs.

\begin{figure*}
\includegraphics[width=1\textwidth]{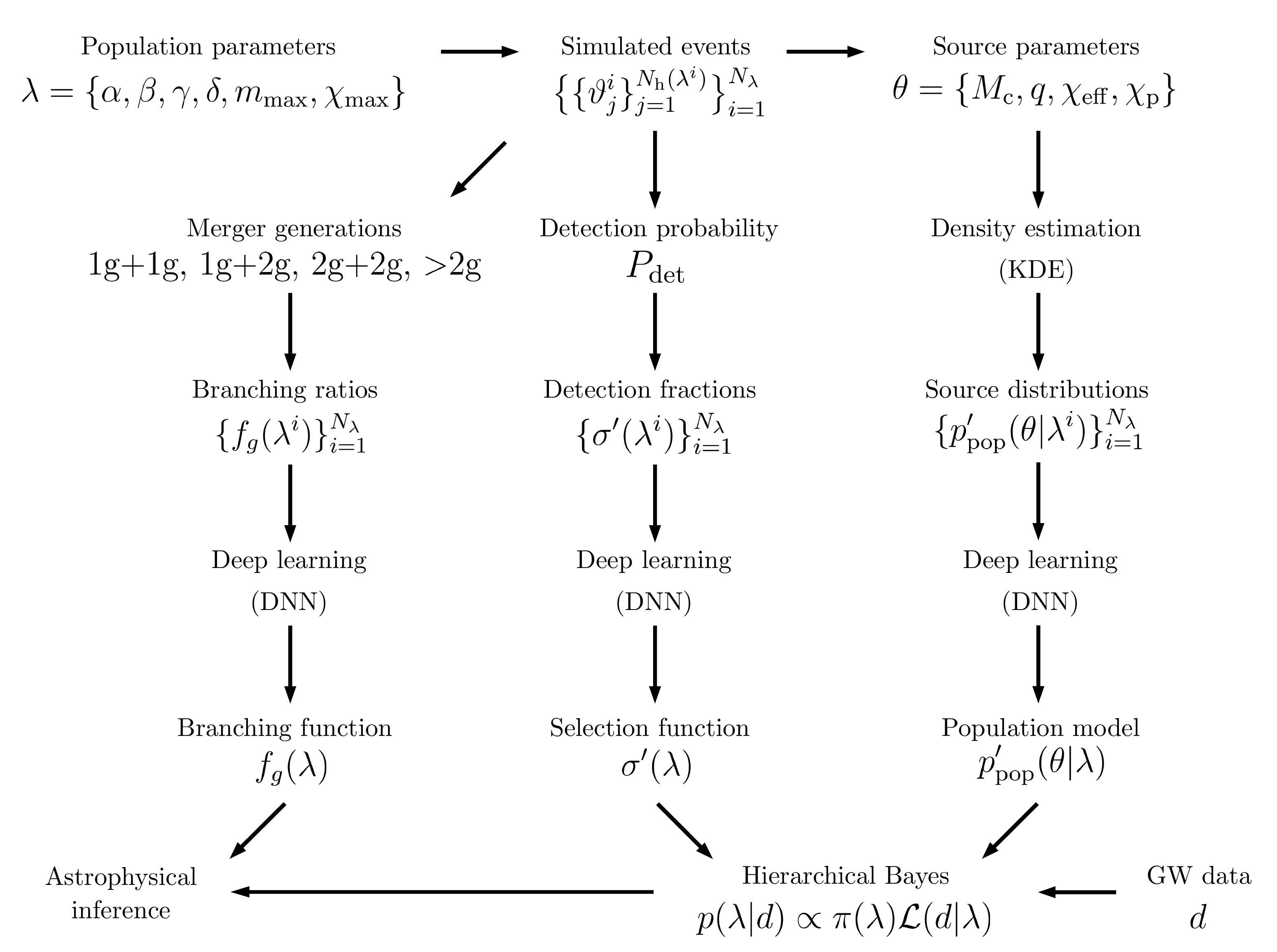}
\caption{Schematic diagram of our population modeling and inference procedure. Arrows indicate information that is passed from one element to another, and elements that occur at the same stage of the pipeline are grouped into rows. The first row represents simulations of binary BH mergers, while the second lists postprocessing applied to the simulated data. We leverage deep learning, shown in the third row, by constructing DNNs to act as functional emulators for key ingredients of GW population inference, indicated within the fourth row. In the final row, the deep-learned selection function and population model are combined with data from GW catalogs to feed into a hierarchical Bayesian inference which, along with a third DNN to predict branching fractions between subpopulations, is used to make conclusions about the underlying distribution of merging stellar-mass binary BHs.}
\label{fig:flowchart}
\end{figure*}

We illustrate our procedure schematically in Fig.~\ref{fig:flowchart}, in which each element represents a single modeling process, arrows direct the one-way flow of information between them, and rows group distinct stages of our pipeline. The first row represents simulations, controlled by population-level parameters, of binary BH mergers characterized by a complete set of source-level parameters that are condensed into those we model. In the second, row we list the postprocessing performed on the simulated data. For each simulation we construct the joint probability density of modeled source-level parameters conditioned on the population-level parameters, the expected fractional number of detectable sources, and the relative contributions from each hierarchical merger generation to the total population. We transform these discrete sets of evaluations into continuous functions using deep learning, as seen in the third row. These DNN functional emulators, listed in the fourth row, are employed in conjunction with data from the GW events detected to date to perform a hierarchical Bayesian inference and ultimately constrain the population of merging stellar-mass binary BHs, as illustrated in the final row. Each ingredient and the relevant symbols are defined throughout the paper.

In Sec.~\ref{sec:populations}, we describe our simple approach to generating sets of simulated hierarchical merger distributions. We lay out the statistical tools of population inference (Sec.~\ref{sec:bayes}), as well as our aforementioned use of DNNs to estimate population models (Sec.~\ref{sec:ppop}), selection biases (Sec.~\ref{sec:sigma}), and population-dependent branching fractions (Sec.~\ref{sec:fractions}). Our deep-learning-enhanced statistical pipeline is validated with mock GW catalogs in Sec.~\ref{sec:mock}. In Sec.~\ref{sec:gwtc-3}, we report the results of our inference on the latest catalog of GW events, discussing the astrophysical implications and comparing to recent related works. We finish with a summary of future extensions to our work in Sec.~\ref{sec:summary} and concluding remarks in Sec.~\ref{sec:conclusions}. The GW events that are included in our analysis and their source parameters are enumerated in Appendix~\ref{app:events}. The inference pipeline established here highlights advancements at the intersection of GW astronomy with statistical analysis and deep learning, and readily accommodates more realistic astrophysical simulations such as binary population synthesis.

\section{Hierarchical merger populations}
\label{sec:populations}

We model the retention and ejection of merger remnants in a ``cluster'', which here simply refers to a collection of BHs in an environment with constant escape speed $\vesc$. We use the setup described in Ref.~\cite{2019PhRvD.100d1301G} (see Refs.~\cite{2020PhRvL.125j1103G,2021ApJ...915...56G} for additional applications). Our model depends on six population parameters, $\lambda\coloneqq\{\alpha,\beta,\gamma,\delta,m_{\rm max},\chi_{\rm max}\}$. These are reported in Table~\ref{tab:limits} and described below. In particular, the quantities $\gamma$, $m_{\rm max}$, and $\chi_{\rm max}$ parametrize the population of first-generation (1g) BHs, while the quantities $\alpha$, $\beta$, and $\delta$ parametrize the pairing and merger process.

This setup is an excellent test bed for our deep-learning explorations because these simulations are not computationally intensive (thus allowing us to explore different DNN architectures) while at the same time providing a binary BH population that ultimately is not parametric (thus making our approach essential).

\subsection{Simulation design}

We generate $\Npop=1000$ sets of population parameters $\lambda$ using Latin hypercube sampling to efficiently cover the higher-dimensional space~\cite{10.2307/1268522,2018PhRvD..98h3017T}. With this design, the hyperparameter space (that is, the space of population-level parameters) is split into $\Npop$ equally probable subintervals in each dimension. From the $\Npop^6$ possible choices, a total of $\Npop$ unique coordinates are drawn such that, for each of the six dimensions, only one of the $\Npop$ subintervals is selected. In general, there are multiple possible realizations of this random draw; we choose to maximize the minimum distance between points, whose values are chosen as the centers of the intervals. Our simulation design is generated with \textsc{pydoe}\footnote{\href{https://pythonhosted.org/pyDOE}{pythonhosted.org/pyDOE}}.

\begin{table}[t]
\label{tab:limits}
\centering
\renewcommand{\arraystretch}{1.3}
\setlength{\tabcolsep}{5.5pt}
\begin{tabular}{c|c|c|c}
& Parameter & Symbol & Range \\
\hline \hline
\multirow{6}*{\rotatebox{90}{Population, $\lambda$}}
& Primary pairing slope & $\alpha$ & $[-10,10]$ \\
& Secondary pairing slope &$\beta$ & $[-10,10]$ \\
& 1g mass slope &$\gamma$ & $[-10,10]$ \\
& Escape-speed slope &$\delta$ & $[-10,10]$ \\
& Maximum 1g mass &$\mmax$ & $[30\Msun,100\Msun]$ \\
& Maximum 1g spin &$\chimax$ & $[0,1]$\\
\hline
\multirow{4}*{\rotatebox{90}{Source, $\theta$}}
& Source-frame chirp mass & $\Mchirp$ & $[5\Msun,105\Msun]$\\
& Mass ratio & $q$ & $[0,1]$ \\
& Effective aligned spin & $\chieff$ & $[-1,1]$ \\
& Effective precessing spin & $\chip$ & $[0,2]$
\end{tabular}
\caption{Parameters in our model of hierarchical binary BH merger populations, the symbols we use to identify them, and their bounds. The population parameters $\lambda=\{\alpha,\beta,\gamma,\delta,\mmax,\chimax\}$ determine the shape of the distribution of first-generation BHs and the properties of the host cluster that can lead to repeated mergers. The bounds on the power-law indices are broad such that the range of training simulations can incorporate more restrictive prior bounds. The source parameters $\theta=\{\Mchirp,q,\chieff,\chip\}$ are measured by LIGO/Virgo when detecting individual GW events. The bounds on chirp mass encompass the extrema of the GW catalog posteriors and are only used when evaluating the population-level likelihood, as described in Sec.~\ref{sec:bayes}.}
\end{table}

\subsection{First-generation black holes}

Each cluster is seeded with $\NBH=5000$ BHs (this number is chosen to ensure  convergence of the resulting merger distributions; see Ref.~\cite{2019PhRvD.100d1301G}). Their masses $\moneg$ are drawn from a simple, truncated, power-law distribution
\begin{align}
p(\moneg | \gamma, \mmax) \propto
\begin{cases}
\moneg^\gamma &\mathrm{if}\quad 5\Msun<\moneg<\mmax \,, \\
0 &\mathrm{otherwise} \,,
\end{cases}
\end{align}
with slope $\gamma\in[-10,10]$, maximum cutoff $\mmax\in[30,100]\Msun$, and a fixed lower boundary of $5\Msun$ (thus only describing black holes and not neutron stars). Pair-instability~\cite{2003ApJ...591..288H} and pulsation pair-instability supernovae (PISN)~\cite{2007Natur.450..390W} prevent the formation of stellar-mass BHs between about 50 and $120\Msun$~\cite{2016A&A...594A..97B,2019ApJ...887...53F, 2021ApJ...912L..31W}. This prediction is supported by current GW observations, which point to a decrease of the merger rate at those masses~\cite{2021arXiv211103634T}. The precise details of the pair-instability mass gap are uncertain and depend on poorly constrained stellar-physics processes such as the nuclear reaction rates~\cite{2019ApJ...887...53F,2020ApJ...902L..36F,2021ApJ...912L..31W}, rotation~\cite{2020A&A...640L..18M,2021ApJ...912L..31W}, accretion~\cite{2021ApJ...908...59R,2020ApJ...903L..21S,2019A&A...632L...8R,2021MNRAS.501.1413N,2020ApJ...897..100V}, winds~\cite{2020ApJ...890..113B,2021MNRAS.504..146V}, envelope retention~\cite{2021ApJ...910...30T,2021MNRAS.502L..40F,2021MNRAS.501L..49K} and dredge-up episodes~\cite{2021MNRAS.501.4514C}. We thus allow for a broad range of values of $\mmax$ and aim to infer it from the GW data.

The BH spin directions are drawn from an isotropic distribution, as expected in dynamical environments. The dimensionless spin magnitudes are uniformly within $[0,\chimax]$, where the maximum natal spin is $\chimax\in(0,1)$. The largest spin formed from stellar collapse is uncertain and difficult to model; see Refs.~\cite{2019ApJ...881L...1F,2020A&A...636A.104B}. The spin model we use for first-generation BHs is therefore not necessarily physically well-motivated but is used for illustrative purposes.

\subsection{Repeated mergers}
\label{sec:repeated}

At each hyperparameter coordinate, we simulate $\Ncl=500$ clusters with escape speeds $\vesc$ drawn according to
\begin{align}
p(\vesc | \delta) \propto
\begin{cases}
\vesc^\delta &\mathrm{if}\quad 0~\kmps < \vesc < 500~\kmps \,, \\
0 &\mathrm{otherwise} \,,
\end{cases}
\end{align}
where $\delta\in[-10,10]$.
Large positive (negative) values of $\delta$ give escape-speed distributions skewed towards the maximum (minimum) value of $\vesc$. For context, the escape speed of a typical globular cluster is 10--100~$\kmps$, while those of nuclear star clusters are up to an order of magnitude larger \cite{2002ApJ...568L..23G,2004ApJ...607L...9M, 2016ApJ...831..187A}; we take an upper limit of 500~$\kmps$ to accommodate these larger escape speeds. Cases with large, negative $\delta$ essentially describe isolated stellar evolution, where repeated mergers do not take place (though we always assume isotropically distributed spins, not partial alignment as expected in isolated binary evolution \cite{2000ApJ...541..319K,2018PhRvD..98h4036G,2021PhRvD.103f3032S}). On the other hand, $\delta=0$ corresponds to a flat $\vesc$ distribution, favoring all environments equally.

The key ingredient in our populations is the presence of so-called ``higher-generation'' BHs that have undergone multiple mergers due to remnant retention in the host cluster. We form circular binary systems by selectively pairing cluster members according to
\begin{align}
\label{eq:pairing}
p(m_1 | \alpha) \propto m_1^\alpha \, ,  \quad p(m_2 | \beta, m_1)\propto
\begin{cases}
m_2^\beta &\mathrm{if}\quad m_2 \leq m_1 \, , \\
0 & \mathrm{otherwise} \, ,
\end{cases}
\end{align}
where $m_1 \geqslant m_2$ are the component BH masses. As for the other power-law indices, we again take $\alpha,\beta\in[-10,10]$; this broad range is taken in each case so that the simulated populations encompass the prior bounds used later in our statistical inference of Sec.~\ref{sec:bayes}. One by one, BH pairs are drawn from the collection according to Eq.~(\ref{eq:pairing}) and the properties of their merger remnants are estimated (assuming a uniform sampling of the orbital phase) with the implementation of Ref.~\cite{2016PhRvD..93l4066G}, which collects various numerical relativity fitting formulas~\cite{2012ApJ...758...63B,2009ApJ...704L..40B,2016ApJ...825L..19H,2007ApJ...659L...5C,2007PhRvL..98w1101G,2008PhRvD..77d4028L,2012PhRvD..85h4015L,2013PhRvD..87h4027L}. Upon merging, the remnant BHs receive a gravitational recoil~\cite{1973ApJ...183..657B,1983MNRAS.203.1049F}. If the magnitude $\vkick$ of this kick velocity exceeds the escape speed of the host cluster, i.e., $\vkick>\vesc$, the remnant BH is removed and does not merge again. Otherwise, it remains inside the cluster and can undergo subsequent mergers. The estimated remnant mass and spin magnitude are retained, while the spin directions are resampled isotropically. This pairing, merger and ejection procedure is iterated until a single BH remains.

For each merger we record the source parameters $\theta\coloneqq\{\Mchirp,q,\chieff,\chip\}$. In particular, $\Mchirp=(m_1m_2)^{3/5}/(m_1+m_2)^{1/5}$ is the chirp mass, $q=m_2/m_1\leq1$ is the mass ratio, $\chieff\in[-1,1]$ is the effective aligned spin~\cite{2008PhRvD..78d4021R}, and $\chip\in[0,2]$ is a suitable parameter encoding the dominant effect of orbital-plane precession; for the latter, we use the augmented definition of Ref.~\cite{2021PhRvD.103f4067G} which consistently averages over the precessional motion including effects from both component spins. While this definition of $\chip$ is still a frequency-dependent quantity over the inspiral timescale, recent work has shown that the influence of the GW reference frequency at the population level is currently subdominant compared to measurement errors~\cite{2022PhRvD.105b4076M}. In the simulated populations we measure $\chip$ at the reference frequency of 20~Hz.

Additionally, we record whether each merger is that of two first-generation BHs (1g+1g) that produces a second-generation (2g) remnant, a first- and second-generation BH (1g+2g), or two second-generation BHs (2g+2g), or whether it contains a component BH of higher generation (>2g). From these, we compute the fraction of mergers in each generation: $f_{\rm 1g+1g}$, $f_{\rm 1g+2g}$, $f_{\rm 2g+2g}$, and $f_{\rm >2g}=1-f_{\rm 1g+1g}-f_{\rm 1g+2g}-f_{\rm 2g+2g}$.

\subsection{Cosmic placement}
\label{sec:cosmic}

The distribution of sources is assumed to be isotropic over the sky, inclination and polarization angle.
We do not infer the redshift distribution of BH binaries but consider it fixed, i.e., independent of the hyperparameters $\lambda$. Each merger is placed at a redshift $z$ according to a distribution that is uniform in comoving volume $V_\mathrm{c}$ and source-frame time, i.e.,
\begin{align}
p(z) \propto \frac{1}{1+z} \frac{{\rm d}V_\mathrm{c}}{{\rm d}z} \,.
\end{align}
An immediate generalization of this work would include taking into account the longer assembly times of higher-generation binaries (e.g., Ref.~\cite{2017PhRvD..95l4046G}) via their redshift distribution. This can be implemented with an additional hyperparameter and will be tackled in future work.

Ostensibly, $z\in(0,\infty)$, but in practice, there is a detector-dependent horizon, $\zmax$, beyond which binary BH mergers are not observable. To find a conservative $\zmax$, we consider a series of binaries with aligned maximal spins, equal masses, and optimal orientation with respect to a single detector (overhead and face on). These are the loudest sources for a given total mass and redshift. We compute signal-to-noise ratios (SNRs) as described in Sec.~\ref{sec:sigma} and find that the entire mass range becomes subthreshold above an upper bound $\zmax=2.3$, which we thus take as the maximum of the redshift distribution (in agreement with Appendix~E of Ref.~\cite{2021ApJ...913L...7A}).

\subsection{Resulting populations}

\begin{figure}
\includegraphics[width=1\columnwidth]{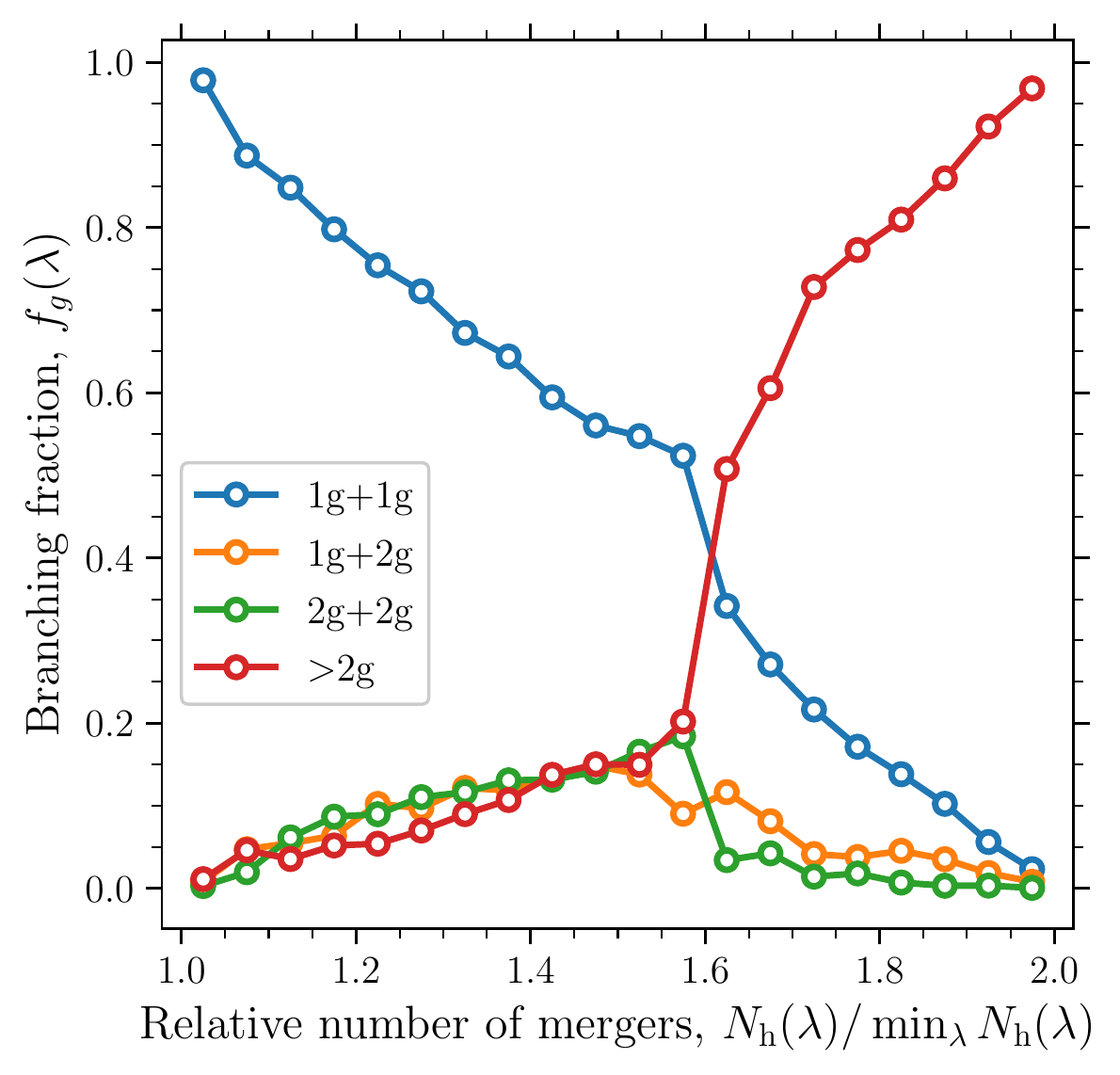}
\caption{Fraction of mergers in our simulations from each binary generation as a function of the total number of mergers. The simulations are separated into bins equally spaced in the total number of mergers and the bin-averaged branching fraction of each binary generation ---1g+1g (blue), 1g+2g (orange), 2g+2g (green), and higher generations (red)--- is plotted. At the lower (upper) end, simulations are dominated by mergers between first- (higher-) generation BHs.}
\label{fig:generations}
\end{figure}

\begin{figure*}
\includegraphics[width=1\textwidth]{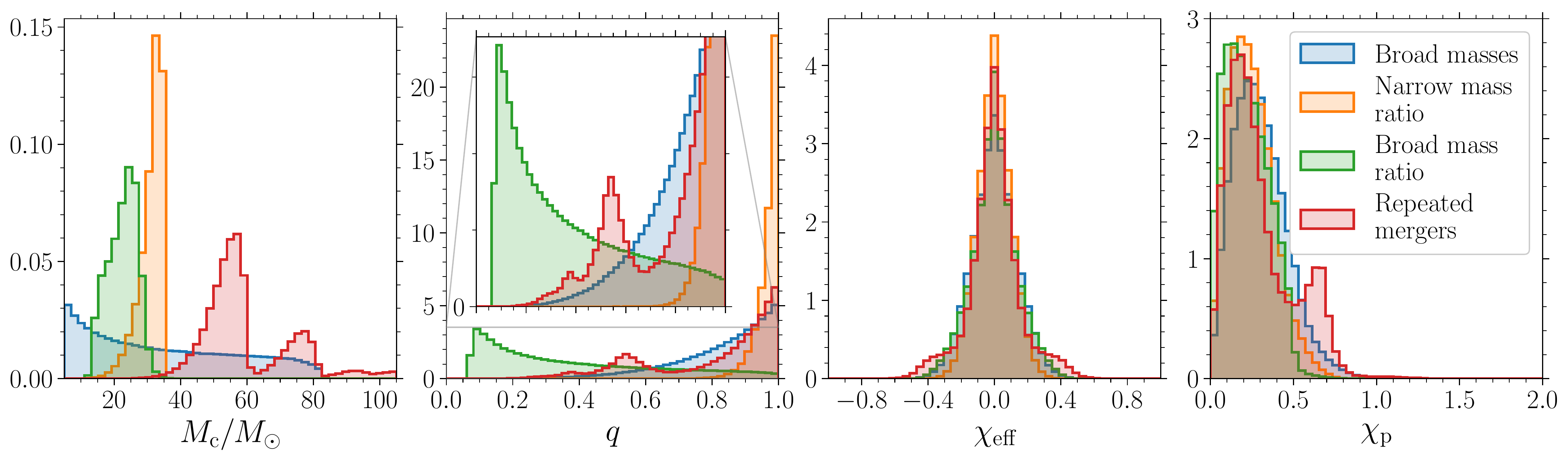}
\caption{Example marginal distributions of chirp mass $\Mchirp$, mass ratio $q$, effective aligned spin $\chieff$, and our precession parameter $\chip$ for different population parameters $\lambda=\{\alpha,\beta,\gamma,\delta,\mmax,\chimax\}$. We select four of our simulations to illustrate different features of the resulting binary BH distributions. In blue, we show broad masses, $\lambda=\{-1.7,1.7,-0.5,-3.4,96\Msun,0.57\}$; this set of hyperparameters results in a large range of binary BH masses due to a high maximum first-generation mass, broad mass function, and broad binary pairing probabilities. In orange, we show narrow mass ratios, $\lambda=\{-8.8,8.3,6.8,-4.1,40\Msun,0.43\}$; binaries are preferentially selected with equal component masses due to pairing probabilities that favor the lightest primary BHs and heaviest secondary BHs. In green, we show broad mass ratios, $\lambda=\{9.2,-9.8,-0.5,-4.0,74\Msun,0.50\}$; the pairing slopes produce binaries with the heaviest primaries and lightest secondaries, resulting in an extended range of mass ratios peaking at lower values. In red, we show repeated mergers, $\lambda=\{4.1,3.1,5.5,5.1,70\Msun,0.37\}$; clusters are preferentially generated with large escape speeds, boosting the presence of repeated mergers, which appear as multiple narrow peaks in the mass distributions. The lower maximum natal spin causes a narrow peak around $\chieff=0$; the occurrence of repeated mergers extends the tails of the $\chieff$ distribution and creates a secondary peak in the $\chip$ distribution.}
\label{fig:popexamples}
\end{figure*}

The above prescription allows us to transform a simple phenomenological description of first-generation BH populations into a complex numerical distribution containing hierarchical mergers. The combined set of hyperparameters $\lambda=\{\alpha,\beta,\gamma,\delta,\mmax,\chimax\}$ are very interdependent, and changes in their values cause large variations in the distributions of source parameters $\theta=\{\Mchirp,q,\chieff,\chip\}$. The total set of simulated events is $\big\{\{\theta_j^i\}_{j=1}^{\Nh(\lambda^i)}\big\}_{i=1}^{\Npop}$, where $\Nh(\lambda^i)$ is the number of mergers occurring in the simulation with hyperparameter coordinate $\lambda^i$. The total number of mergers occurring at a given hyperparameter coordinate depends on the distribution of escape speeds, determined by $\delta$. For the numerical setup adopted here, it ranges from $\min_\lambda\Nh(\lambda)=\Ncl\NBH/2=1.25\times10^6$ (when each remnant BH is ejected so only first-generation mergers occur) to $\max_\lambda\Nh(\lambda)=\Ncl\NBH-1\approx 2.5\times10^6$ when BHs are repeatedly paired with the same single retained remnant, i.e., a cluster catastrophe) and the upper range is populated by simulations with larger numbers of repeated mergers. This is demonstrated in Fig.~\ref{fig:generations}, where we plot the branching fractions of different merger generations as a function of the total number of mergers. Four representative cases among the set of $\Npop=1000$ simulations we performed are illustrated in Fig.~\ref{fig:popexamples} and labeled based on the qualitative properties of the resulting source distributions: broad masses, narrow mass ratio, broad mass ratio, and repeated mergers.

If clusters are preferentially formed with larger escape speeds, many remnants are retained and proceed to take part in hierarchical mergers, leading to multiple modes in the mass distributions. This is the case for the red curves (repeated mergers) in Fig.~\ref{fig:popexamples}, where $\delta=5.1$. Since the sharp initial mass function (IMF) ($\gamma=5.5$) forms first-generation BHs with masses that are all very close to the maximum $\mmax=70\Msun$, hierarchical mergers appear as distinct peaks in the mass distributions. The first generation of mergers has $m_1\approx m_2\approx\mmax$, giving $\Mchirp\approx50\Msun$. Cross-generational mergers also occur. For example, there is a 1g+2g peak; the peak does not occur at $q=0.5$ because a fraction $1-\epsilon\approx5\%$ of mass is lost via GWs~\cite{2005PhRvL..95l1101P} such that second-generation BHs have mass of approximately $2\epsilon\mmax$, implying $q=1/(2\epsilon)\approx0.53$ and $\Mchirp\approx80\Msun$. Similarly, for a 1g+3g merger, one has $q\approx1/[\epsilon (2\epsilon+1)]\approx0.36$, which explains the third peak observed in the red curves of Fig.~\ref{fig:popexamples}.

When more first-generation BHs are born with large spins, set by $\chimax$, fewer second-generation mergers occur due to the larger imparted recoils~\cite{2018PhRvD..97j4049G}. On the other hand, if natal spins are small and repeated mergers do occur, the distribution of effective spins features a sharp peak at $\chieff=0$ from first-generation mergers as well as extended tails from high-generation mergers, as is the case for the red curve in the third column of Fig.~\ref{fig:popexamples}. The $\chieff$ distributions are always symmetric about zero due to the assumption of spin isotropy. For the 1g+2g populations, the 2g BH spin is approximately $0.7$ \cite{2021NatAs...5..749G} and, because in this case $\chimax=0.37$, is typically higher than the spin of the 1g BH. In this limit, one has $\chip\approx \sqrt{0.7^2 - 4\chieff^2}\approx 0.7$ \cite{2021PhRvD.104h4002B}, thus explaining the secondary peak in the $\chip$ distribution.

Whether higher-generation BHs pair with other BHs of equal generation or form cross-generational binaries (e.g., 1g+2g) depends on the pairing slopes $\alpha$ and $\beta$. If $\alpha,\beta,\gamma\approx0$, then the first-generation mass distribution is broad and binary components are selected with uniform probabilities leading to an extended range of mass ratios, as seen in the blue ``broad masses'' curves of Fig.~\ref{fig:popexamples}. If $\alpha,\beta\gg 0$ ($\alpha,\beta\ll 0$), the heaviest (lightest) BHs are preferentially selected for both binary components, leading to a heavier (lighter) first generation of approximately equal-mass binaries. If $\alpha \ll 0$ and $\beta \gg 0$ ($\alpha\gg 0$ and $\beta\ll 0$), then the lightest (heaviest) primaries and heaviest (lightest) secondaries are paired, leading to mass-ratio distributions that are sharply peaked at unity (broad and peaked at lower values), as seen in the orange ``narrow mass ratio'' (green ``broad mass ratio'') curve of Fig.~\ref{fig:popexamples}. In the case of narrow mass ratios, given the maximum first-generation mass $\mmax\approx40\Msun$ and since $q\approx1$, the chirp mass peak is located at $\Mchirp\approx35\Msun$.

\section{Deep-learning-enhanced population inference}
\label{sec:neural}

Although challenging to treat, a set of highly degenerate hyperparameters makes our simplified population model  indicative of realistic applications where GW events are modeled using, e.g., stellar population-synthesis codes. As shown below, deep learning is the ideal tool for such a complex scenario. First, we review the key ingredients that enter hierarchical Bayesian inference to recover the hyperparameters of a population model given GW data from a catalog of mergers (Sec.~\ref{sec:bayes}). We then present our method to model the population prior (Sec.~\ref{sec:ppop}) and selection effects (Sec.~\ref{sec:sigma}) using deep learning. We use similar techniques to model the branching fractions between different merger generations (Sec.~\ref{sec:fractions}).

\subsection{Hierarchical Bayesian inference}
\label{sec:bayes}

Given observational data $d=\{d_n\}_{n=1}^{\Nobs}$ of $\Nobs$ independent GW events and a population model $\ppop$, our goal is to infer the parameters $\lambda$ governing the shape of the underlying distribution of binary BH source parameters $\vartheta$.  The distribution of predicted sources is given by $dN/d\vartheta = N \ppop(\vartheta|\lambda)$, where $\int \ppop(\vartheta|\lambda) \dd\vartheta = 1$ is normalized over the entire domain of source parameters. Here, we have separated the parameters that determine the shape $\lambda$ and overall scale $N$ of the population. Note also that $\vartheta\supset\theta$ is a superset of the source parameters $\theta=\{\Mchirp,q,\chieff,\chip\}$ we wish to model and additionally contains, e.g., redshift, sky location, inclination, etc. In our case, the extra parameters do not depend on the population-level parameters, such that $\ppop(\vartheta|\lambda) = \ppop(\theta|\lambda) \ppop(\thetab)$, where $\thetab \coloneqq \vartheta\ssetminus\theta$, and the normalization over $\vartheta$ implies that $\int \ppop(\theta|\lambda) \dd\theta = \int \ppop(\thetab) \dd\thetab = 1$.

\subsubsection{Selection effects}
\label{sec:selection}

We wish to infer the observable population of merging BHs from the small subset that we have observed. This requires modeling detector selection effects. The expected number of detectable sources for a given population model is
\begin{align}
\Ndet(\lambda) \coloneqq \iint \frac{d^2N}{d\vartheta dt} \Pdet(\vartheta,t) \dd\vartheta \dd t
\, ,
\label{eq:Ndet}
\end{align}
where $\Pdet(\vartheta,t)$ is the probability that a binary BH with source parameters $\vartheta$ is detectable at an observation time $t$ (this is a probability and not a probability density, as distinguished by the use of capital $P$). We describe the calculation of $\Pdet$ in Sec.~\ref{sec:pdet}. The detection efficiency ---i.e., the fraction of detectable events given the population model--- is given by
\begin{align}
\sigma(\lambda) \coloneqq \frac{\Ndet(\lambda)}{N(\lambda)}
= \frac{1}{T} \iint \ppop(\vartheta|\lambda) \Pdet(\vartheta,t) \dd\vartheta \dd t
\, ,
\end{align}
where we have assumed equally likely arrival times of GW signals at the detectors over the observing period of duration $T$. The integral over time indicates that we must account for the detector duty cycle and change in sensitivity over observing epochs. We approximate the sensitivity as constant within each observation period: the combined first and second run (O1+O2) and the third run (O3). The corresponding two-detector observing periods are $T_\mathrm{O1+O2}\approx166$ days~\cite{2016PhRvX...6d1015A,2019PhRvX...9c1040A} and $T_\mathrm{O3}\approx275$ days~\cite{2021PhRvX..11b1053A,2021arXiv211103606T}, respectively. With this approximation, the time integral reduces to the weighted average~\cite{2022hgwa.bookE..45V}
\begin{align}
\sigma(\lambda) = \sum_r \frac{T_r}{T} \int \ppop(\vartheta|\lambda) \Pdet(\vartheta, r) \dd\vartheta
\, ,
\label{eq:sigma}
\end{align}
where $r\in\{\mathrm{O1+O2},\mathrm{O3}\}$ indicates the observing run and corresponding instrument sensitivity, and $T=T_\mathrm{O1+O2}+T_\mathrm{O3}$ is the total observing time.

\subsubsection{Population likelihood}
\label{sec:likelihood}

Including selection effects, the likelihood of the GW data $\{d_n\}_{n=1}^{\Nobs}$ given the parameters $\lambda$ of our population model is (see, e.g., Refs.~~\cite{2019MNRAS.486.1086M,2022hgwa.bookE..45V})
\begin{align}
\L(d|\lambda,N) = e^{-\Ndet(\lambda)} \prod_{n=1}^{\Nobs} \int \frac{dN}{d\vartheta_n} \L(d_n|\vartheta_n) \dd\vartheta_n
\, .
\label{eq:poplkl}
\end{align}
The single-event likelihoods $\L(d_n|\vartheta_n)$ may be rewritten using Bayes's theorem as $\L(d_n|\vartheta_n) \propto p(\vartheta_n|d_n)/\pi(\vartheta_n)$, where $p(\vartheta_n|d_n)$ is the posterior on the source parameters for the $n$th event as inferred by parameter estimation, and $\pi(\vartheta_n)$ is the prior used in that analysis (which may differ event to event). Using Bayes's theorem again, the posterior distribution of population parameters is given by
\begin{align}
p(\lambda|d) \propto \pi(\lambda) \prod_{n=1}^{\Nobs} \frac{1}{\sigma(\lambda)} \int \frac {\ppop(\vartheta_n|\lambda) p(\vartheta_n|d_n)} {\pi(\vartheta_n)} \dd\vartheta_n
\, ,
\label{eq:poppos}
\end{align}
where we have marginalized over the rate parameter $N$ with a scale-independent prior $\pi(N) \propto 1/N$~\cite{2018ApJ...863L..41F} and $\pi(\lambda)$ is the prior over the remaining shape parameters. The priors on the parameters $\mmax$ and $\chimax$ are uniform over the ranges listed in Table~\ref{tab:limits}. The priors of the power-law indices $\alpha$, $\beta$, $\gamma$, and $\delta$ are uniform over $[-8,8]$; these prior bounds lie within the training data range and we checked that resulting posteriors are robust to more stringent constraints.

\subsubsection{Factorization of the observed volume}

While the integrals in Eqs.~(\ref{eq:sigma}) and (\ref{eq:poppos}) are formally defined over the entire domain of source parameters, in practice they can be safely performed within the observable volume $\Vh \coloneqq \{\vartheta : z<\zmax=2.3\}$, beyond which the detection probability is zero, as discussed in Sec.~\ref{sec:cosmic}. Even if $\ppop$ models the binary BH population outside of $\Vh$, as it appears in both the numerator and through $\sigma(\lambda)$ in the denominator of Eq.~(\ref{eq:poppos}), one can safely assume that $\int_{\Vh} \ppop(\vartheta|\lambda) \dd\vartheta = 1$.

Since it will be useful in Sec.~\ref{sec:ppop}, we define the observed volume $\Vp \coloneqq \{\vartheta : p(\vartheta|d_n)>0 \,\forall n\} \subset \Vh$ as the subset of the observable volume beyond which all single-event posteriors $p(\vartheta_n|d_n)$ vanish. For the events considered in this work (see Sec.~\ref{sec:events} and Appendix~\ref{app:events}), we find that $\Vp$ corresponds to $\Mchirp\in[5,105]\Msun$, while for $q$, $\chieff$, and $\chip$, we maintain their natural bounded domains ($[0,1]$, $[-1,1]$, and $[0,2]$, respectively). It will also be useful to define the population prior of our modeled parameters $\theta$ normalized over the observed volume,
\begin{align}
\ppop'(\theta|\lambda) \coloneqq \frac {\ppop(\theta|\lambda)} {\int_{\Vp} \ppop(\theta|\lambda) \dd\theta}
\, .
\end{align}
Since $\ppop$ is normalized over $\Vh$, we can write this extra normalization factor as
\begin{align}
\int_{\Vp} \ppop(\theta|\lambda) \dd\theta = \frac{\Np(\lambda)}{\Nh(\lambda)} \leq 1
\, ,
\end{align}
where $\Nh(\lambda)$ is the number of mergers occurring within the horizon volume $\Vh$ and $\Np(\lambda)$ is the number of mergers occurring within the subset $\Vp\subset\Vh$, given population parameters $\lambda$. For convenience, we refactor this term into the detection efficiency by defining the selection function
\begin{align}
\label{sigmaprime}
\sigma'(\lambda) \coloneqq \frac{\Nh(\lambda)}{\Np(\lambda)} \sigma(\lambda)
\, .
\end{align}

By separating the source parameters and noting that our population prior and the parameter estimation prior over the unmodeled parameters are equal, i.e., $\ppop(\thetab)/\pi(\thetab)\equiv1$, the hyperposterior in Eq.~(\ref{eq:poppos}) may be written as
\begin{align}
p(\lambda|d) \propto \pi(\lambda) \prod_{n=1}^{\Nobs} \int_{\Vp} \frac {\ppop'(\theta_n|\lambda) p(\theta_n,\bar\theta_n|d_n)} {\sigma'(\lambda) \pi(\theta_n|\bar\theta_n)} \dd\theta_n\dd\bar\theta_n
\, .
\label{eq:poppos'}
\end{align}
Since the parameter estimation prior is placed on detector-frame masses, we must convert the prior on detector-frame chirp mass $\Mchirp^\mathrm{det}$ to the source frame. In particular, we have $\pi(\theta|z) = \pi(\Mchirp^\mathrm{det},q,\chieff,\chip|z) |\partial\Mchirp^\mathrm{det}/\partial\Mchirp|$. Since the Jacobian is $|\partial\Mchirp^\mathrm{det}/\partial\Mchirp| = 1+z$ and the prior on detector-frame masses is independent of the prior on redshift for the parameter estimation results we use below, we have $\pi(\theta|z) = \pi(\Mchirp^\mathrm{det},q,\chieff,\chip) (1+z)$.

\subsubsection{Event samples}
\label{sec:events}

Given discrete samples $\{\{\vartheta_{n,k}\}_{k=1}^{S_n} \sim p(\vartheta_n|d_n)\}_{n=1}^{\Nobs}$ from the individual event posteriors, where $S_n$ is the number of samples in the posterior for the $n$th event, and since these samples lie, by definition, within the posterior volume $\Vp$, Eq.~(\ref{eq:poppos'}) can be evaluated with the Monte Carlo summation
\begin{align}
p(\lambda|d) \propto \pi(\lambda) \prod_{n=1}^{\Nobs} \frac{1}{\sigma'(\lambda)S_n} \sum_{k=1}^{S_n} \frac {\ppop'(\theta_{n,k}|\lambda)} {\pi(\theta_{n,k}|z_{n,k})}
\, .
\label{eq:poppos samples}
\end{align}
For each event we draw prior samples for $\{\Mchirp^\mathrm{det},q,\chieff,\chip\}$ and compute $\pi(\Mchirp^\mathrm{det},q,\chieff,\chip)$ using Gaussian kernel density estimates (KDEs) as implemented in \textsc{scipy}~\cite{2020NatMe..17..261V}, modified to enforce reflective boundary conditions~\cite{1986desd.book.....S}. Each KDE is then evaluated on the single-event posterior samples. Equation~(\ref{eq:poppos samples}) is sampled using \textsc{dynesty}~\cite{2020MNRAS.493.3132S} and \textsc{bilby}~\cite{2019ApJS..241...27A}.

We select the confident binary BH detections made during the first (O1), second (O2) and third (O3) observing runs, employing a threshold minimum false alarm rate (FAR) of less than $1~\mathrm{yr}^{-1}$ across all search analyses. This results in a catalog of $\Nobs=69$ binary BH events. For the events in O1 and O2, we use samples\footnote{\href{https://dcc.ligo.org/LIGO-P2000193/public}{dcc.ligo.org/LIGO-P2000193/public}} from the reanalysis of Ref.~\cite{2020MNRAS.499.3295R} because the precession parameter $\chip$ depends on the azimuthal spin angles whose posteriors were not released in GWTC-1~\cite{2019PhRvX...9c1040A}. For the events in O3, we take the posterior samples combining analyses with waveforms including both precession and higher-order modes as provided by the GWTC-2\footnote{\href{https://www.gw-openscience.org/GWTC-2/}{gw-openscience.org/GWTC-2}}~\cite{2021PhRvX..11b1053A}, GWTC-2.1\footnote{\href{https://www.gw-openscience.org/GWTC-2.1/}{gw-openscience.org/GWTC-2.1}}~\cite{2021arXiv210801045T} (\texttt{PrecessingSpinIMRHM}), and GWTC-3\footnote{\href{https://www.gw-openscience.org/GWTC-3/}{gw-openscience.org/GWTC-3}}~\cite{2021arXiv211103606T} (\texttt{C01:Mixed}) data releases. We list all of the events that enter our analysis in Appendix~\ref{app:events}.

\subsection{Population model}
\label{sec:ppop}

The results of our simulations are lists of binary BH mergers, characterized by source parameters $\theta=\{\Mchirp,q,\chieff,\chip\}$, at each of the $\Npop=1000$ population parameter coordinates, $\lambda=\{\alpha,\beta,\gamma,\delta,\mmax,\chimax\}$. Our approach to modeling the resulting population distribution $\ppop'(\theta|\lambda)$ employs a combination of probability density estimation and regression algorithms.

\subsubsection{Density estimation}
\label{sec:ppop kde}

At each of the hyperparameter locations $\{\lambda^i\}_{i=1}^{\Npop}$, we evaluate the conditional population density $\ppop'(\theta|\lambda^i)$ with a Gaussian KDE. To efficiently evaluate $\ppop'$ with sufficient resolution in the four-dimensional space of source parameters, we use a version of the convolution-based implementation in \textsc{kdepy}~\cite{tommy_odland_2018_2392268}, which we modify to enforce the parameter limits (Table~\ref{tab:limits}) with reflective boundary conditions~\cite{1986desd.book.....S}. With this method, density estimations of multivariate data with millions of samples evaluated on millions of points take seconds on a standard, off-the-shelf machine, compared to hours with standard KDE routines (the evaluation points must, however, lie on a linearly spaced Cartesian grid that bounds the data extrema). Each dimension is individually scaled with bandwidths determined by the Improved Sheather Jones (ISJ) plug-in selection rule~\cite{2010arXiv1011.2602B,10.2307/2345597}. The ISJ algorithm does not make the assumption of normality on the underlying distribution and, as such, is more robust when determining optimal bandwidths for non-Gaussian multimodal distributions. We evaluate each of the $\Npop$ KDEs on a linearly spaced Cartesian grid, including the parameter bounds, with 21 points in each axis.

\subsubsection{Regression with a deep neural network}
\label{sec:ppop nn}

Elucidating the scale of the regression problem, there are $21^4\approx 2\times 10^5$ KDE evaluations estimating $\ppop'(\theta|\lambda)$ over the combined ten-dimensional vectors of source and population parameters $(\theta,\lambda)$ at each of the $\Npop=1000$ hyperparameter locations. While the KDEs approximate the $\Npop$ functions $\{\theta\mapsto\ppop'(\theta|\lambda^i)\}_{i=1}^{\Npop}$, we must also interpolate over the population parameters to find an accurate mapping $(\theta,\lambda)\mapsto\ppop'(\theta|\lambda)$.

To achieve this result, we make use of a fully connected DNN implemented with Google's \textsc{tensorflow} deep-learning library~\cite{abadi2016tensorflow}. The network performs a regression of the KDE values of $\ppop'$ over the space of $(\theta,\lambda)$ coordinates. As a preprocessing step, we normalize all coordinates $(\theta,\lambda)$ to a unit hypercube using the limits given in Table~\ref{tab:limits}, while the values of $\ppop'$ are similarly scaled between zero and their maximum. The input layer has $\mathrm{dim}(\theta)+\mathrm{dim}(\lambda)=10$ neurons, while the output layer has one neuron with enforced non-negativity corresponding to the predicted value of the probability density. Between the input and output layers, the network architecture consists of five hidden layers, each with 128 neurons. We summarize the network architecture in Table~\ref{tab:nn population}. The number of parameters in a given layer is given by the number of weights (equal to the product of the number of neurons with that of the preceding layer) plus the number of biases (equal to the number of neurons).

\begin{table}[]
\label{tab:nn population}
\centering
\renewcommand{\arraystretch}{1.3}
\setlength{\tabcolsep}{5.5pt}
\begin{tabular}{c|c|c|c}
Layer & Neurons & Activation & Parameters \\
\hline \hline
Input & 10 &  ... &0 \\
Dense 1 & 128 & RReLU & 1,408 \\
Dense 2 & 128 & RReLU & 16,512 \\
Dense 3 & 128 & RReLU & 16,512 \\
Dense 4 & 128 & RReLU & 16,512 \\
Dense 5 & 128 & RReLU & 16,512 \\
Output & 1 & Absolute value & 129 \\
\hline
\multicolumn{1}{c}{} & \multicolumn{2}{r|}{Total} & \multicolumn{1}{c}{67,585} \\
\end{tabular}
\caption{Architecture of the DNN that emulates the simulated populations by predicting the conditional density $\ppop'(\theta|\lambda)$ of the source parameters $\theta$ given population-level parameters $\lambda$. Each row represents a single layer of the network and lists the number of neurons in the layer, the activation function of those neurons (RReLU for the hidden layers and absolute value for the final output), and the corresponding number of free parameters.}
\end{table}

We use randomized leaky rectified linear units (RReLUs)~\cite{2015arXiv150500853X}  in each layer. This modifies the standard rectified linear unit (ReLU) activation function, given by $\mathrm{ReLU}(x) \coloneqq \max(0,x)$, in two ways. First, leaky ReLU activation functions are maps $x \mapsto \max(0,x)+\min(0,ax)$, where $a\in[l,u]$ is a parameter fixed to a small number; i.e., the positive region is linear with unit slope while the negative region is linear with slope $a$. Second, the randomized leaky variant RReLU samples $a$ uniformly in $[l,u]$ during training and fixes $a=(l+u)/2$ when making predictions (we keep the default values of $l=1/8$ and $u=1/3$~\cite{2015arXiv150500853X}). Empirically, we find that, among other ReLU variants and nonlinear activations, RReLU gives the best predictive performance while reducing overfitting to the training data.

We split the $\Npop=1000$ simulations into a training data set of 900 runs and a validation set of 100 runs. The validation sample is unseen by the training process except to assess the network performance. The training data input to the network, which is randomly shuffled at each iteration, thus consists of approximately $1.75\times10^8$ values of the ten-dimensional vector $(\theta,\lambda)$ and the corresponding KDE estimates of $\ppop'(\theta|\lambda)$. The network is trained using the Adam optimizer~\cite{2014arXiv1412.6980K}, the mean absolute error (MAE) loss function, a learning rate of $10^{-4}$, and batch size equal to $0.01\%$ of the total number of training data points. Training is performed for $10^4$ epochs on an NVIDIA A100 Tensor Core GPU, taking about four days. With this setup, the number of training epochs is sufficient to ensure convergence of the MAE; the average gradient of the (smoothed) validation MAE over the penultimate 100 epochs is less than $0.1\%$ that of the first 100.

When making predictions with the trained NN, the values are first rescaled from the unit interval to the probability density parameter space. While the predictions are approximately normalized, the network does not enforce unit normalization. Therefore, we estimate normalization factors $\int \ppop'(\theta|\lambda)\dd\theta$ by numerically integrating the predicted distributions.

\begin{figure}
\centering
\includegraphics[width=1\columnwidth]{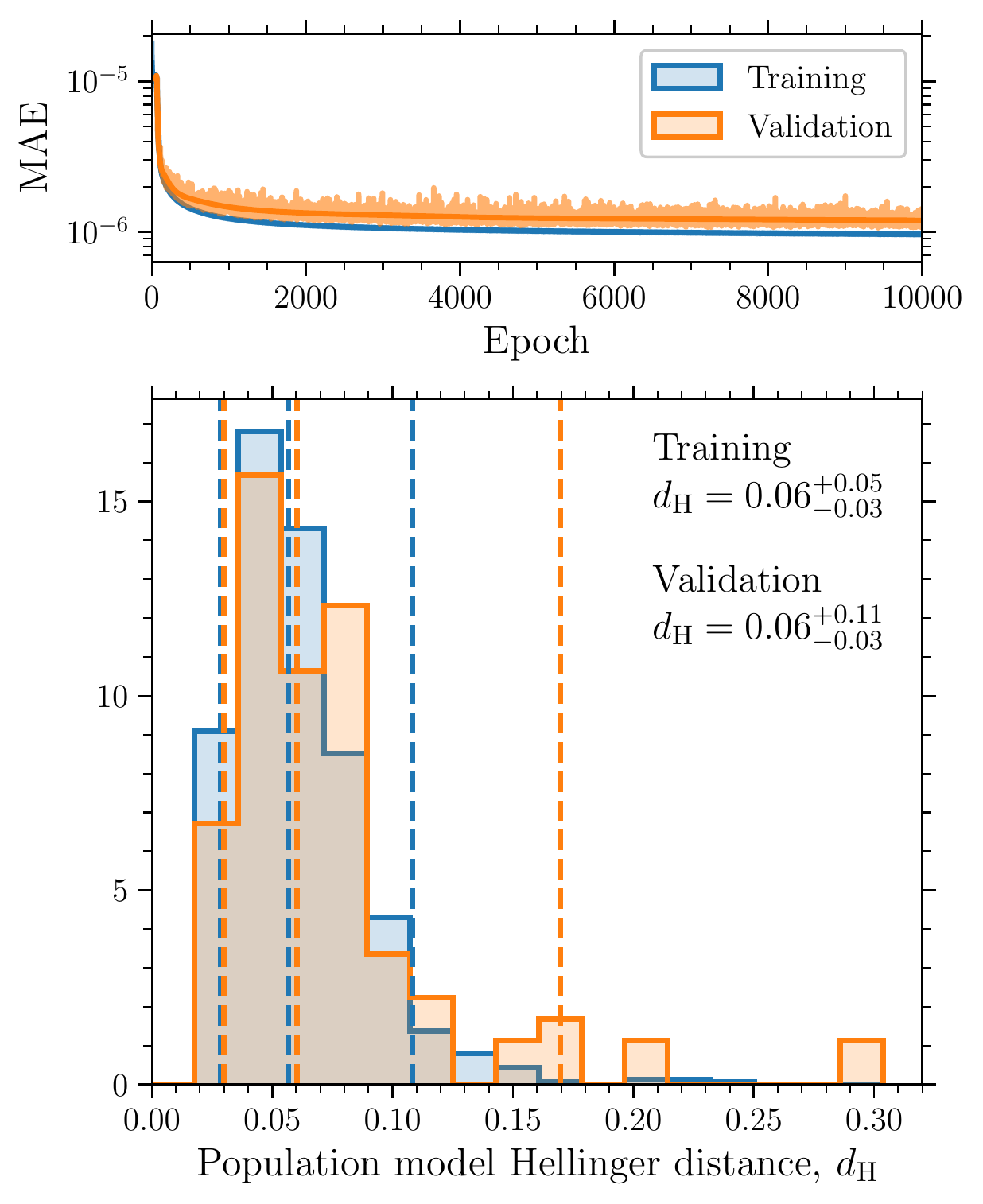}
\caption{Top panel: loss functions versus epoch for the training (blue) and validation (orange) data of the population density NN $\ppop'(\theta|\lambda)$. Smoothed versions are overplotted in bold. Bottom panel: distribution across all simulations of the Hellinger distances $\dH$ between the true KDE evaluations of $\ppop'(\theta|\lambda)$ and those predicted by the NN. The medians and 90\% intervals of $\dH$ are plotted as vertical dashed lines and listed explicitly.}
\label{fig:ppop_loss}
\end{figure}

In Fig.~\ref{fig:ppop_loss}, we summarize the training procedure and predictive performance of our NN population model. The convergence of the MAE loss function for the training and validation samples is plotted in the top panel. The NN fits slightly better to the training data ---the validation MAE being, on average, about $1.2$ times larger--- but there is no significant overfitting. In the bottom panel of Fig.~\ref{fig:ppop_loss}, we quantify this statement by comparing the predictive accuracy of the trained population model using the Hellinger distance~\cite{Hellinger+1909+210+271}, a metric $\dH$ over the space of probability densities that measures the ``distance'' between two distributions. For two probability densities $p$ and $q$, it is given by
\begin{align}
\dH(p,q)^2
=
\frac{1}{2} \int \left[ \sqrt{p(x)} - \sqrt{q(x)} \right]^2 \dd x
\, .
\end{align}
Here, $\dH$ has the desirable properties of being symmetric and bounded in $[0,1]$, with $\dH(p,q)=0$ only when $p\equiv q$ and $\dH(p,q)=1$ when $p$ and $q$ have disjoint supports (see Appendix~C of Ref.~\cite{2021PhRvD.104h3008M} for a physics-oriented summary of the properties of the Hellinger distance). For each of our simulations we compute the distance between the KDE evaluation and the NN prediction for the probability density. While the mild overfitting presents itself as a small number of outliers at larger values of $\dH$ in the validation distribution, both the training and validation subsets have median values of approximately $0.06$ and are consistent with each other.

\begin{figure*}
\includegraphics[width=1\textwidth]{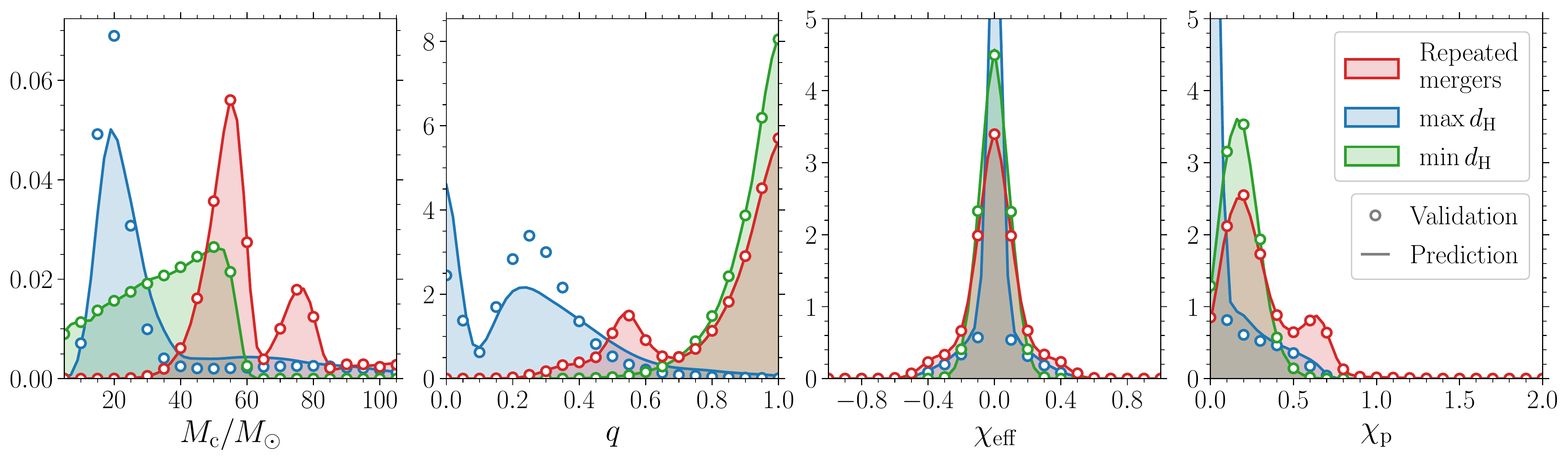}
\caption{True KDE evaluations (circle markers) of the population density $\ppop'(\theta|\lambda)$ compared against the NN population model predictions (solid lines) for three validation simulations. The full four-dimensional distributions are marginalized to each one-dimensional event-level parameter (left to right: chirp mass $\Mchirp$, mass ratio $q$, effective aligned spin $\chieff$, and effective precessing spin $\chip$) for the purpose of visualization. In blue, we show the validation simulation that has the worst predictive accuracy, with a Hellinger distance of $\dH=0.30$ and population-level parameters $\alpha=6.3$, $\beta=-7.3$, $\gamma=1.8$, $\delta=8.7$, $\mmax=46\Msun$, and $\chimax=0.01$. In green, we show the validation simulation with the smallest Hellinger distance $\dH=0.02$ and $\alpha=-1.9$, $\beta=5.2$, $\gamma=0.5$, $\delta=-9.4$, $\mmax=67\Msun$, $\chimax=0.35$. In red, we show a validation simulation (as in Fig.~\ref{fig:popexamples}) with $\dH=0.10$ and whose distribution contains distinct features due to repeated mergers.}
\label{fig:ppop_examples}
\end{figure*}

In Fig.~\ref{fig:ppop_examples}, we illustrate example predictions from our deep-learned population model. For a given set of  population-level parameters $\lambda$, the NN predicts the value of the joint four-dimensional probability density over the source parameters $\theta=\{\Mchirp,q,\chieff,\chip\}$. For three validation simulations, we plot the predicted values of $\ppop'(\theta|\lambda)$ (solid lines) along with the true KDE evaluations for comparison (circle markers), numerically marginalizing to one-dimensional distributions for the purpose of visualization.

The first example (red) has good predictive accuracy, with $\dH=0.10$. Here, we use the same distribution labeled ``repeated mergers'' in Fig.~\ref{fig:popexamples}, with parameters $\alpha=4.1$, $\beta=3.1$, $\gamma=5.5$, $\delta=5.1$, $\mmax=70\Msun$, and $\chimax=0.37$. Here, the larger escape velocities and sharp mass function and pairing probabilities lead to distinct peaks due to higher-generational mergers. Even though the Hellinger distance of this simulation is greater than the median value, the one-dimensional marginal predictions present excellent matches to the true validation data, accurately capturing all sharp features.

The second case ($\max\dH$, in blue) is a very conservative bound on the performance of our NN, taking the validation simulation with the largest value of the Hellinger distance $\dH=0.30$ (i.e., that with the worst predictive accuracy). The population parameters are $\alpha=6.3$, $\beta=-7.3$, $\gamma=1.8$, $\delta=8.7$, $\mmax=46\Msun$, and $\chimax=0.01$. While the distributions of the spin parameters $\chieff$ and $\chip$ are still fairly well captured, the predictions in the mass distributions suffer from larger errors, though the main features have been learned. The small value of the maximum natal spin $\chimax=0.01$ leads to sharply peak effective spins $\chieff,\chip\approx0$, while the pairing process generates smaller mass ratios. We stress that this is the worst case among the entire validation set and a rather extreme outlier (cf. Fig.~\ref{fig:ppop_loss}). Figure~\ref{fig:ppop_examples} presents the marginalized distributions, while the model predicts the full four-dimensional density, meaning errors over the full source parameter are propagated to the one-dimensional marginals.

The third case ($\min\dH$, in green) represents the best predictive accuracy of our population model, with $\dH=0.02$. In this validation simulation, the hyperparameters are $\alpha=-1.9$, $\beta=5.2$, $\gamma=0.5$, $\delta=-9.4$, $\mmax=67\Msun$, and $\chimax=0.35$, which produces equal masses and a unimodal distribution in the joint four-dimensional space of source parameters. Unsurprisingly, distributions with a simple feature set like this are easier to learn by our DNN population model.

\subsection{Selection function}
\label{sec:sigma}

\subsubsection{Detection probability}
\label{sec:pdet}

We assume sources are distributed uniformly in sky location, inclination, and polarization angle. We estimate $\Pdet$ with the widely used single-detector semianalytic approximation of Refs.~\cite{1993PhRvD..47.2198F,1996PhRvD..53.2878F}, as implemented in the Python package \textsc{gwdet}~\cite{davide_gerosa_2017_889966}, which relies on computing the SNR of optimally oriented sources with the same intrinsic parameters. This is estimated using \textsc{pycbc}~\cite{alex_nitz_2021_4556907}, the \textsc{imrphenompv2} waveform approximant~\cite{2014PhRvL.113o1101H,2016PhRvD..93d4006H,2016PhRvD..93d4007K}, and noise curves representative of the LIGO detector performance during O1O2\footnote{Early high from \href{https://dcc.ligo.org/LIGO-P1200087-v47/public}{dcc.ligo.org/LIGO-P1200087-v47/public}.} and O3\footnote{LIGO Livingston from \href{https://dcc.ligo.org/LIGO-T2000012/public}{dcc.ligo.org/LIGO-T2000012/public}.}~\cite{2018LRR....21....3A}. While the analytic marginalization of Refs.~\cite{1993PhRvD..47.2198F,1996PhRvD..53.2878F} is,  strictly speaking, only valid if one neglects spin precession and higher-order GW modes, the impact of these additional effects is subdominant~\cite{2020PhRvD.102j3020G}. Their inclusion requires further modeling, which has also been recently tackled using machine-learning techniques~\cite{2020PhRvD.102j3020G,2022ApJ...927...76T}; we plan to include these refinements in future versions of our population inference pipeline. We employ a SNR threshold of 8~\cite{2016ApJS..227...14A} and  thus set $\Pdet=0$ for all subthreshold binaries.

At each location in the population parameter space $\{\lambda^i\}_{i=1}^{\Npop}$, we  compute $\Pdet$ for all the binaries in the simulation. They have parameters $\{\vartheta_j^i\}_{j=1}^{\Nh(\lambda^i)} \sim \ppop(\vartheta|\lambda^i)$, allowing us to approximate the refactored detection efficiency of Eq.~(\ref{sigmaprime}) as
\begin{align}
\sigma'(\lambda^i) = \sum_r \frac{T_r}{T} \left[ \frac{1}{\Np(\lambda^i)} \sum_{j=1}^{\Nh(\lambda^i)} \Pdet(\vartheta_j^i,r) \right]
\, ,
\end{align}
where the term in brackets is the Monte Carlo approximation of the integral in Eq.~(\ref{eq:sigma}).

\subsubsection{Regression with a deep neural network}
\label{sec:sigma nn}

To evaluate the (refactored) detection efficiency at arbitrary values of the population parameters, the function $\sigma'(\lambda)$ must be emulated using the discrete evaluations at $\lambda^i$. Here, we also use a DNN with \textsc{tensorflow}~\cite{abadi2016tensorflow}. The network architecture consists of an input layer with $\dim(\lambda)=6$ neurons and a linear output layer with one, corresponding to the predicted value of $\ln\sigma'(\lambda)$. We add three hidden layers with 128 neurons each and RReLU activation. This network architecture is summarized in Table~\ref{tab:nn selection}.

\begin{table}[]
\label{tab:nn selection}
\centering
\renewcommand{\arraystretch}{1.3}
\setlength{\tabcolsep}{5.5pt}
\begin{tabular}{c|c|c|c}
Layer & Neurons & Activation & Parameters \\
\hline \hline
Input & 6 &  ... &0 \\
Dense 1 & 128 & RReLU & 896 \\
Dense 2 & 128 & RReLU & 16,512 \\
Dense 3 & 128 & RReLU & 16,512 \\
Output & 1 & ... & 129 \\
\hline
\multicolumn{1}{c}{} & \multicolumn{2}{r|}{Total} & \multicolumn{1}{c}{34,049} \\
\end{tabular}
\caption{Architecture of the DNN that predicts the logarithmic selection function $\ln\sigma'(\lambda)$ as a function of the population-level parameters $\lambda$. Each row represents a single layer and lists its number of neurons, the activation function used, and the corresponding number of free parameters. All hidden layers employ RReLU nonlinearities.}
\end{table}

We split the hyperparameter coordinates into the same $90\%$ training and $10\%$ validation simulations as in Sec.~\ref{sec:ppop}, though note that the training data here consist only of the hyperparameters $\lambda$ rather than the joint vector $(\theta,\lambda)$. As a preprocessing stage, we again normalize the input values of $\{\lambda^i\}_{i=1}^{\Npop}$ to a unit hypercube and train on the output values of $\{\ln\sigma'(\lambda^i)\}_{i=1}^{\Npop}$, which are normalized to the unit interval according to the extrema across the simulations. Predictions are rescaled back to the relevant parameter space. We use Adam optimization~\cite{2014arXiv1412.6980K} with a learning rate of $10^{-3}$ to minimize the mean squared error (MSE) loss function. At each epoch, the training data are shuffled into batches containing 1\% of the training data. We train the network for 2000 epochs on a single Intel Core i5-8365U CPU, which took approximately $4$~minutes. The training of this DNN is significantly quicker than that of $\ppop'$ since it has input dimensionality $\dim(\lambda)=6$, corresponding to a much smaller training sample size of 900 and a smaller network architecture.

\begin{figure}
\centering
\includegraphics[width=1\columnwidth]{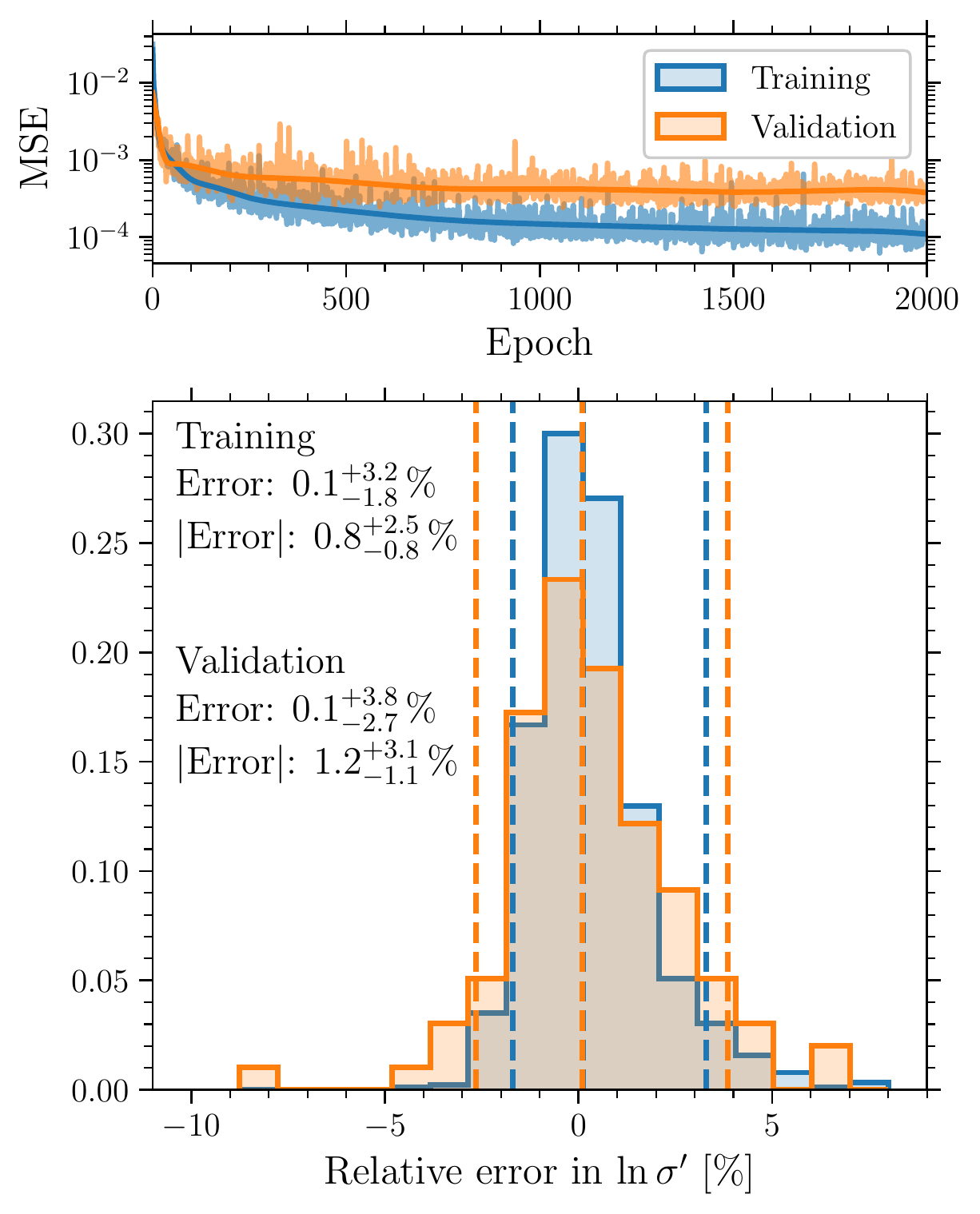}
\caption{Top panel: loss function over epochs for the training (blue) and validation (orange) data of the DNN predicting the refactored detection efficiency $\sigma'(\lambda)$. Bottom panel: relative error between the true and predicted values of $\ln\sigma'$. The medians and $90\%$ intervals of the errors are plotted as vertical dashed lines. They are also listed explicitly, as are the magnitudes of the relative errors.}
\label{fig:sigma_matches}
\end{figure}

\begin{figure*}
\centering
\includegraphics[width=1\textwidth]{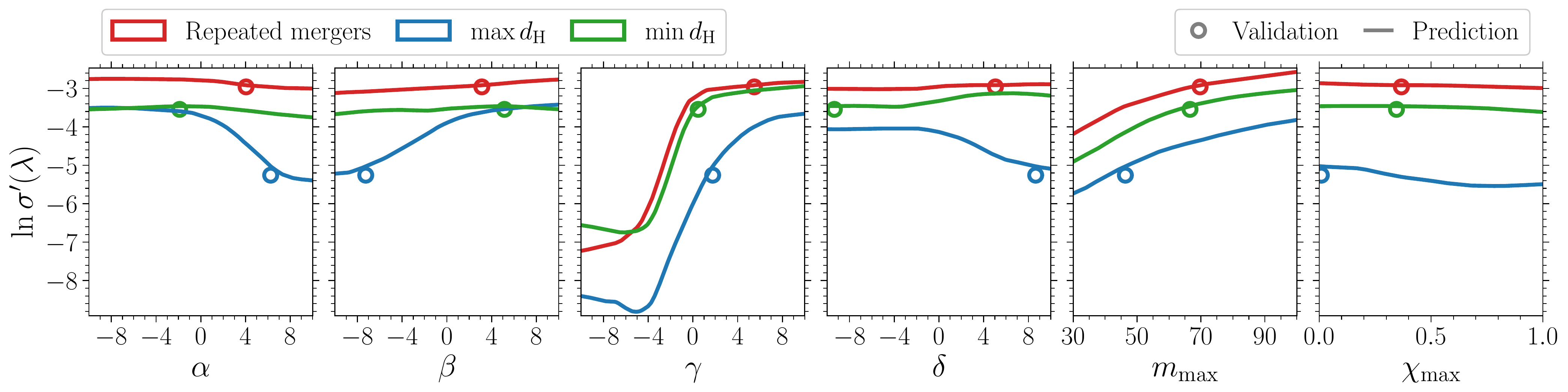}
\caption{Example evaluations of the DNN selection function $\ln\sigma'(\lambda)$ at the same three hyperparameter coordinates $\lambda$ displayed in Fig.~\ref{fig:ppop_examples}: a simulation containing repeated mergers (red), and those with the least (blue) and most (green) accurate predictions for the population model DNN. The true value for each simulation is displayed as a circle marker, while predictions made by the DNN are solid lines. In each panel, we vary a single hyperparameter, while the others are fixed to  values in the three simulations.}
\label{fig:sigma_examples}
\end{figure*}

The performance of our DNN to predict $\ln\sigma'$ is reported in Figs.~\ref{fig:sigma_matches} and \ref{fig:sigma_examples}.  In the top panel of Fig.~\ref{fig:sigma_matches}, we display the convergence of the loss function over the training epochs; the average gradient of the (smoothed) validation MSE over the final 100 epochs is less than or close to $0.5\%$ that over the first 100. While there is some overfitting to the training data, we verify the effect is mild, as follows. In the bottom panel, we display the relative error between the true and DNN-predicted values of $\ln\sigma'$. Since the median and $90\%$ symmetric interval for the validation and training errors are $0.1_{-1.8}^{+3.2}\%$ and $0.1_{-2.7}^{+3.8}\%$, respectively, both are consistent with being centered on and symmetric about zero, i.e., the DNN introduces no systematic biases. The magnitudes of the relative errors for the validation and training sets are consistent with each other and typically less than or close to $5\%$; the medians and $90\%$ intervals are $0.8_{-0.8}^{+2.5}\%$ and $1.2_{-1.1}^{+3.1}\%$, respectively.

In Fig.~\ref{fig:sigma_examples}, we show the dependence of the DNN selection function on each of the hyperparameters for the same three example simulations as in Fig.~\ref{fig:ppop_examples}. The true values of $\ln\sigma'(\lambda)$ are shown with circle markers. The predictions of the DNN are given by the solid lines, where in each panel we vary a single hyperparameter while keeping the others fixed to the values corresponding to each simulation. For all simulations, $\sigma'$ is an increasing function of both $\gamma$ ---the power-law index of 1g BHs--- and $\mmax$ ---the maximum 1g mass; larger $\gamma$ implies a greater number of BHs born with masses closer to the maximum $\mmax$, while heavier sources emit louder signals and are thus easier to detect (though there is also a compromise with the frequency-dependent ---and therefore, mass-dependent--- detector sensitivity). The simulation containing repeated mergers (red) consistently features higher values ---implying a larger fraction of detectable mergers in the underlying population--- due to the larger average binary mass.

The mismatch for the least accurate hyperparameter coordinate of the population model ($\max\dH$, blue) is visible in the offset between truths and predictions. Here, the selection function also depends on $\alpha$ and $\beta$, which determine the primary and secondary pairing probabilities. For this simulation, the first-generation mass slope, $\gamma=1.8$ is quite broad. A wider range of masses implies that a wider range of mass ratios are possible when selecting the BHs in the binary pairing procedure. Higher (lower) values of $\alpha$ ($\beta$) lead to higher (lower) primary (secondary) masses nd more extreme mass ratios, thus decreasing the detectability. Though repeated mergers occur due to high $\delta=8.7$, they are preferentially of mixed generations, and therefore, larger $\delta$ also leads to lower detectability.

For the validation simulation with the highest population model accuracy ($\min\dH$, green), first-generation masses are broad since $\gamma=0.5$ and larger since $\mmax=67\Msun$ (compared to $\mmax=46\Msun$ for the $\max\dH$ case). Masses are paired equally since $\alpha=-1.9$ selects the lightest primaries and $\beta=5.2$ selects the heaviest secondaries. The greater prevalence of higher-mass sources with unity mass ratios results in a selection function that is higher (corresponding to increased detectability in the binary BH population) and flatter (with respect to all hyperparameters except $\gamma$ and $\mmax$, as discussed).

\subsection{Merger-generation fractions}
\label{sec:fractions}

\begin{figure}
\centering
\includegraphics[width=1\columnwidth]{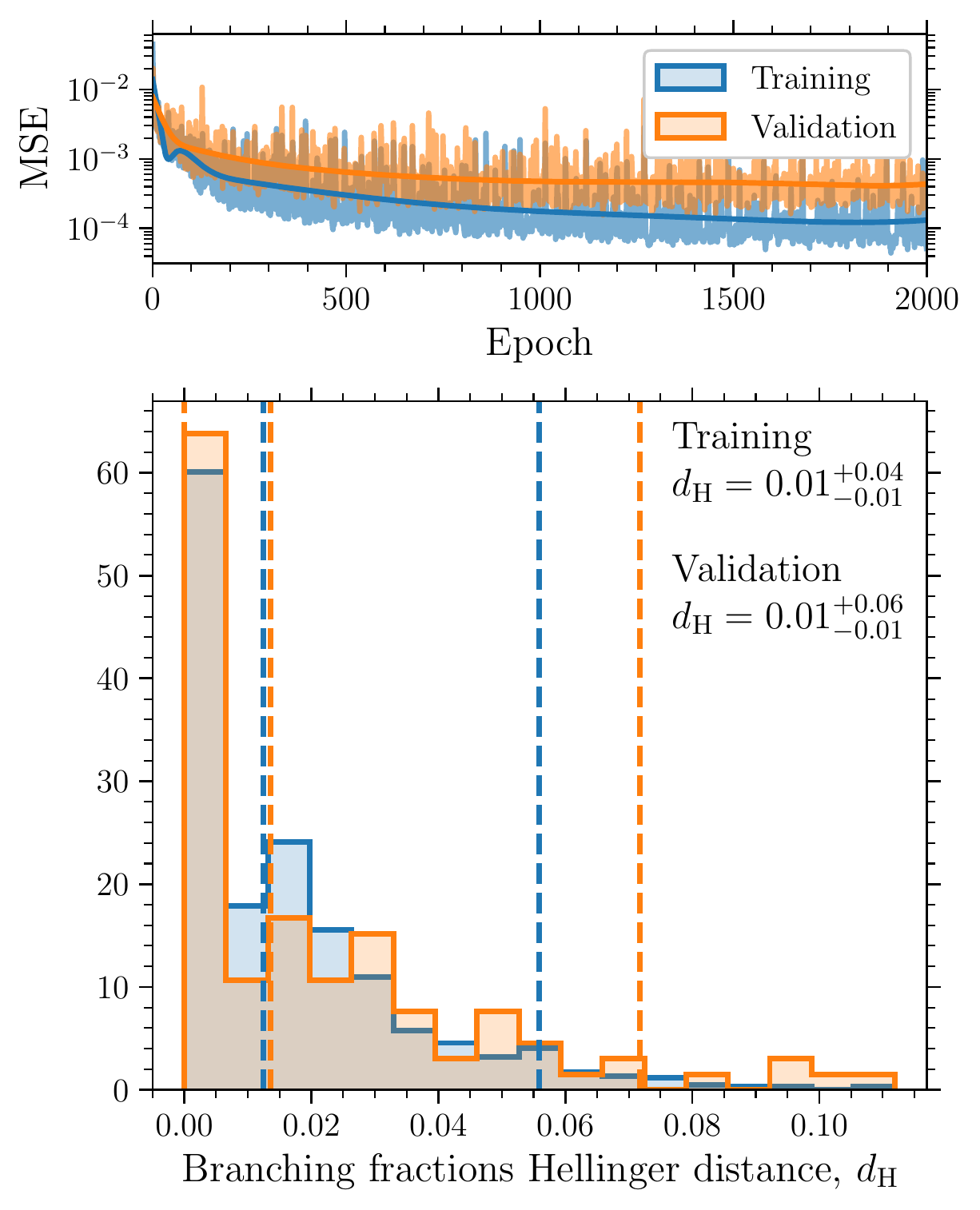}
\caption{Top panel: loss functions over training epochs for the training (blue) and validation (orange) data of the DNN predicting the branching fractions $f_{\rm 1g+1g}$, $f_{\rm 1g+2g}$, $f_{\rm 2g+2g}$, and $f_{\rm >2g}$. The actual loss curves are plotted with shading and smoothed versions are overplotted in bold. Bottom panel: Hellinger distances between the discrete distributions of the true and DNN-predicted merger-generation branching fractions. The medians and $90\%$ confidence intervals are plotted as vertical dashed lines and listed explicitly.}
\label{fig:fg loss}
\end{figure}

\begin{figure}
\centering
\includegraphics[width=1\columnwidth]{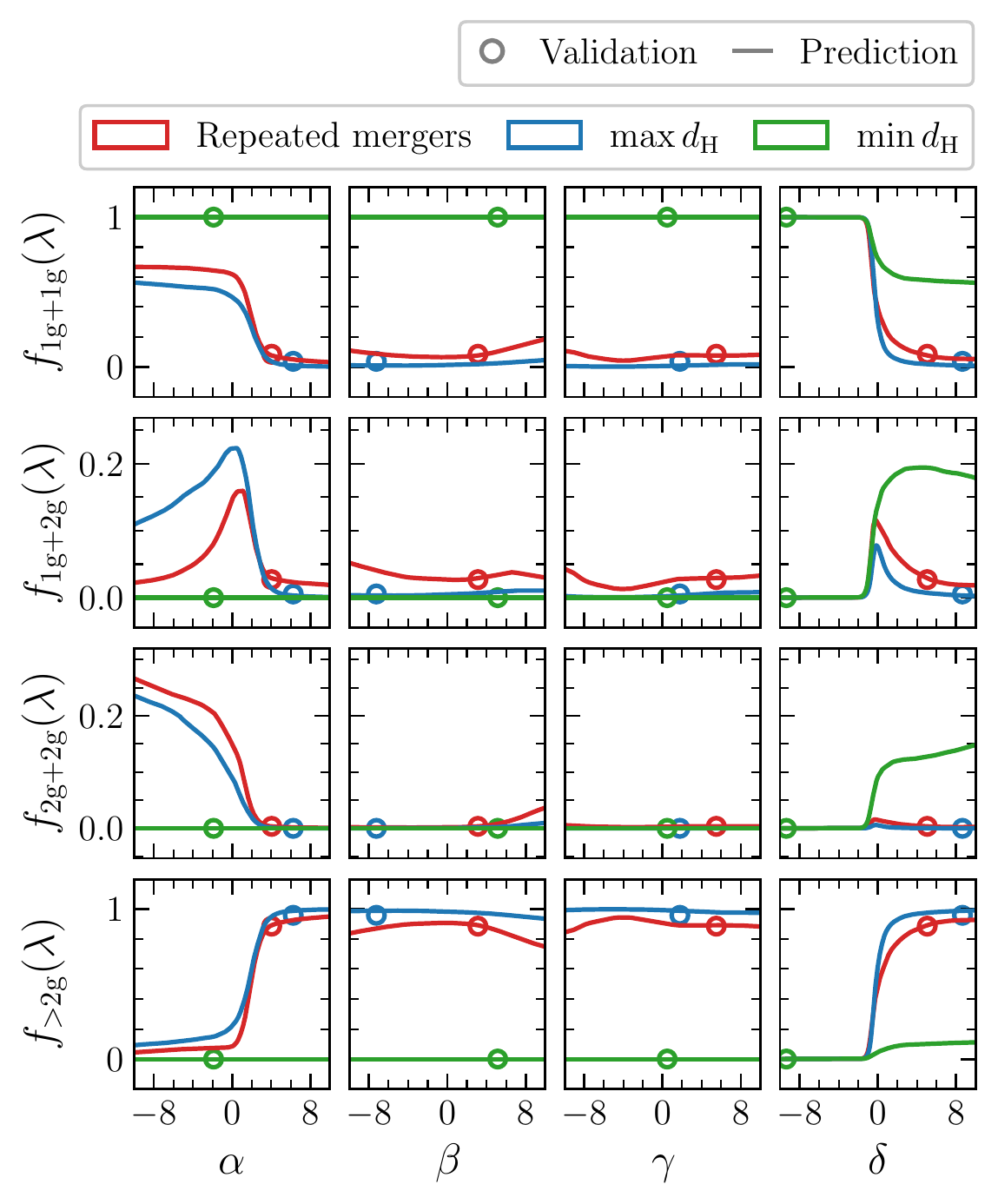}
\caption{Example evaluations of the DNN predicting the binary merger-generation branching fractions $f_g(\lambda)$, $g\in\{\mathrm{1g+1g,1g+2g,2g+2g,>2g}\}$ (from top to bottom rows), as a function of the hyperparameters $\lambda$. The results for hyperparameters taken from three illustrative simulations as in Fig.~\ref{fig:ppop_examples} ---repeated mergers (red), and least and most accurate population predictions ($\max\dH$ in blue and $\min\dH$ in green, respectively)--- are presented. The true values of the generation fractions are plotted as circle markers, whereas DNN predictions are given by solid lines. In each column, a single population-level parameter is varied while keeping the others fixed to those from the simulations.}
\label{fig:fg_examples}
\end{figure}

As a final demonstration of deep-learning techniques within GW population inference, we train a DNN to infer the branching fractions $f_g$ of the merger generations $g\in\{$1g+1g, 1g+2g, 2g+2g, $>$2g$\}$, as defined in Sec.~\ref{sec:repeated}. It is important to note that, unlike the case of branching ratios in mixture population models (e.g., Refs.~\cite{2011PhRvD..83d4036S,2017PhRvD..95l4046G, 2019ApJ...886...25B,2021PhRvD.103h3021W,2021PhRvD.104h3027T}), these fractions are not hyperparameters themselves but are functions of the model hyperparameters $\lambda$. In particular, $f_g(\lambda) = \int \ppop(\theta|\lambda) \mathcal{I}_g(\theta,\lambda) \dd\theta$, where $\mathcal{I}_g(\theta,\lambda)$ is a selector function that labels the merger generation, such that $\sum_g f_g(\lambda) \equiv 1$. Our application to the fraction of systems in each hierarchical generation is an example of the more generic problem of constraining formation subchannels that enter a single population.

We use the same training process and network architecture as in Sec.~\ref{sec:sigma nn}, with one modification. Since the four branching fractions form a discrete distribution with unit sum, the output layer here has four neurons and employs the activation function $\mathrm{softmax}(\boldsymbol{x})_i \coloneqq \exp(x_i) / \sum_{j}\exp(x_{j})$, where $x_i$ are the components of the input vector $\boldsymbol{x}$. The architecture of this DNN is summarized in Table~\ref{tab:nn generation}.

\begin{table}[]
\label{tab:nn generation}
\centering
\renewcommand{\arraystretch}{1.3}
\setlength{\tabcolsep}{5.5pt}
\begin{tabular}{c|c|c|c}
Layer & Neurons & Activation & Parameters \\
\hline \hline
Input & 6 &  ... &0 \\
Dense 1 & 128 & RReLU & 896 \\
Dense 2 & 128 & RReLU & 16,512 \\
Dense 3 & 128 & RReLU & 16,512 \\
Output & 4 & Softmax & 516 \\
\hline
\multicolumn{1}{c}{} & \multicolumn{2}{r|}{Total} & \multicolumn{1}{c}{34,436} \\
\end{tabular}
\caption{Structure of the DNN that models the branching fractions $f_\mathrm{1g+1g}$, $f_\mathrm{1g+2g}$, $f_\mathrm{2g+2g}$, and $f_\mathrm{>2g}$ between the binary merger generations, where, e.g., 1g (2g) denotes a first- (second-) generation component BH. The rows illustrate each layer of the network and report the number of neurons in each, their activation functions (RReLU for the hidden layers and softmax for the output layer), and the number of free parameters.}
\end{table}

In Fig.~\ref{fig:fg loss}, we plot the converged MSE loss curves. Once more, we assess the accuracy of the DNN predictions against the true generation fractions on the training and validation data sets  using the Hellinger distance, which, for discrete probability densities $p$ and $q$, is given by
\begin{align}
\dH(p,q)^2 = 1 - \sum\nolimits_i \sqrt{p_i q_i} \, .
\end{align}
The performances on training and validation subsets are consistent with each other, representing a lack of overfitting. Both have median Hellinger distances of $\dH\approx0.01$ with $\dH\lesssim0.1$ for most simulations. The enforced unit summation implies the branching fraction emulator in fact has only three independent outputs despite predicting four contributions, and in many of our simulations, one or more of the generation labels has zero contribution (e.g., no higher-generation mergers when all remnants are ejected from the host cluster). Both of these considerations produce a tendency for small values of the Hellinger distance, which explains the skew towards $\dH\lesssim0.01$.

In Fig.~\ref{fig:fg_examples}, we display the dependence of the DNN to predict the branching fractions $f_g(\lambda)$ on the hyperparameters $\lambda$ for the same validation simulations reported in Figs.~\ref{fig:ppop_examples} and \ref{fig:sigma_examples}. As in Fig.~\ref{fig:sigma_examples}, the true values computed from the simulated data are given by circle markers, while predictions made by the DNN are plotted as solid lines where a single hyperparameter is varied while keeping the others fixed. We only display the variation with the power-law indices $\{\alpha,\beta,\gamma,\delta\}$ as we found each $f_g(\lambda)$ to be independent of the maximum first-generation mass $\mmax$ and spin $\chimax$ in these cases. Each branching fraction depends most strongly on the distribution of escape speeds ---as determined by the power-law index $\delta$--- and the primary binary component pairing probability index $\alpha$, whereas the indices of the first-generation mass distribution $\gamma$ and the secondary component pairing $\beta$ are less impactful.

When $\delta<0$, the host clusters all have small escape speeds, and therefore, the branching fractions of sources with a remnant BH are close to zero, i.e., $f_\mathrm{1g+1g}\approx1$, as seen in the rightmost column of Fig.~\ref{fig:fg_examples}. With a fixed negative $\delta$, as in the case of the green simulation, the branching fractions become independent of the other hyperparameters as no repeated mergers take place. On the other hand, when $\delta$ becomes positive, escape speeds are typically larger and repeated mergers can occur, so the contribution to the population from first-generation-only binaries decreases, i.e., $f_\mathrm{1g+1g}<1$.

Which binary generation then begins to dominate the population depends on the BH pairing process. When heavier (lighter) primary components form binaries due to a fixed $\alpha>0$ ($\alpha<0$), as is the case for the red and blue (green) simulations, >2g higher-generation (equal second-generation 2g+2g) binaries preferentially populate the merger distributions. For small positive $\delta$, the contribution from binaries of mixed first and second generations increases, but it is reduced at larger $\delta$ in favor of higher-generation mergers. For fixed positive $\delta$ (red and blue simulations), larger primary pairing indices $\alpha$ select, with increasingly strong preference, the heaviest remnant BHs in the population to form new binaries, thus increasing the fraction of greater-than-second-generation binaries, i.e., $f_\mathrm{>2g}\approx1$, while reducing the prevalence of other generations, as seen in the leftmost column of Fig.~\ref{fig:fg_examples}. The branching fractions are flat for $\alpha<0$ if, as for the blue simulation, $\gamma<0$ because all first-generation BHs are typically lighter; therefore, reducing low-mass primary selection bias (i.e., making $\alpha$ less negative) has little effect. In contrast, when $\gamma>0$ as in the red simulation, first-generation BHs are heavier, and increasing $\alpha$ while keeping $\beta$ fixed will select heavier primaries relative to the secondaries, therefore favoring 1g+2g binaries.

\section{Validation with mock catalogs}
\label{sec:mock}

\begin{figure*}
\centering
\includegraphics[width=1\columnwidth]{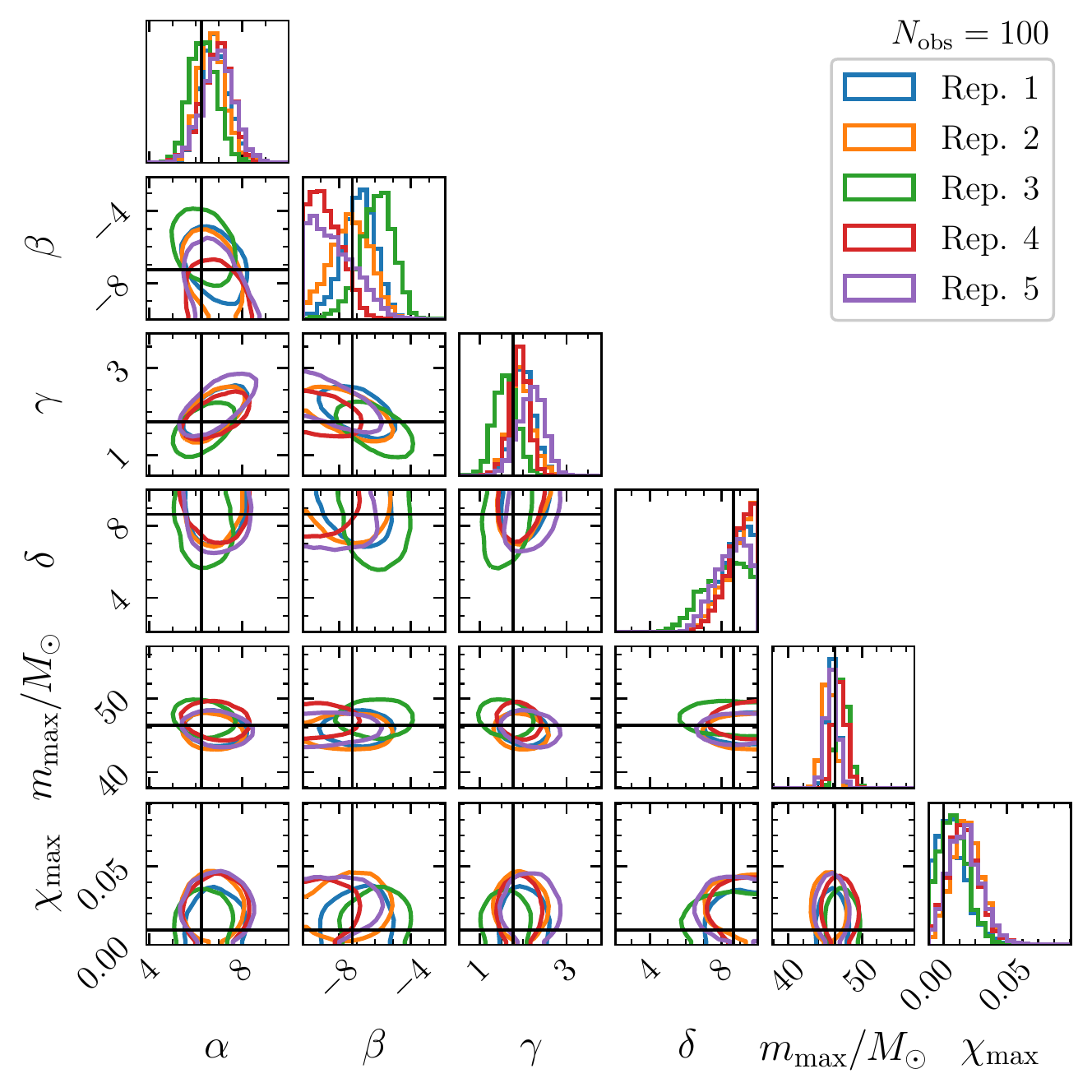}
\includegraphics[width=1\columnwidth]{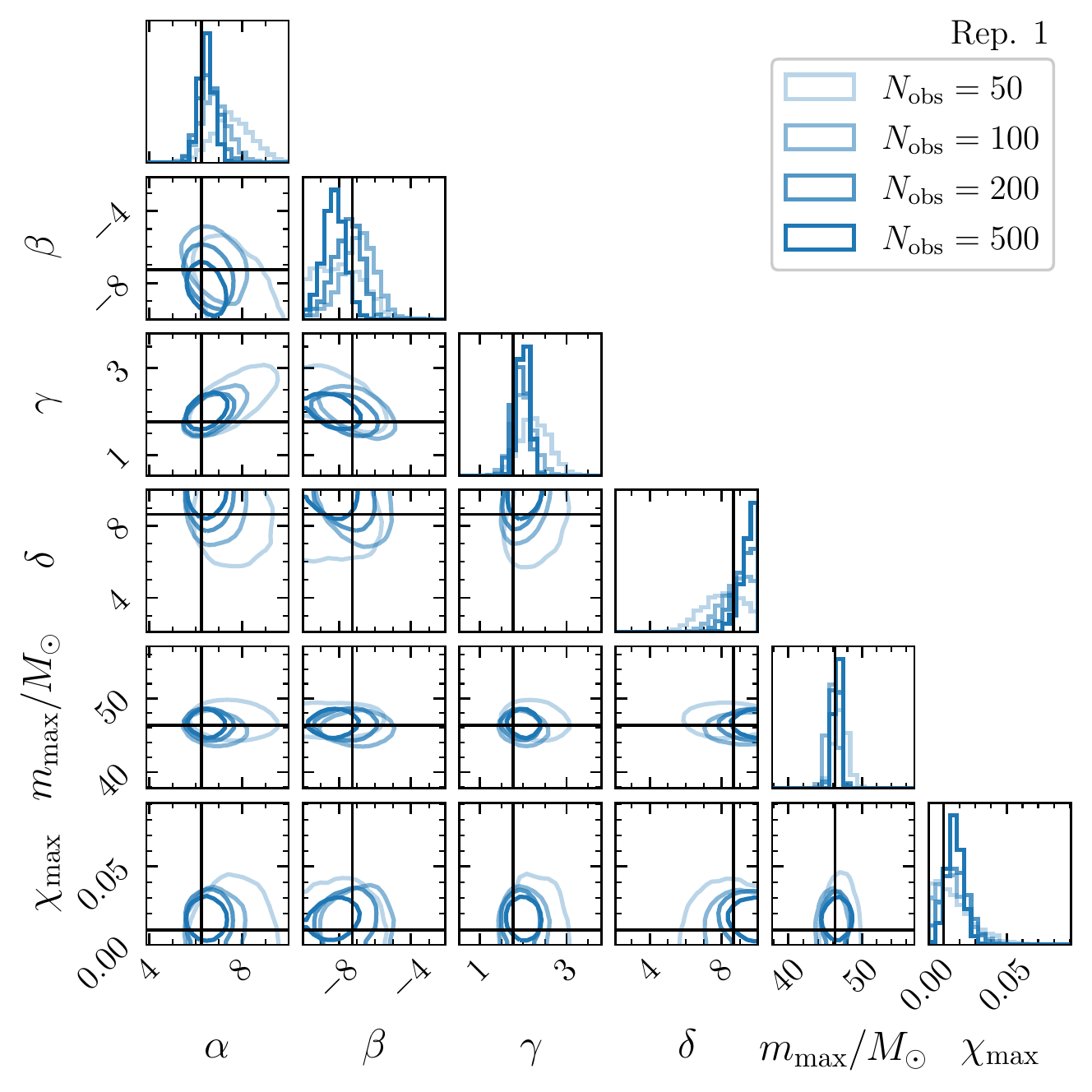}
\caption{One- and two-dimensional marginalized posteriors of the population-level parameters $\lambda = \{\alpha,\beta,\gamma,\delta,\mmax,\chimax\}$, corresponding to the $\max\dH$ simulation of Fig.~\ref{fig:ppop_examples}, as measured from inference runs without measurement errors and selection biases (corresponding to the high SNR limit), and systematics from the DNN population model by drawing mock GW catalogs directly from $\ppop'$. For the joint two-dimensional panels, each contour encloses the 90\% credible region for a single analysis. Injected values are marked with black lines. Left panel: number of observations in the catalog fixed to $\Nobs=100$ and five independent realizations of the inference with distinct events performed, each represented with a different colored curve (Reps.~1--5). Poisson fluctuations emerge as variations in the Bayesian measurements of the population-level parameters. Right panel: hyperposteriors for mock catalogs drawn for Rep.~1 on the left presented for an increasing number of observed events, $\Nobs=50,100,200,500$ (light to dark shading). Larger catalogs break degeneracies between parameters, and the resulting posteriors converge upon the true hyperparameters with tighter constraints.}
\label{fig:mock}
\end{figure*}

To test the inference pipeline in the absence of detection biases and single-event measurement uncertainties (equivalent to the limit of large SNRs) and without systematics due to the DNN population, we generate mock GW catalogs by drawing binary BH mergers from our DNN population model $\ppop'$. Since for the technical reasons discussed in Sec.~\ref{sec:bayes} this distribution is bounded in chirp mass, these draws are inherently taken from that range (listed in Table~\ref{tab:limits}). This also means that the selection function constructed in Sec.~\ref{sec:sigma} cannot be used in this mock inference; $\sigma'(\lambda)$ is defined over the entire range of source parameters, not just the observed range, and also accounts for the required missing factor between $\ppop'(\theta|\lambda)$ and $\ppop(\theta|\lambda)$. Including selection effects would require training a different model for the detection efficiency and thus our tests would include ingredients that do not enter the actual inference of Sec.~\ref{sec:gwtc-3}. Another technical difficulty is that we model two effective spins, $\chieff$ and $\chip$, while the detection probability $\Pdet$, in principle, depends on all six spin degrees of freedom. Creating a mock catalog of observable GW events, i.e., taking samples from the detection-weighted population $\Pdet(\theta)\ppop(\theta|\lambda)$, would require assuming an effective lower-dimensional dependence or resampling full spin vectors consistent with the sampled values of $\chieff$ and $\chip$ (cf. Ref.~\cite{2019PhRvD.100h3015W} for a more in-depth exploration of these issues). However, correctly including spin information in selection biases has a measurable effect at the population level~\cite{2022PhRvD.105b4076M}.

For testing purposes, we consider the high SNR limit, in which all events are detectable and their source parameters are measured exactly. This corresponds to a selection function $\sigma\equiv1$ and a single-event likelihood $\mathcal{L}(d_n|\theta)=\delta(\theta-\theta_n)$ for the $n$th GW event in the catalog. From Eq.~(\ref{eq:poplkl}) the population-level likelihood is thus given by $\mathcal{L}(d|\lambda) \propto \prod_{n=1}^{\Nobs} \ppop'(\theta_n|\lambda)$ (where the statistical details are otherwise equivalent to Sec.~\ref{sec:bayes}). We draw $\Nobs=50,100,200,500$ events to create increasingly large catalogs (and in going from, e.g., 50 to 100 events, the first 50 are added to when increasing the catalog size) with source parameters $\theta_n$ ($n=1,...,\Nobs$) using rejection sampling of $\ppop'$. We repeat the analysis 5 times with new catalogs to assess the impact of population Poisson fluctuations on the inference. To enable a conservative mock catalog test we fix the true hyperparameters to those of the validation simulation with the lowest predictive accuracy for the DNN population model ($\max\dH$ in Figs.~\ref{fig:ppop_examples}, \ref{fig:sigma_examples}, and \ref{fig:fg_examples}): $\alpha=6.3$, $\beta=-7.3$, $\gamma=1.8$, $\delta=8.7$, $\mmax=46\Msun$, and $\chimax=0.01$.

We present the results of our mock inference runs in Fig.~\ref{fig:mock}. The one- and two-dimensional marginal posterior distributions of the hyperparameters are plotted, where the two-dimensional panels display the 90\% contours. The solid black lines denote the true values listed above. In the left panel, we fix the number of observations in the catalog to $\Nobs=100$ and perform five independent repetitions of the analysis with five different mock catalogs, given by the different colored curves. Each run is consistent with both the injected hyperparameter values and each other at the 90\% level, though there are significant fluctuations between realizations. Recall that the single-event likelihoods neglect measurement errors; relaxing this assumption and including nonzero widths in those posteriors would decrease the overall accuracy of the hyperparameter measurements and thus blend the results from independent realizations to distributions with greater consistency. Increasing the number of observations in the catalog improves the hyperparameter measurement error and reduces the statistical fluctuations between realizations; we take $\Nobs=100$ here to approximate the current size of real catalogs~\cite{2021arXiv211103606T}.

The impact of the growing size of the catalog is illustrated in the right panel of Fig.~\ref{fig:mock}. Here, we choose one particular realization and analyze the catalog as increasing numbers of events are added incrementally (light- to dark-blue curves). We recover the expected result: The posterior constraints become tighter as $\Nobs$ increases from 50 to 500 while remaining consistent with the true hyperparameter values at the 90\% level. Larger catalog sizes also break degeneracies between parameter pairs, e.g., the $\beta$-$\gamma$ correlations, and remove posterior support in regions far from the truth, e.g., in the column for $\alpha$.

If the events from the mock catalogs are instead drawn from the simulated populations used as validation samples when training the $\ppop'$, one may expect a systematic bias in the recovered hyperposteriors as the number of observations increases due to mismodeling in the trained NN. Indeed, when repeating the above analysis but injecting from simulated validation data while recovering with the NN, we find hyperposteriors that can exclude the injected values at 90\% confidence for the lowest accuracy ($\max\dH$) simulation considered above when $\Nobs\geq100$. However, this point is a considerable outlier in terms of accuracy (see Fig.~\ref{fig:ppop_loss}). For most regions in the hyperparameter space the mismodeling between injection and recovery remains consistent at 90\% confidence. In particular, we verify that this is true for the validation simulation whose hyperparameters are closest to the recovered medians in Sec.~\ref{sec:gwtc-3}, suggesting our inference on the GWTC-3 catalog below is robust within the measurement uncertainties. While the tests performed here are admittedly limited in scope, they allow us to assess the renormalization and sampling capabilities in the pipeline.

\section{Population from GWTC-3}
\label{sec:gwtc-3}

\begin{figure*}
\label{fig:corner}
\includegraphics[width=0.9\textwidth]{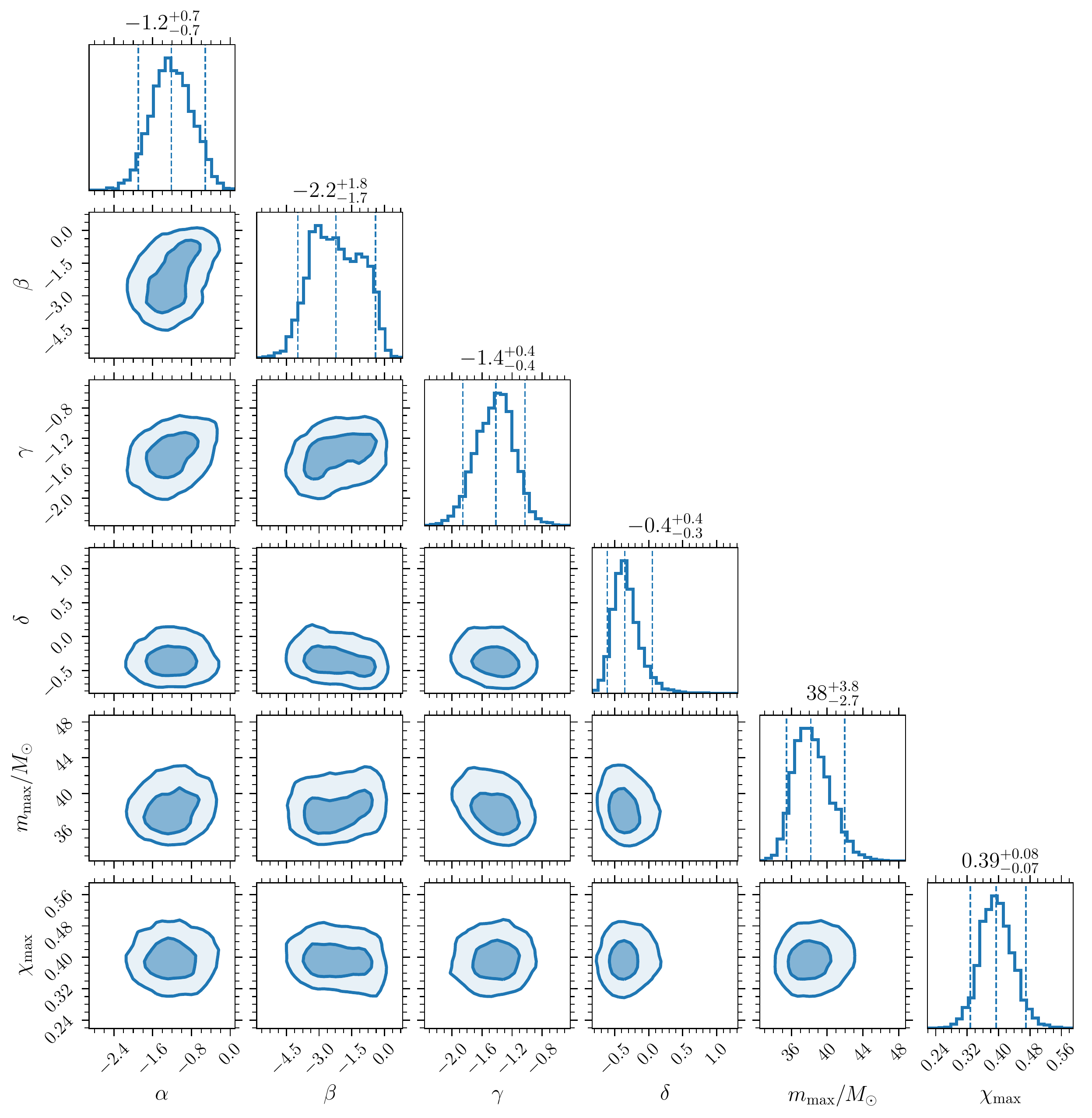}
{\caption{One- and two-dimensional marginal distributions of the population-level parameters $\lambda=\{\alpha,\beta,\gamma,\delta,\mmax,\chimax\}$ in our model of hierarchical mergers as measured using the real GW data from the confident ($\mathrm{FAR}<1~\mathrm{yr}^{-1}$) binary BH events through GWTC-3. In each two-dimensional distribution, the contours enclose the 50\% (dark shading) and 90\% (light shading) confidence regions. The one-dimensional median and symmetric 90\% intervals are reported above each diagonal and are plotted as vertical dashed lines in the
corresponding
panels.}}
\end{figure*}

In the following, we infer the population properties of the binary BHs in GWTC-3 given our deep-learned population model of hierarchical mergers. In Fig.~\ref{fig:corner}, we present the result of our population inference ---the posterior distribution of the hyperparameters $\lambda$. Along the diagonal is the one-dimensional marginalization of each hyperparameter, while the other panels display the 50\% and 90\% confidence intervals of each two-dimensional distribution.

\subsection{Host escape speeds}
\label{sec:vesc}

\begin{figure}
\label{fig:ppd_vesc}
\includegraphics[width=1\columnwidth]{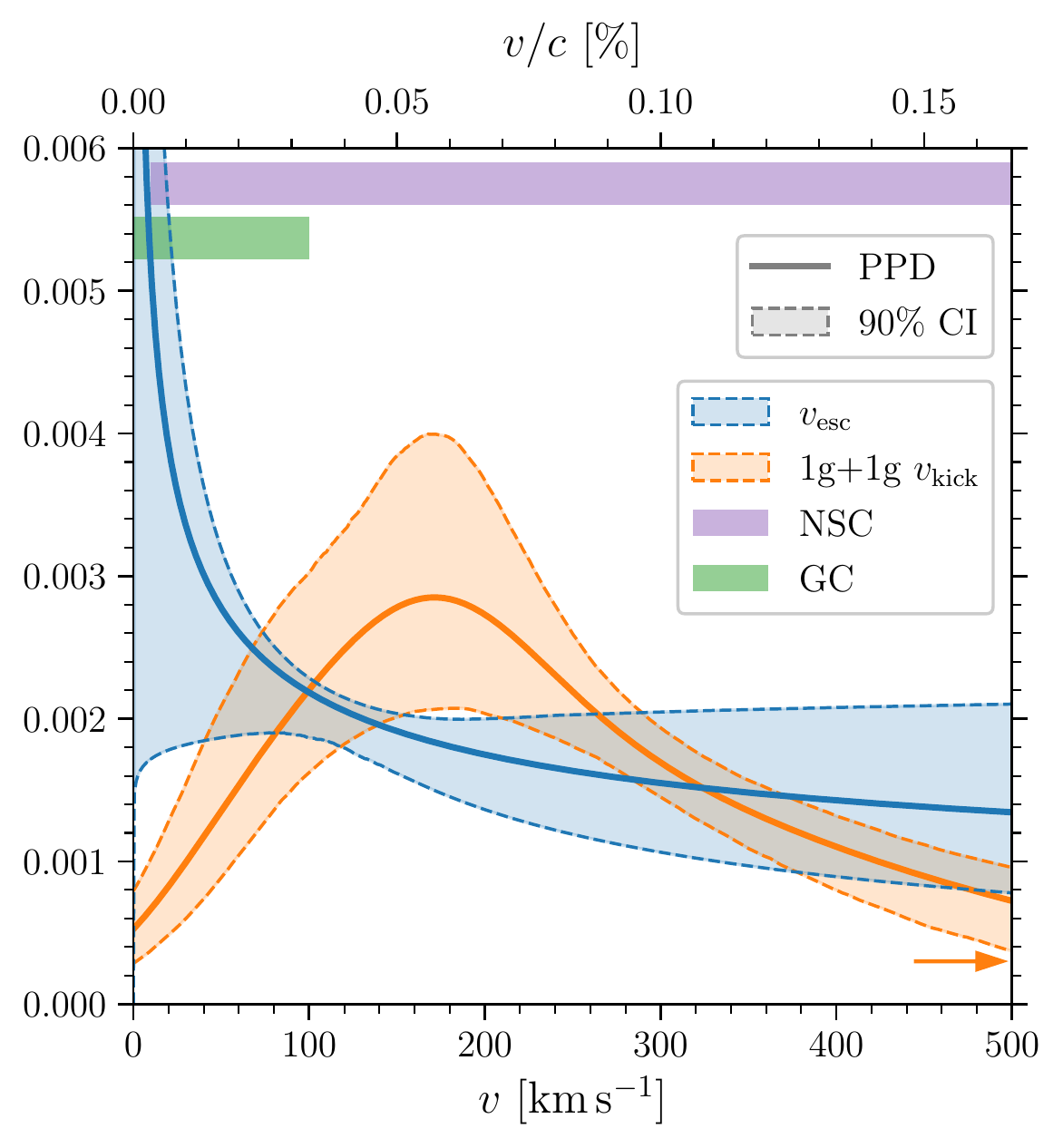}
{\caption{PPDs of cluster escape velocities $\vesc$ (blue), and of gravitational recoils $\vkick$ for binaries consisting of two first-generation BHs (orange). The orange distributions are normalized over a range extending beyond the upper $\vesc$ limit, as indicated by the arrow. The colored shaded bands contextualize the velocity scale by denoting the typical escape speeds of globular clusters (green) and nuclear star clusters (purple).}}
\end{figure}

We begin with the properties of host environments. We consider clusterlike hosts ---simply collections of individual BHs that may be paired to form binaries--- which are solely characterized by their escape speeds $\vesc$. Recall that the simulated clusters are distributed according to a truncated power law $p(\vesc|\delta)\propto\vesc^\delta$, with $0~\kmps < \vesc < 500~\kmps$. The value of each cluster's escape speed controls whether repeated mergers take place since sources receiving larger gravitational kicks are ejected. Though a power-law distribution is a simplified model, it is indicative of a preference (or lack thereof) towards either edge of the domain.

The marginal distribution of the escape-speed index $\delta$ is displayed in the fourth diagonal entry in Fig.~\ref{fig:corner}. Negative (positive) values of $\delta$ indicate an escape-speed distribution favoring lower (higher) values, while $\delta=0$ corresponds to a uniform distribution in $\vesc$. We report a median and symmetric 90\% interval of $\delta=-0.4_{-0.3}^{+0.4}$, corresponding to an escape-speed distribution biased toward smaller values though consistent with uniformity within the 90\% credible bounds.

In Fig.~\ref{fig:ppd_vesc}, we display the distribution of escape speeds reconstructed from the hyperposterior $p(\delta|d)$. Marginalizing the escape-speed model $p(\vesc|\delta)$ over the uncertainty in the hyperparameter $\delta$ returns the posterior population distribution (PPD)
\begin{align}
\PPD(\vesc) = \int p(\vesc|\delta) p(\delta|d) \dd\delta
\, ,
\end{align}
whereas the posterior uncertainty is displayed interior to the 5\% and 95\% quantiles of $p(\vesc|\delta)$, with $\delta \sim p(\delta|d)$. For context, order-of-magnitude estimates of the escape speeds of globular clusters (GCs; $\lesssim100~\kmps$) and nuclear star clusters (NSCs; $\lesssim500~\kmps$) are shown as horizontal colored bands \cite{2004ApJ...607L...9M,2002ApJ...568L..23G,2016ApJ...831..187A}. Additionally, we display the PPD of gravitational kicks, $\vkick$, received by the 1g+1g sources implied by our population model and hyperparameter constraints. Recall that, since the first-generation and binary pairing distributions have parametric forms (see Sec.~\ref{sec:populations}), the 1g+1g distribution does also (i.e., power-law mass distributions, uniform dimensionless spin magnitudes, isotropic spin directions; we present the measurements of the population-level parameters governing this distribution in the following sections). Though GW kicks peak at about $200~\kmps$, the distribution of escape speeds features support across the defined range up to $\vesc=500~\kmps$. For these 1g+1g sources, we find that $P(\vkick<500~\kmps)=0.85_{-0.08}^{+0.06}$ and $P(\vkick<\vesc)=0.37_{-0.12}^{+0.13}$, implying that host environments can retain the kicked remnants of a portion of first-generation mergers and support a population of hierarchical BHs.

\subsection{Mass distribution}
\label{sec:mass}

\begin{figure*}
\label{fig:ppd_mass}
\includegraphics[width=1\textwidth]{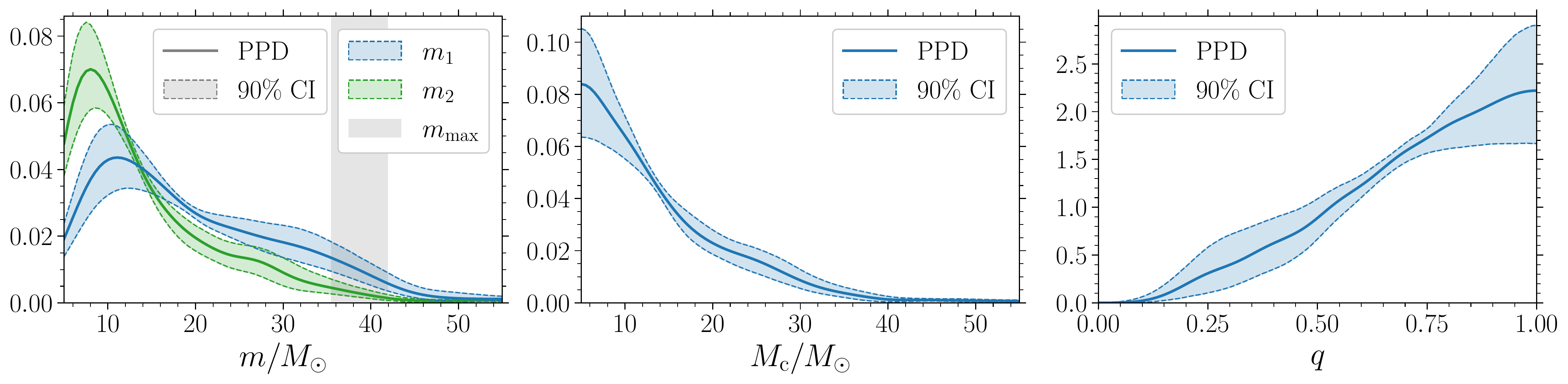}
{\caption{Astrophysical distributions of the modeled chirp mass $\Mchirp$ (middle panel) and mass ratio $q$ (right panel), as well as the implied distributions of primary and secondary masses, $m_1$ and $m_2$, respectively (left panel), as determined by our DNN population model and Bayesian analysis of the binary BH merger events in GWTC-3. The solid blue lines represent the PPDs, while the dashed lines enclose the 90\% symmetric confidence intervals (shaded). In the left panel, the vertical gray band encloses the 90\% confidence interval for the maximum mass of first-generation BHs, $\mmax$.}}
\end{figure*}

\begin{figure}
\label{fig:ppd_mass1g}
\includegraphics[width=1\columnwidth]{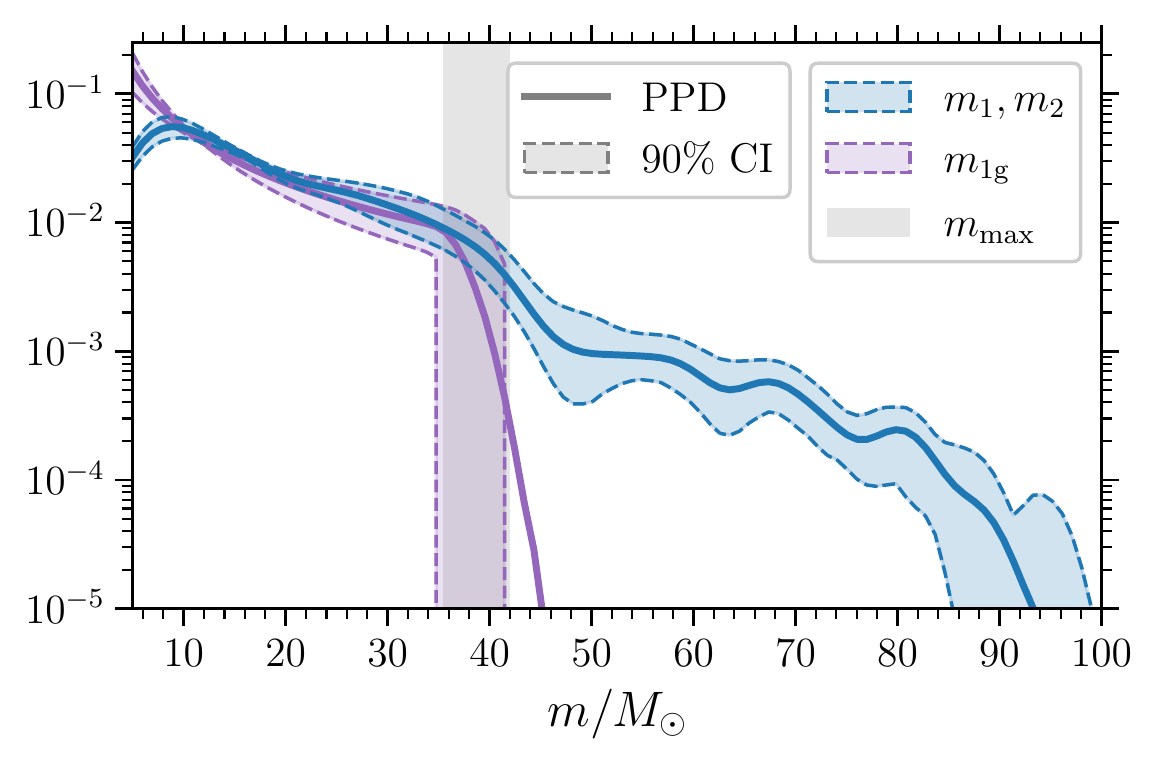}
{\caption{PPDs (in logarithmic scale) of the first-generation BH masses $m_\mathrm{1g}$ (purple) and the combined distribution of all components masses $m_1,m_2$ (blue). The solid lines denote the means, while the dashed lines bound the shaded 90\% symmetric confidence regions. The vertical gray band encloses the 90\% constraint on the maximum first-generation BH mass, $\mmax$.}}
\end{figure}

First-generation BHs ---those born in stellar collapse--- are drawn according to $p(\moneg|\gamma,\mmax) \propto \moneg^\gamma$, with $5\Msun<\moneg<\mmax$ providing a mass limit corresponding to the lower (upper) edge of the purported upper (lower) mass gap.  We recover $\gamma=-1.4_{-0.4}^{+0.4}$, implying lighter BHs closer in mass to $5\Msun$ (chosen here to conservatively rule out NS and ambiguous source classifications) preferentially populate the underling population. Negative 1g mass power-law exponents are expected, and they reflect the stellar IMF~\cite{2001MNRAS.322..231K}.

If first-generation BHs are drawn from this single power-law prescription, we ind a first-generation upper mass limit of $\mmax=38_{-2.7}^{+3.8}\Msun$. The presence of events in the GW catalog with component masses greater than $50\Msun$ already points to the possibility of hierarchical mergers. Theoretical and simulated estimates of a mass gap location due to PISN typically predict $\mmax\sim50\Msun$, but they range within $40\Msun \lesssim \mmax \lesssim 70\Msun$ (or even higher~\cite{2020ApJ...905L..15B}) due to varying assumptions on key uncertain parameters~\cite{2019ApJ...887...53F, 2021ApJ...912L..31W}. For comparison, taking the \textsc{power law + peak} GWTC-3 analysis of Ref.~\cite{2021arXiv211103634T} ---which features a Gaussian peak with mean $\mu_m$ to model mass buildup, potentially due to PISN--- we find consistency within 90\% credible bounds between the inferred $\mu_m=34_{-4.0}^{+2.6}\Msun$ and $\mmax$. Note, however, that while $\mmax$ is a sharp cut specifically characterizing the first-generation BH mass limit, the model of Ref.~\cite{2021arXiv211103634T} parametrizes all BHs in a single distribution and $\mu_m$ is only the mean of a broadened feature, such that $\mmax$ and $\mu_m$ are not directly equivalent. In our case, BH with masses larger than $\mmax$ are accommodated with hierarchical mergers.

Here, we point out a key distinction of our modeling procedure: We assume all first-generation component BHs are drawn from a shared distribution (above) and then binary formation is separately modeled with component pairing probabilities $p(m_1|\alpha) \propto m_1^\alpha$ and $p(m_2|\beta,m_1) \propto m_2^\beta$ ($m_2<m_1$). This choice differs from, e.g., \textsc{power law + peak}~\cite{2021arXiv211103634T}, which models each component mass distribution with multiple features superimposed on a power-law distribution. One may be tempted to think that, e.g., the primary mass distribution is equivalent to $p(m_1|\alpha) p(m_1|\gamma,\mmax) \propto m_1^{\alpha+\gamma}$ (and similarly for secondary masses), however this applies only to 1g+1g binaries. Our DNN population model additionally captures the interdependence between binary pairing and remnant retention. In short, the power-law indices parametrizing the distributions in this work are not directly comparable to such models. We infer $\alpha=-1.2_{-0.7}^{+0.7}$ and $\beta=-2.2_{-1.7}^{+1.8}$, such that both component pairing probabilities are bottom heavy with positive power-law indices ruled out at the 90\% confidence level.

Having reported the inferred population-level parameters governing the binary BH distributions, we now turn to the implied source-parameter PPDs given by
\begin{align}
\label{eq:ppd}
\PPD(\theta) = \int \ppop'(\theta|\lambda) p(\lambda|d) \dd\lambda
\, ,
\end{align}
such that the astrophysical distribution of each source parameter is given by the one-dimensional marginalizations of $\PPD(\theta)$. In Fig.~\ref{fig:ppd_mass}, we present the inferred source distributions of the modeled mass parameters ---chirp mass $\Mchirp$ and mass ratio $q$--- and the implied distributions of primary and secondary masses, $m_1=\Mchirp(1+q)^{1/5}/q^{3/5}$ and $m_2=qm_1$, respectively. Each PPD is plotted as a bold solid line, while the symmetric 90\% confidence region of each marginal $\ppop'(\theta|\lambda)$ with $\lambda \sim p(\lambda|d)$ is represented by shaded bands. The chirp mass distribution peaks at the minimum value $5\Msun$ allowed by our model before an approximately exponential decline, with $\Mchirp\lesssim40\Msun$. Equal-mass binaries are preferred in the underlying population, the mass-ratio distribution having a peak at $q=1$ but with a broader linear decline down to $q\gtrsim0.1$.

Substructure is apparent in the distributions of component source masses, corroborating the findings of Refs.~\cite{2021arXiv211103634T,2021ApJ...913L..19T}. Tighter constraints at $\Mchirp\approx13\Msun$ and $q\approx0.6$ result in a cusp in the primary (secondary) mass distribution around $m_1\approx20\Msun$ ($m_2\approx12\Msun$) between two features: the peak of the distribution at $m_1\approx12\Msun$ ($m_2\approx8\Msun$) and a buildup-following decline at the first-generation mass limit $\mmax\approx40\Msun$. This suggests two contributions to the mass distribution in the range $20\Msun \lesssim m_1 \lesssim 40\Msun$: (1) first-generation BHs with masses above the peak of the distribution, and (2) higher-generation BHs with masses still smaller than $\mmax$ but whose parents originally had masses in the peak 10--20$\Msun$ region. While high-mass outliers above $\mmax$ might be considered as clear indicators of repeated mergers, the bottom-heavy nature of the stellar IMF implies that hierarchical mergers may be prominent also for sources with masses below $\mmax$.

The first-generation and combined component mass distributions are compared in Fig.~\ref{fig:ppd_mass1g}. In purple, we show the reconstructed distribution of first-generation masses,
\begin{equation}
\PPD(\moneg) = \int p(\moneg|\gamma,\mmax) p(\gamma,\mmax|d) \dd\gamma\dd\mmax\,,
\end{equation}
and in blue, we show the joint distribution of all primary and secondary masses. The gray shaded band represents the 90\% constraint on the mass limit of first-generation BHs, $\mmax$. Note the logarithmic scale, and that the PPD is a set of expectation values (i.e., means) and, as such, can lie outside the region bounded by given quantiles. Though declining above the first-generation cutoff, the mass distribution features an extended spectrum above $\mmax$ which cannot result from 1g+1g mergers. We find that 99\% of all BHs have masses less than $59_{-6.5}^{+7.8}\Msun$. The spectrum ultimately abates at $m_1>80\Msun$ ---roughly $2\mmax$, implying a lack of greater-than-2g mergers with parent components from the upper end of the 1g mass spectrum--- and features multiple small-scale modes in the intervening region. These observations again point to hierarchical mergers in the underlying population.

\subsection{Spin distribution}
\label{sec:spin}

\begin{figure*}
\label{fig:ppd_spin}
\includegraphics[width=1\textwidth]{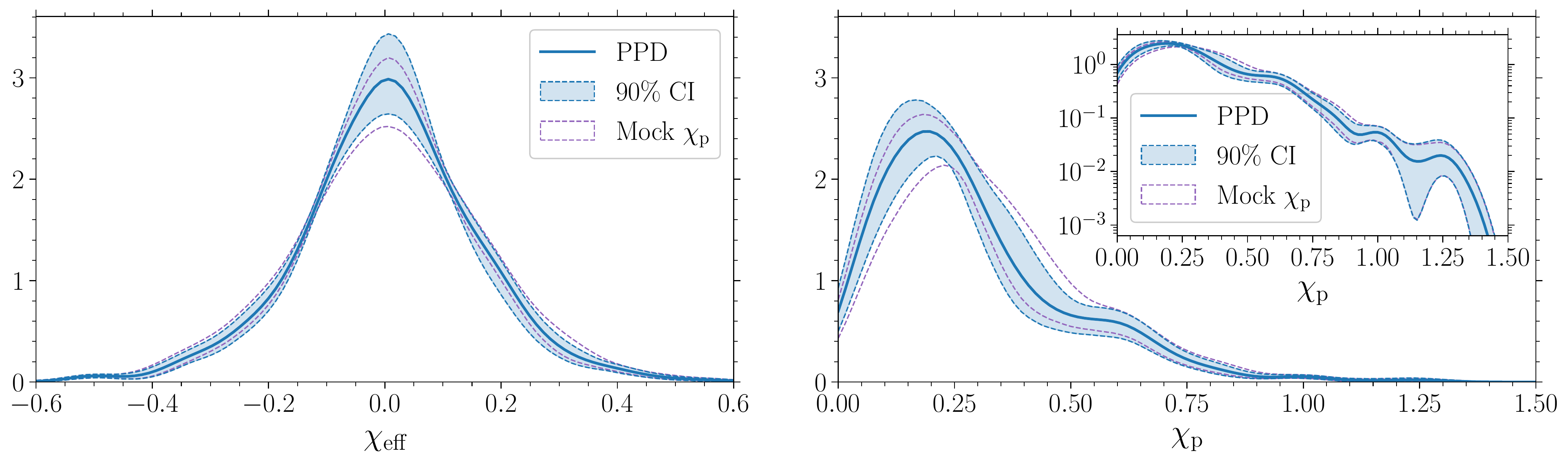}
{\caption{PPDs of the effective aligned ($\chieff$, left panel) and precessing ($\chip$, right panel) spins derived from our DNN population inference of binary BHs in GWTC-3. The means of the distributions are plotted with solid blue lines, and the symmetric 90\% confidence intervals are given by the shaded bands bounded by dashed lines. The inset for the $\chip$ panel shows the same distribution with logarithmic scaling to highlight smaller-scale features. The dashed purple lines bound the 90\% confidence region of the distributions that are measured when replacing the true parameter estimation results for $\chip$ with mock samples from the prior for each event in the catalog.}}
\end{figure*}

Moving to binary BH spins, recall that the first generation of BHs are modeled with isotropic spins whose dimensionless magnitudes are distributed uniformly up to a maximum $\chimax\in(0,1)$, representing the maximum natal spin a BH may be born with in stellar collapse. We infer a value $\chimax=0.39_{-0.07}^{+0.08}$. With limited constraining power in the spin observables, the precise constraints reported here are likely to be very model dependent. We opt for a uniform distribution of 1g spin magnitudes because of the large uncertainties surrounding the spin of compact objects following core collapse (e.g., Refs.~ \cite{2018MNRAS.473.4174Z,2019ApJ...881L...1F,2020A&A...635A..97B,2020A&A...636A.104B,2021PhRvD.103f3032S}); this is an area where more accurate observations and more constraining predictions are very much needed. The overall distribution of spins is determined jointly by the first-generation distribution, the binary pairing procedure (as inferred above), the general-relativistic mapping of binary to remnant properties, and the ejection or retention of merger remnants in host environments. While we account for the dimensionless spin magnitudes of higher-generation binaries in our population modeling, the spin directions are resampled isotropically.

A more solid finding we report is that the spins of 1g+1g binary BHs are limited below the typical dimensionless spin of merger remnants, approximately $0.7$~\cite{2021NatAs...5..749G}. Hierarchical BHs with much lower spins are extremely rare \cite{2021ApJ...915...56G}, yet another indication that some higher-generation binary BHs are required to fit the data with our model (cf. Sec.~\ref{subsecmergen}). We measure spins using two effective parameters: The effective aligned spin $\chieff$ measures the binary spin component parallel to the orbital plane~\cite{2008PhRvD..78d4021R}, and the effective precessing spin $\chip$ measures the in-plane, two-spin projection~\cite{2021PhRvD.103f4067G}. For sources with negligible, misaligned, or (equal-mass) oppositely aligned spins, we have $\chieff\approx0$, while large positive (negative) values indicate high aligned (antialigned) spins. Similarly, $\chip\approx0$ for spins that are small, aligned with the orbital angular momentum, or oppositely aligned in the orbital plane. Nonzero values of $\chip$ indicate the presence of spin precession, with $\chip>1$ being a region exclusively occupied by binaries with precessing spin contributions from both BH components.

Figure~\ref{fig:ppd_spin} displays the PPDs of these two modeled effective spin parameters. In the left panel, we show the distribution of effective aligned spins $\chieff$. Here, the assumption of isotropic spins leads to an overly tight constraint. This mismodeling enforces a distribution that is symmetric about and centered on $\chieff=0$, in contrast with more generic spin models that infer asymmetric distributions skewed to positive $\chieff$~\cite{2021arXiv211103634T} (and thus favoring alignment) or those that rule out negative $\chieff$~\cite{2021PhRvD.104h3010R,2021ApJ...921L..15G}. However, we find that, typically, $|\chieff|\lesssim0.4$, in agreement with the results of Ref.~\cite{2021arXiv211103634T} (\textsc{gaussian spin} model); in particular, we report $|\chieff|<0.46_{-0.06}^{+0.04}$ for 99\% of the population.

On the other hand, the right panel of Fig.~\ref{fig:ppd_spin} shows the distribution of precessing spins measured with $\chip$, where, unlike Ref.~\cite{2021arXiv211103634T}, we observe substructure; note that, although they use the earlier $\chip$ definition of Ref.~\cite{2015PhRvD..91b4043S}, for the majority of events, the two measurements are indistinguishable~\cite{2021PhRvD.103f4067G,2022CQGra..39l5003H}. We note that, like for $\chieff$, the uncertainty is likely also underestimated here due to our modeling assumptions. The distribution features two prominent modes. The primary one appears at $\chi\approx0.2$. A peak at $\chip>0$ is determined by the model, given isotropic spin directions (as is the case for all merger generations in our model) and uniform nonzero spin magnitudes (as for the first-generation binaries). A single-peaked distribution essentially corresponds to the implied $\chip$ prior used in parameter estimation analyses~\cite{2021PhRvD.103f4067G}. If this feature is astrophysical in origin rather than due to our model choices, however, it may imply that sources with at least moderately misaligned spins ---and thus undergoing spin precession--- make up a sizable portion of the population. The shape and location of this mode are in broad agreement with the results of Ref.~\cite{2021arXiv211103634T}; see their Fig.~16.

However, in contrast to their finding that $\chip$ measurements can be explained by \emph{either} a narrow distribution with peak $\chip\approx0.2$ \emph{or} a broad distribution centered on $\chip=0$ (which results in multimodality when marginalized over the posterior uncertainty), we find that individual distributions drawn according to the hyperposterior \emph{always} decrease at $\chip=0$, peak at $\chip\approx0.2$, \emph{and} feature a secondary mode typically around $\chip\approx0.6$. While our population model naturally accommodates such multimodal structure, the \textsc{gaussian spin} model employed in Ref.~\cite{2021arXiv211103634T} only allows for a single peak and is thus unable to jointly capture the narrow $\chip\approx0.2$ peak in addition to the extended distribution above $\chip\gtrsim0.5$, instead favoring one or the other. Indeed, a single Gaussian distribution cannot fit the distribution of $\chip$ within the 90\% credible bounds.

The inset in the $\chip$ panel of Fig.~\ref{fig:ppd_spin} shows the same distribution with logarithmic scaling to highlight smaller-scale features. The distribution falls off above the feature at $\chip\approx0.6$ before a tertiary buildup at $\chip\approx1$ and a final minor mode at $\chi\approx1.25$, with a large decline in between and eventual declivity beyond. We find minor evidence for a population of sources occupying the exclusive two-spin region $\chip>1$; the 99\% quantile lies at $\chip=0.95_{-0.13}^{+0.07}$ while $P(\chip>1)=0.8_{-0.4}^{+0.6}\%$. There is no support in the population for $\chip\gtrsim1.5$.

Turning to the origins of these spin features, the precessing spin posterior we measure differs from a population prior with uniform spin magnitudes and directions due to the inferred constraint $\chimax<1$, leading to a shift towards lower values and, more importantly, the feature at $\chip\approx0.6$, which is not explainable with such a model.

To test whether the posterior constraints are really due to measurements of precession or correlations with other parameters ---primarily the best-measured spin $\chieff$ and the mass ratio $q$--- we repeat the hierarchical inference but replace the $\chip$ posterior for each event in Eq.~(\ref{eq:poppos'}) with samples from the parameter estimation prior. The measured 90\% confidence intervals for the effective spin posteriors are shown by the dashed purple lines in Fig.~\ref{fig:ppd_spin}. The constraints are qualitatively the same, with only small differences in the 90\% credible regions. The purple $\chip$ distribution favors slightly larger values in the region $\chip<0.6$, suggesting the real precession measurements from GW data offer \emph{some} information beyond the parameter estimation prior, but the differences are minor. The most informative constraints originate from the aforementioned better-measured parameters.

\begin{figure}
\label{fig:ppd_chip}
\includegraphics[width=1\columnwidth]{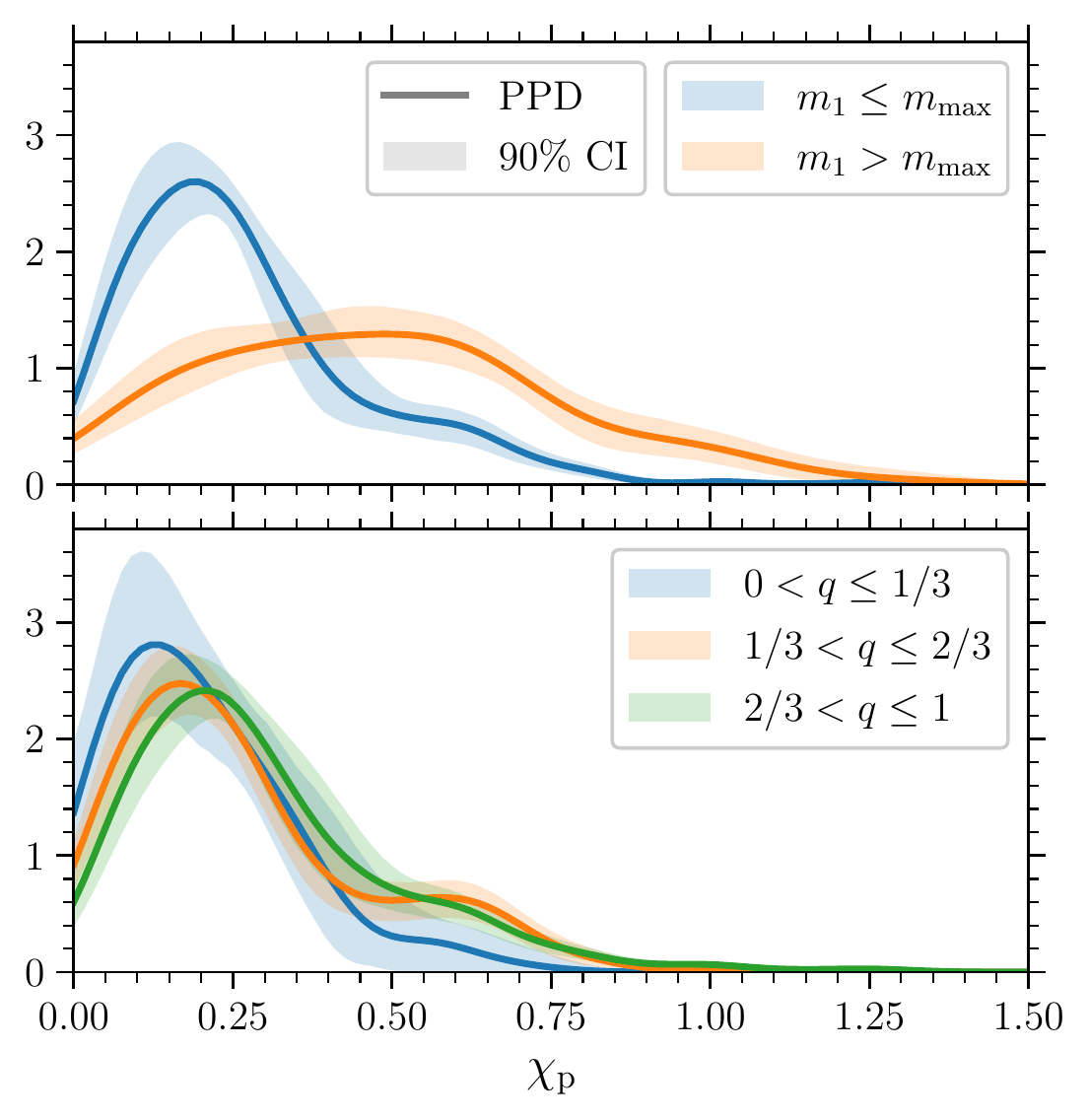}
{\caption{PPD of the effective precessing spin $\chip$ as a function of the primary mass $m_1$ (top panel) and mass ratio $q$ (bottom panel). The mean distributions are given by the solid lines and the symmetric 90\% confidence intervals are given by the shaded bands. In the top panel, we split the $\chip$ posterior for primary masses below (blue) and above (orange) the maximum first-generation mass $\mmax$. In the bottom panel, we split the $\chip$ posterior for mass ratios $0<q\leq1/3$ (blue), $1/3<q\leq2/3$ (orange), and $2/3<q\leq1$ (green).}}
\end{figure}

In our single-channel model, a BH with mass above $\mmax$ is necessarily a merger remnant. Since merger remnants have large spins, this model requires heavy BHs to have large spins if there are masses above $\mmax$ in the catalog and natal spins are small, as inferred above. Our DNN model naturally allows for correlations between parameters, unlike simple phenomenological priors, so we can assess which masses contribute to the $\chip\approx0.6$ feature. In the top panel of Fig.~\ref{fig:ppd_chip}, we split the inferred $\chip$ population posterior into contributions from primary masses $m_1\leq\mmax$, which can be both first- or higher-generation BHs, and definitely higher-generation sources with $m_1>\mmax$. Though the latter, heavier population of sources necessarily has a preference for larger spins, the contribution to the $\chip$ distribution from sources with $m_1\leq\mmax$ still contains the feature at $\chip\approx0.6$, implying that this inference is not solely driven by the requirement for heavy BHs to have large spins. This is a consequence of the previous conclusion that hierarchical mergers in our model also populate the region $m_1\leq\mmax$ due to the bottom-heavy mass function.

In the bottom panel of Fig.~\ref{fig:ppd_chip}, we similarly observe the population distribution of $\chip$ as a function of the mass ratio $q$. Larger precessing spins are suppressed for more unequal masses $q\leq1/3$ with the peak lowered to $\chip\approx0.15$, while for mass ratios $q>2/3$ closer to unity, it increases to the slightly larger value $\chip\approx0.25$. Spins around $\chip\approx0.6$ are present for these mass ratios, but the distinct feature is most prominent for $1/3<q\leq2/3$. This region of the parameter space is prominently occupied by mixed-generation mergers, e.g., 1g+2g. We verify that this secondary structure is consistent with repeated mergers in our model as follows. Starting with the 1g+1g PPD and binary pairing measurements to compute the distribution of 2g remnant masses and dimensionless spin magnitudes (computed with Ref.~\cite{2016PhRvD..93l4066G} as in Sec.~\ref{sec:repeated}), the distributions of $\chip$ for binaries formed either with a 1g BH and a remnant BH (i.e., 1g+2g) or two remnant BHs (i.e., 2g+2g) both feature peaks at $\chip\approx0.6$. This is because, for 1g+2g sources, the dominant contribution to $\chip$ is from the primary, which is more likely to be 2g, while for 2g+2g sources, the primary and secondary are more likely to contribute equally such that their average is similar to the 1g+2g case.

\subsection{Merger generations}
\label{subsecmergen}

\begin{figure*}
\label{fig:fg}
\includegraphics[width=1\textwidth]{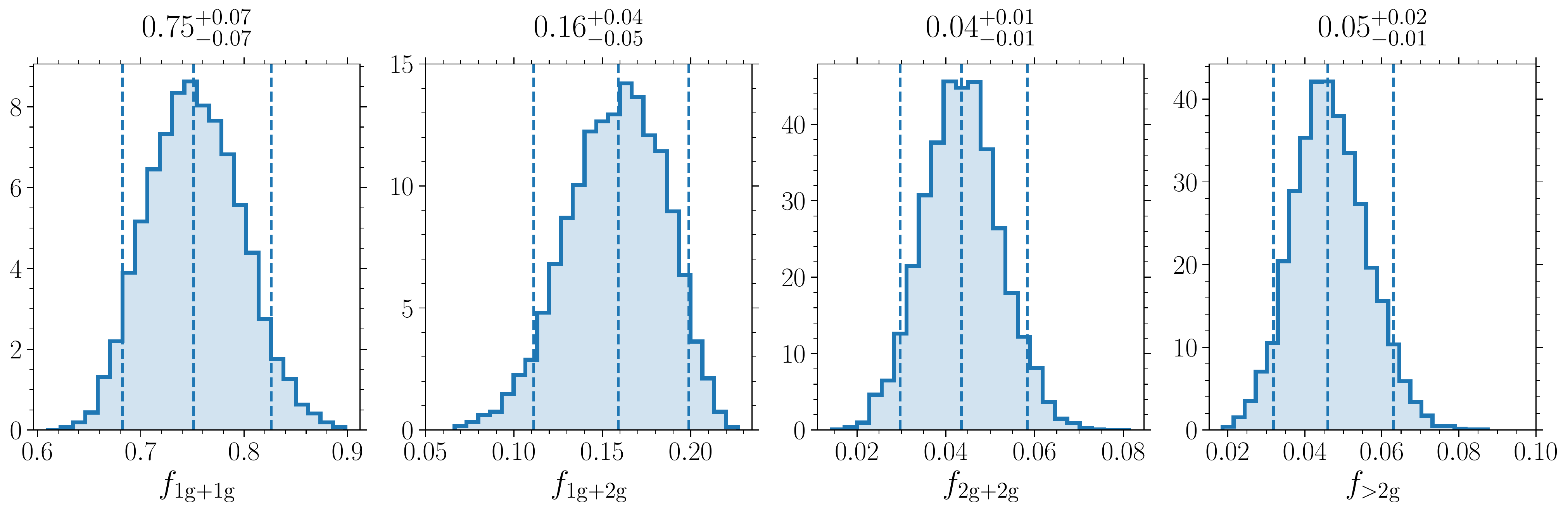}
{\caption{Distributions of the branching fractions (left to right: $f_\mathrm{1g+1g}$, $f_\mathrm{1g+2g}$, $f_\mathrm{2g+2g}$, $f_\mathrm{>2g}$) for merger generations in the astrophysical distribution of merging stellar-mass binary BHs, as measured with our deep-learning approach to population inference on the GWTC-3 catalog. The median and symmetric 90\% confidence region for each generation fraction is reported above ---and plotted as vertical dashed lines within--- the corresponding panel.}}
\end{figure*}

Given our DNN population model, the observations of the previous sections suggest the presence of hierarchical mergers in the underlying population of merging stellar-mass binary BHs. Taking samples $\lambda \sim p(\lambda|d)$ from the posterior distribution of population parameters in Eq.~(\ref{eq:poppos samples}), the corresponding draws from the posterior of merger-generation fractions can be derived as $f_g(\lambda) \sim p(f_g|d)$, where $f_g$ is given by the DNN described in Sec.~\ref{sec:fractions}.

Figure~\ref{fig:fg} presents the posterior distributions of the fractional contribution to the population from each binary merger generation; the medians and 90\% symmetric intervals are quoted and indicated as vertical dashed lines. In the underlying distribution, $75_{-7}^{+7}\%$ of sources contain only first-generation BHs (1g+1g), which implies around 25\% contain a component that is the remnant of a previous merger, with 90\% (99\%) one-sided support for $1-f_\mathrm{1g+1g} \gtrsim 0.19$ (0.14). Mixed-generation binaries with both a first- and second-generation component make up the second-largest portion of the population, with $f_\mathrm{1g+2g}=0.16_{-0.05}^{+0.04}$, while binaries containing two second-generation BHs or any component of even higher generation contribute equally at about the $5\%$ level ($f_\mathrm{2g+2g}=0.04_{-0.01}^{+0.01}$ and $f_\mathrm{>2g}=0.05_{-0.01}^{+0.02}$, respectively).

Previous studies of older GW catalogs found weak evidence for the presence of hierarchical mergers~\cite{2020ApJ...893...35D,2020ApJ...900..177K}. However, further detections through GWTC-2 brought the addition of events whose properties, including higher masses and mass ratios, hinted at higher-generation origins. Reference~\cite{2021ApJ...915L..35K} presented a population analysis based on a phenomenological model of globular clusters, implying the presence of at least one second-generation BH in the GWTC-2 events with greater than $96\%$ probability, rising to greater than $99.99\%$ when considering their highest Bayes factor model corresponding to an escape speed $\vesc\sim300\kmps$. In this case, they found median relative merger rates of 0.15 and 0.01 when comparing 1g+2g and 2g+2g binaries to the 1g+1g case, respectively, with 99\% upper limits of 0.29 and 0.04. Equivalently, in our GWTC-3 analysis, we find broadly consistent relative branching fractions $f_\mathrm{1g+2g}/f_\mathrm{1g+1g}=0.21_{-0.08}^{+0.08}$ and $f_\mathrm{2g+2g}/f_\mathrm{1g+1g}=0.06_{-0.02}^{+0.03}$ (reporting medians and symmetric 90\% confidence intervals). Given the disparity of the underlying model assumptions between the two analyses and the addition of new detections in GWTC-3, our results jointly point to the fact that, if admitted in the fitted population, a modest number of binary BHs with hierarchical origin appears necessary to best explain the data.

\section{Summary and future prospects}
\label{sec:summary}

Our findings were made possible by advances in the treatment of the GW data, in particular, exploiting deep-learning techniques. These aspects are summarized in Sec.~\ref{sec:summary astro} and \ref{sec:summary nn}, respectively.

\subsection{Astrophysics summary}
\label{sec:summary astro}

We fit current LIGO/Virgo data assuming a population of sources that generalizes current phenomenological functional forms while consistently allowing for the occurrence of hierarchical mergers. Therefore, the crucial feature of our model is the separation of first- and higher-generation merger populations, the latter of which is not phenomenological. This feature allows us to place constraints directly on the properties of those BHs born as stellar remnants in addition to the population as a whole. We summarize our key results as follows (quoting medians and 90\% credible regions):

\begin{itemize}
\item The distribution of escape speeds of environments hosting binary BH mergers is relatively flat, though lower values are preferred; modeled as a power law in the range $0<\vesc<500~\kmps$, the index is $\delta=-0.4_{-0.3}^{+0.4}$. Such environments may retain merger remnants since $37_{-12}^{+13}\%$ of 1g+1g remnants receive GW recoils $\vkick<\vesc$.
\item When parametrized as a truncated power law (whose minimum is fixed to $5\Msun$), the distribution of first-generation masses has index $\gamma=-1.4_{-0.4}^{+0.4}$ and thus favors lighter BHs. First-generation BHs have an upper mass limit $\mmax=38_{-2.7}^{+3.8}\Msun$.
\item Negative power-law slopes are recovered for the binary-pairing probability distributions, indicating both components are selected with a preference for lighter BHs, though this preference is stronger for secondaries; the primary (secondary) pairing index is $\alpha=-1.2_{-0.7}^{+0.7}$ ($\beta=-2.2_{-1.7}^{+1.8}$). This finding is inconsistent with uniform binary pairing ($\alpha=\beta=0$) at the 90\% level.
\item This results in a primary (secondary) mass distribution that peaks around $m_1\approx12\Msun$ ($m_2\approx8\Msun$), with a buildup and then decline before the first-generation upper mass limit. Mass ratios peak at unity but extend to $q\gtrsim0.1$. While 99\% of the population has masses less than $<59_{-6.5}^{+7.8}\Msun$, there is an extended spectrum beyond the first-generation mass distribution due to repeated mergers.
\item Assuming a distribution of first-generation BH spins that is isotropic in direction and uniform in magnitude, we find that the maximum spin formed in stellar collapse is $\chimax=0.39_{-0.07}^{+0.08}$. The distribution of effective aligned spins features support within $|\chieff|<0.46_{-0.06}^{+0.04}$. The effective precessing spins are multimodal, with a maximum at $\chip\approx0.2$ and a secondary peak due to repeated mergers at $\chip\approx0.6$, but they fall off in the two-spin region with less than or about $1\%$ of the distribution at $\chip>1$ and vanishing support for $\chip\gtrsim1.5$.
\item Approximately 25\% of binaries in the underlying population contain a higher-generation BH, with 99\% one-sided support for a fraction greater than or about $14\%$.
\end{itemize}

While we are able to highlight some key insights into the astrophysics of stellar-mass black-hole mergers, we stress that our DNN population model is based on simulations that are simplified by, e.g., employing various power-law parametrizations. This work serves as a test case to demonstrate the efficacy of the modeling procedure by bridging the gap between phenomenological and accurate simulated models.

The complexity of our simulated populations can be increased in various ways. First, we model the spin distributions of first-generation binary BHs as uniform and isotropic, and while we account for the spin magnitude of merger remnants, we continue the assumption of isotropicity in higher-generation mergers. Employing more sophisticated spin magnitude models and adding an additional hyperparameter to control the degree of first-generation spin alignment, we can better capture the behavior of a wider class of host environments, e.g., isolated evolution~\cite{2018PhRvD..98h4036G} or the disks of active galactic nuclei~\cite{2020MNRAS.494.1203M}. More generally, allowing for contributions from a mixture of distinct formation channels would lead to a more realistic fit. In particular, $\chieff$ has been shown to favor positive values, which may indicate a significant contribution from isolated binary formation to the merger rate~\cite{2021arXiv211103634T}, but we only consider a single-channel dynamical formation model. By underestimating the location of the $\chieff$ distribution, the fraction of hierarchical mergers may be overestimated~\cite{2022ApJ...935L..26F}. The added complexity of multichannel modeling is beyond the scope of this first study, and we aim to address it in future work. Capturing more realistic distributions of higher-generation spins requires both retaining information on post-merger spin directions and characterizing any changes in relative orientation during binary formation.

Second, we model redshifts with a fixed distribution corresponding to a rate of events that is uniform in comoving volume and source-frame time. A simple extension would be to include a parametrized redshift model \cite{2018ApJ...863L..41F}, though this would also increase the dimensionality of the hyperparameter space. We consider the mergers in our simulated populations as an ensemble and do not account for dynamical assembly of hierarchical mergers, i.e., for the fact that a remnant BH can only form a new binary at later times than its parent system \cite{2017PhRvD..95l4046G}. In practice, one should include the effect of time delays between formation and merger and thus more realistically model the merger rate; we leave such explorations to future work.

\subsection{Deep-learning summary}
\label{sec:summary nn}

Compared to previous work~\cite{2018PhRvD..98h3017T,2019PhRvD.100h3015W}, the approach presented here replaces the approximation of simulated binary BH distributions via histograms with Gaussian KDEs, and the emulation of these distributions across both the source- and population-level parameter spaces via GPR with DNNs. While GPR has been shown to be an ineffective approach in higher dimensions~\cite{2021arXiv211206707C}, alternative deep-learning techniques such as normalizing flows~\cite{2017arXiv170507057P} have proven successful~\cite{2020PhRvD.101l3005W,2021PhRvD.103b3026W,2021PhRvD.103h3021W}.

Rather than training on probability density evaluations (the required number of which scales exponentially with the dimensionality) as in this work, normalizing flows are trained directly on samples from the true distribution (thus scaling linearly with dimensionality), making the latter more effective in high-dimensional spaces. Further, our methodology requires truncating the population model in the unbounded chirp mass parameter in order to generate training data and employ numerical normalization, introducing a refactoring term in the population-level likelihood. This issue may be solved with domain compactification by, e.g., modeling the inverse of the mass scale instead. Normalizing flows also have the advantage of being generative models (i.e., from which new predictive samples can be drawn) that additionally provide density estimation with correctly enforced normalization. However, typically either just the forward (density prediction) or inverse (sample generation) model is efficient to evaluate~\cite{2017arXiv170507057P,2017arXiv171110433V}. On the other hand, neither deep-learning approach provides any estimation of modeling uncertainty, whereas GPR does; this is an area where Bayesian deep learning may prove fruitful (see, e.g., Ref.~\cite{2020arXiv200706823V} for a hands-on approach).

That said, our DNN framework has some advantages. The separation of density estimation and emulation adds a level of flexibility not otherwise available. For example, outputs of current state-of-the-art stellar-physics codes provide evolutionary tracks that act as proxies for a given contribution to the merger rate, in practice outputting a set of weighted binary BHs. This can be trivially implemented in our formalism without any modification to the underlying computational framework by including sample weights when fitting the KDEs. The KDE-first approach also allows for sufficient smoothing of the training distributions prior to the learning stage. This distinction is important for simulations with modeling choices that lead to nonphysical numerical discretization of outputs or low sampling densities; e.g., despite employing normalizing flows, Ref.~\cite{2021PhRvD.103h3021W} required kernel density resampling to boost the set of simulated BH mergers.

More importantly, in a follow-up study, we will also explore the potential for emulating labeled subpopulations within a given model. Taking the case at hand, one may wish to individually model the distributions of each merger generation rather than the combined distribution as a whole. Each of these are themselves not probability densities due to potential degeneracies in the hyperparameter space and must be modeled correspondingly; e.g., for low escape speeds, the entire population will be contained within the first-generation label while the others will have empty supports. Since each distribution is formed by the same generative process, a single model with multidimensional output should be used to predict the partitioned populations; this requirement can be readily satisfied in our DNN framework with the inclusion of additional output neurons whose activation functions, combined with the overall numerical normalization, constrain predictions to the required unit summation.

\section{Conclusions}
\label{sec:conclusions}

In this work, we have made use of multiple DNNs to tackle several aspects of the inference of GW catalogs. We focused on stellar-mass binary BH mergers, which are target sources of current ground-based GW observatories. In particular, we considered populations of binary BHs containing repeated mergers ---systems in which individual BHs may be born not only from stellar collapse (i.e., as first-generation BHs) but as the remnants of (potentially multiple) previous binary mergers (higher-generation BHs).

Starting from simple phenomenological parametrizations of clusterlike progenitor host environments, the mass and spin distributions of first-generation BHs, and binary pairing, we constructed a suite of simulated mergers, using NR fitting formulas to estimate the properties ---mass, spin magnitude, and kick--- of remnants, self-consistently accounting for their ejection from (or retention in) the host clusters due to gravitational recoil. The resulting hierarchical merger populations are complex and cannot be represented with closed-form expressions, as is precisely the case for more realistic progenitor modeling (e.g., population synthesis simulations). This a textbook case where machine-learning methods can show their full potential.

We trained a high-dimensional DNN to act as an emulator for our population model, interpolating across parameters at both a four-dimensional source level and six-dimensional population level. By approximating the detection probabilities and recording the merger generations of individual sources in the simulated populations, we also constructed DNNs to predict, respectively, the fraction of detectable events and the generational branching ratios across the population-level parameter space. These applications of deep learning are then combined with (rate-marginalized) hierarchical Bayesian inference of the events in GWTC-3, performed here with nested sampling, to make measurements of the population-level parameters and reconstruct the astrophysical distribution of merging stellar-mass binary BHs.

This work serves as a showcase of the developments made possible by combining advanced techniques from the fields of deep learning and statistical analysis, applied within the context of GW astrophysics. Our deep-learning population pipeline, which we applied to the case study of simple simulations of hierarchical stellar-mass BH mergers, is thus ready to be used in conjunction with more sophisticated simulated populations. Combined with the state of the art in population synthesis, we will be able to constrain the properties of progenitor formation environments by directly comparing GW data with higher-dimensional models of binary evolution.

\acknowledgements

We thank Daria Gangardt, Nicola Giacobbo, Nathan Steinle,
Chris Moore, Geraint Pratten, Riccardo Buscicchio, Alberto Vecchio, Tom Callister, Mike Zevin, and Kaze Wong for useful discussions. M.M. and D.G. are supported by European Union's H2020 ERC Starting Grant No. 945155-GWmining, Cariplo Foundation Grant No. 2021-0555, Leverhulme Trust Grant No. RPG-2019-350, and Royal Society Grant No. RGS-R2-202004.
M.M. acknowledges networking support by the European COST Action CA16104--GWverse. S.R.T. acknowledges support from NSF Grant No. AST-2007993, the NANOGrav Physics Frontier Center NSF Grant No. PHY-2020265, an NSF CAREER Award No. PHY-2146016, and a Vanderbilt University College of Arts \& Science Dean's Faculty Fellowship.
Computational work was performed on the University of Birmingham BlueBEAR cluster, the Baskerville Tier 2 HPC service funded by EPSRC Grant No. EP/T022221/1, and at CINECA with allocations through INFN, Bicocca, and ISCRA Type-B Project No. HP10BEQ9JB.

\appendix

\section{EVENT SELECTION}
\label{app:events}

In Table~\ref{tab:events}, we report all of the GW events that enter our population analysis. We include only the binary BH mergers with $\mathrm{FAR} < 1~\mathrm{yr}^{-1}$ in at least one of the detection pipelines. The 69 events are equivalent to the 76 listed in Table~I of Ref.~\cite{2021arXiv211103634T}, excluding those that potentially contain a neutron star (GW170817, GW190425\_081805, GW190426\_152155, GW190814, GW190917\_114630, GW200105\_162426, and GW200115\_042309).

\setlength{\LTcapwidth}{\textwidth}
\rowcolors{2}{gray!15}{white}
\renewcommand{\arraystretch}{1.3}
\setlength{\tabcolsep}{10pt}
\begin{longtable*}{ccccccc}
\label{tab:events}\\
\hline
Event & Catalog & $\min\mathrm{FAR}~(\mathrm{yr}^{-1})$ & $\Mchirp/M_\odot$ & $q$ & $\chieff$ & $\chip$ \\
\hline
GW150914 & GWTC-1 & $<1\times10^{-5}$ & $28.38_{-1.47}^{+1.56}$ & $0.86_{-0.19}^{+0.12}$ & $-0.04_{-0.11}^{+0.11}$ & $0.34_{-0.27}^{+0.50}$ \\
GW151012 & GWTC-1 & $7.92\times10^{-3}$ & $15.27_{-1.11}^{+1.55}$ & $0.71_{-0.37}^{+0.26}$ & $0.01_{-0.17}^{+0.22}$ & $0.34_{-0.26}^{+0.46}$ \\
GW151226 & GWTC-1 & $<1\times10^{-5}$ & $8.90_{-0.30}^{+0.32}$ & $0.66_{-0.33}^{+0.30}$ & $0.17_{-0.06}^{+0.13}$ & $0.44_{-0.27}^{+0.29}$ \\
GW170104 & GWTC-1 & $<1\times10^{-5}$ & $21.62_{-1.84}^{+2.04}$ & $0.70_{-0.25}^{+0.26}$ & $-0.05_{-0.20}^{+0.16}$ & $0.39_{-0.30}^{+0.39}$ \\
GW170608 & GWTC-1 & $<1\times10^{-5}$ & $7.95_{-0.18}^{+0.19}$ & $0.75_{-0.34}^{+0.22}$ & $0.04_{-0.05}^{+0.13}$ & $0.32_{-0.25}^{+0.40}$ \\
GW170729 & GWTC-1 & $1.80\times10^{-1}$ & $35.23_{-4.87}^{+6.33}$ & $0.66_{-0.28}^{+0.30}$ & $0.32_{-0.26}^{+0.22}$ & $0.44_{-0.30}^{+0.46}$ \\
GW170809 & GWTC-1 & $<1\times10^{-5}$ & $24.93_{-1.63}^{+2.19}$ & $0.73_{-0.25}^{+0.25}$ & $0.05_{-0.15}^{+0.17}$ & $0.37_{-0.27}^{+0.41}$ \\
GW170814 & GWTC-1 & $<1\times10^{-5}$ & $24.17_{-1.23}^{+1.45}$ & $0.86_{-0.22}^{+0.13}$ & $0.07_{-0.12}^{+0.12}$ & $0.50_{-0.40}^{+0.50}$ \\
GW170818 & GWTC-1 & $<1\times10^{-5}$ & $26.67_{-1.75}^{+1.99}$ & $0.78_{-0.24}^{+0.19}$ & $-0.10_{-0.21}^{+0.17}$ & $0.51_{-0.34}^{+0.42}$ \\
GW170823 & GWTC-1 & $<1\times10^{-5}$ & $29.29_{-3.25}^{+4.49}$ & $0.77_{-0.27}^{+0.20}$ & $0.05_{-0.21}^{+0.20}$ & $0.46_{-0.34}^{+0.50}$ \\
GW190408\_181802 & GWTC-2 & $<1\times10^{-5}$ & $18.32_{-1.23}^{+1.87}$ & $0.75_{-0.25}^{+0.21}$ & $-0.03_{-0.19}^{+0.13}$ & $0.38_{-0.30}^{+0.39}$ \\
GW190412 & GWTC-2 & $<1\times10^{-5}$ & $13.26_{-0.33}^{+0.40}$ & $0.28_{-0.06}^{+0.12}$ & $0.25_{-0.10}^{+0.08}$ & $0.32_{-0.16}^{+0.18}$ \\
GW190413\_052954 & GWTC-2 & $8.17\times10^{-1}$ & $24.70_{-4.03}^{+5.43}$ & $0.68_{-0.26}^{+0.28}$ & $-0.01_{-0.35}^{+0.28}$ & $0.42_{-0.31}^{+0.46}$ \\
GW190413\_134308 & GWTC-2 & $1.81\times10^{-1}$ & $33.00_{-5.29}^{+8.35}$ & $0.69_{-0.32}^{+0.27}$ & $-0.03_{-0.29}^{+0.24}$ & $0.56_{-0.42}^{+0.48}$ \\
GW190421\_213856 & GWTC-2 & $2.83\times10^{-3}$ & $31.18_{-4.32}^{+5.90}$ & $0.79_{-0.31}^{+0.18}$ & $-0.05_{-0.27}^{+0.22}$ & $0.47_{-0.35}^{+0.51}$ \\
GW190503\_185404 & GWTC-2 & $<1\times10^{-5}$ & $30.09_{-4.23}^{+4.52}$ & $0.66_{-0.24}^{+0.28}$ & $-0.03_{-0.26}^{+0.21}$ & $0.40_{-0.30}^{+0.44}$ \\
GW190512\_180714 & GWTC-2 & $<1\times10^{-5}$ & $14.60_{-1.00}^{+1.27}$ & $0.53_{-0.17}^{+0.37}$ & $0.03_{-0.14}^{+0.12}$ & $0.22_{-0.17}^{+0.35}$ \\
GW190513\_205428 & GWTC-2 & $<1\times10^{-5}$ & $21.57_{-1.92}^{+3.78}$ & $0.52_{-0.19}^{+0.41}$ & $0.11_{-0.17}^{+0.28}$ & $0.31_{-0.23}^{+0.38}$ \\
GW190517\_055101 & GWTC-2 & $3.47\times10^{-4}$ & $26.58_{-3.93}^{+3.78}$ & $0.68_{-0.28}^{+0.27}$ & $0.52_{-0.20}^{+0.19}$ & $0.51_{-0.30}^{+0.37}$ \\
GW190519\_153544 & GWTC-2 & $<1\times10^{-5}$ & $44.43_{-7.28}^{+6.26}$ & $0.61_{-0.19}^{+0.26}$ & $0.31_{-0.22}^{+0.20}$ & $0.46_{-0.29}^{+0.38}$ \\
GW190521 & GWTC-2 & $<1\times10^{-5}$ & $69.11_{-10.55}^{+17.31}$ & $0.75_{-0.34}^{+0.22}$ & $0.03_{-0.39}^{+0.31}$ & $0.71_{-0.47}^{+0.55}$ \\
GW190521\_074359 & GWTC-2 & $1.00\times10^{-2}$ & $32.01_{-2.44}^{+3.29}$ & $0.77_{-0.20}^{+0.19}$ & $0.09_{-0.13}^{+0.10}$ & $0.39_{-0.28}^{+0.34}$ \\
GW190527\_092055 & GWTC-2 & $2.28\times10^{-1}$ & $24.30_{-4.26}^{+9.13}$ & $0.64_{-0.32}^{+0.32}$ & $0.11_{-0.27}^{+0.27}$ & $0.45_{-0.34}^{+0.50}$ \\
GW190602\_175927 & GWTC-2 & $<1\times10^{-5}$ & $48.97_{-8.72}^{+8.88}$ & $0.70_{-0.33}^{+0.25}$ & $0.07_{-0.24}^{+0.27}$ & $0.42_{-0.30}^{+0.53}$ \\
GW190620\_030421 & GWTC-2 & $1.12\times10^{-2}$ & $38.22_{-6.45}^{+8.10}$ & $0.62_{-0.27}^{+0.33}$ & $0.32_{-0.24}^{+0.23}$ & $0.44_{-0.29}^{+0.41}$ \\
GW190630\_185205 & GWTC-2 & $<1\times10^{-5}$ & $24.98_{-2.16}^{+2.01}$ & $0.68_{-0.22}^{+0.27}$ & $0.09_{-0.12}^{+0.12}$ & $0.32_{-0.23}^{+0.34}$ \\
GW190701\_203306 & GWTC-2 & $5.71\times10^{-3}$ & $40.25_{-5.12}^{+5.40}$ & $0.77_{-0.32}^{+0.21}$ & $-0.07_{-0.30}^{+0.22}$ & $0.41_{-0.30}^{+0.51}$ \\
GW190706\_222641 & GWTC-2 & $<1\times10^{-5}$ & $42.81_{-7.17}^{+10.20}$ & $0.58_{-0.25}^{+0.34}$ & $0.28_{-0.29}^{+0.25}$ & $0.40_{-0.29}^{+0.45}$ \\
GW190707\_093326 & GWTC-2 & $<1\times10^{-5}$ & $8.47_{-0.42}^{+0.63}$ & $0.73_{-0.24}^{+0.22}$ & $-0.03_{-0.08}^{+0.09}$ & $0.26_{-0.21}^{+0.35}$ \\
GW190708\_232457 & GWTC-2 & $3.09\times10^{-4}$ & $13.15_{-0.66}^{+0.89}$ & $0.75_{-0.27}^{+0.22}$ & $0.02_{-0.07}^{+0.09}$ & $0.30_{-0.24}^{+0.39}$ \\
GW190719\_215514 & GWTC-2 & $6.31\times10^{-1}$ & $23.47_{-4.02}^{+6.62}$ & $0.58_{-0.30}^{+0.36}$ & $0.32_{-0.31}^{+0.28}$ & $0.43_{-0.30}^{+0.40}$ \\
GW190720\_000836 & GWTC-2 & $<1\times10^{-5}$ & $8.79_{-0.77}^{+0.59}$ & $0.63_{-0.28}^{+0.31}$ & $0.18_{-0.11}^{+0.13}$ & $0.30_{-0.20}^{+0.35}$ \\
GW190725\_174728 & GWTC-2.1 & $4.58\times10^{-1}$ & $7.44_{-0.54}^{+0.56}$ & $0.57_{-0.31}^{+0.37}$ & $-0.04_{-0.14}^{+0.26}$ & $0.38_{-0.29}^{+0.49}$ \\
GW190727\_060333 & GWTC-2 & $<1\times10^{-5}$ & $28.61_{-3.83}^{+5.33}$ & $0.79_{-0.31}^{+0.18}$ & $0.11_{-0.25}^{+0.25}$ & $0.48_{-0.37}^{+0.50}$ \\
GW190728\_064510 & GWTC-2 & $<1\times10^{-5}$ & $8.62_{-0.33}^{+0.52}$ & $0.69_{-0.30}^{+0.27}$ & $0.12_{-0.06}^{+0.13}$ & $0.29_{-0.20}^{+0.30}$ \\
GW190731\_140936 & GWTC-2 & $3.35\times10^{-1}$ & $29.55_{-5.19}^{+6.68}$ & $0.72_{-0.31}^{+0.25}$ & $0.05_{-0.23}^{+0.25}$ & $0.41_{-0.31}^{+0.51}$ \\
GW190803\_022701 & GWTC-2 & $7.32\times10^{-2}$ & $27.26_{-3.97}^{+5.48}$ & $0.74_{-0.31}^{+0.23}$ & $-0.03_{-0.27}^{+0.24}$ & $0.45_{-0.33}^{+0.54}$ \\
GW190805\_211137 & GWTC-2.1 & $6.28\times10^{-1}$ & $33.68_{-7.23}^{+9.75}$ & $0.68_{-0.32}^{+0.27}$ & $0.35_{-0.36}^{+0.30}$ & $0.55_{-0.36}^{+0.47}$ \\
GW190828\_063405 & GWTC-2 & $<1\times10^{-5}$ & $24.95_{-2.15}^{+3.48}$ & $0.82_{-0.22}^{+0.15}$ & $0.19_{-0.16}^{+0.16}$ & $0.43_{-0.31}^{+0.45}$ \\
GW190828\_065509 & GWTC-2 & $<1\times10^{-5}$ & $13.36_{-0.97}^{+1.20}$ & $0.43_{-0.16}^{+0.38}$ & $0.08_{-0.16}^{+0.16}$ & $0.29_{-0.23}^{+0.39}$ \\
GW190910\_112807 & GWTC-2 & $2.87\times10^{-3}$ & $34.25_{-3.96}^{+4.21}$ & $0.82_{-0.23}^{+0.15}$ & $0.02_{-0.18}^{+0.18}$ & $0.40_{-0.31}^{+0.40}$ \\
GW190915\_235702 & GWTC-2 & $<1\times10^{-5}$ & $25.05_{-2.59}^{+3.02}$ & $0.79_{-0.29}^{+0.19}$ & $0.02_{-0.23}^{+0.19}$ & $0.57_{-0.41}^{+0.41}$ \\
GW190924\_021846 & GWTC-2 & $<1\times10^{-5}$ & $5.76_{-0.21}^{+0.26}$ & $0.60_{-0.25}^{+0.33}$ & $0.02_{-0.08}^{+0.14}$ & $0.22_{-0.17}^{+0.33}$ \\
GW190925\_232845 & GWTC-2.1 & $7.20\times10^{-3}$ & $15.78_{-0.95}^{+1.06}$ & $0.74_{-0.29}^{+0.22}$ & $0.11_{-0.14}^{+0.16}$ & $0.40_{-0.29}^{+0.42}$ \\
GW190929\_012149 & GWTC-2 & $1.55\times10^{-1}$ & $34.09_{-6.67}^{+9.34}$ & $0.35_{-0.17}^{+0.36}$ & $0.01_{-0.28}^{+0.27}$ & $0.38_{-0.28}^{+0.43}$ \\
GW190930\_133541 & GWTC-2 & $1.23\times10^{-2}$ & $8.51_{-0.48}^{+0.49}$ & $0.66_{-0.36}^{+0.28}$ & $0.14_{-0.13}^{+0.19}$ & $0.34_{-0.24}^{+0.34}$ \\
GW191103\_012549 & GWTC-3 & $4.58\times10^{-1}$ & $8.34_{-0.57}^{+0.65}$ & $0.67_{-0.36}^{+0.29}$ & $0.21_{-0.10}^{+0.16}$ & $0.41_{-0.26}^{+0.40}$ \\
GW191105\_143521 & GWTC-3 & $1.18\times10^{-2}$ & $7.81_{-0.45}^{+0.61}$ & $0.72_{-0.30}^{+0.24}$ & $-0.02_{-0.10}^{+0.12}$ & $0.30_{-0.24}^{+0.45}$ \\
GW191109\_010717 & GWTC-3 & $1.80\times10^{-4}$ & $47.08_{-7.33}^{+9.52}$ & $0.72_{-0.23}^{+0.22}$ & $-0.30_{-0.29}^{+0.39}$ & $0.60_{-0.36}^{+0.67}$ \\
GW191127\_050227 & GWTC-3 & $2.49\times10^{-1}$ & $29.73_{-9.16}^{+11.70}$ & $0.47_{-0.36}^{+0.46}$ & $0.17_{-0.36}^{+0.34}$ & $0.52_{-0.41}^{+0.44}$ \\
GW191129\_134029 & GWTC-3 & $<1\times10^{-5}$ & $7.30_{-0.28}^{+0.43}$ & $0.64_{-0.29}^{+0.30}$ & $0.06_{-0.07}^{+0.16}$ & $0.27_{-0.20}^{+0.36}$ \\
GW191204\_171526 & GWTC-3 & $<1\times10^{-5}$ & $8.56_{-0.27}^{+0.38}$ & $0.69_{-0.26}^{+0.26}$ & $0.16_{-0.05}^{+0.08}$ & $0.40_{-0.26}^{+0.35}$ \\
GW191215\_223052 & GWTC-3 & $<1\times10^{-5}$ & $18.34_{-1.65}^{+2.20}$ & $0.73_{-0.27}^{+0.24}$ & $-0.04_{-0.21}^{+0.17}$ & $0.52_{-0.38}^{+0.43}$ \\
GW191216\_213338 & GWTC-3 & $<1\times10^{-5}$ & $8.33_{-0.19}^{+0.22}$ & $0.62_{-0.28}^{+0.32}$ & $0.11_{-0.06}^{+0.13}$ & $0.24_{-0.16}^{+0.33}$ \\
GW191222\_033537 & GWTC-3 & $<1\times10^{-5}$ & $33.82_{-5.01}^{+7.04}$ & $0.80_{-0.32}^{+0.18}$ & $-0.04_{-0.25}^{+0.19}$ & $0.41_{-0.30}^{+0.49}$ \\
GW191230\_180458 & GWTC-3 & $5.02\times10^{-2}$ & $36.37_{-5.55}^{+8.13}$ & $0.76_{-0.33}^{+0.22}$ & $-0.06_{-0.30}^{+0.27}$ & $0.53_{-0.39}^{+0.51}$ \\
GW200112\_155838 & GWTC-3 & $<1\times10^{-5}$ & $27.34_{-2.01}^{+2.59}$ & $0.80_{-0.26}^{+0.17}$ & $0.06_{-0.14}^{+0.15}$ & $0.39_{-0.30}^{+0.39}$ \\
GW200128\_022011 & GWTC-3 & $4.29\times10^{-3}$ & $32.04_{-5.53}^{+7.55}$ & $0.80_{-0.31}^{+0.17}$ & $0.12_{-0.25}^{+0.24}$ & $0.60_{-0.42}^{+0.54}$ \\
GW200129\_065458 & GWTC-3 & $<1\times10^{-5}$ & $27.16_{-2.28}^{+2.07}$ & $0.85_{-0.41}^{+0.12}$ & $0.11_{-0.16}^{+0.11}$ & $0.50_{-0.35}^{+0.47}$ \\
GW200202\_154313 & GWTC-3 & $<1\times10^{-5}$ & $7.49_{-0.20}^{+0.23}$ & $0.72_{-0.32}^{+0.25}$ & $0.04_{-0.06}^{+0.14}$ & $0.29_{-0.22}^{+0.41}$ \\
GW200208\_130117 & GWTC-3 & $3.11\times10^{-4}$ & $27.71_{-3.05}^{+3.55}$ & $0.73_{-0.28}^{+0.24}$ & $-0.07_{-0.26}^{+0.21}$ & $0.38_{-0.29}^{+0.46}$ \\
GW200209\_085452 & GWTC-3 & $4.64\times10^{-2}$ & $26.77_{-4.17}^{+5.93}$ & $0.79_{-0.32}^{+0.19}$ & $-0.12_{-0.30}^{+0.24}$ & $0.52_{-0.39}^{+0.57}$ \\
GW200216\_220804 & GWTC-3 & $3.50\times10^{-1}$ & $32.93_{-8.51}^{+9.15}$ & $0.61_{-0.40}^{+0.34}$ & $0.10_{-0.38}^{+0.34}$ & $0.46_{-0.35}^{+0.48}$ \\
GW200219\_094415 & GWTC-3 & $9.94\times10^{-4}$ & $27.72_{-3.93}^{+5.74}$ & $0.77_{-0.32}^{+0.20}$ & $-0.08_{-0.29}^{+0.23}$ & $0.47_{-0.34}^{+0.50}$ \\
GW200224\_222234 & GWTC-3 & $<1\times10^{-5}$ & $31.09_{-2.56}^{+3.18}$ & $0.82_{-0.26}^{+0.16}$ & $0.10_{-0.15}^{+0.14}$ & $0.49_{-0.35}^{+0.46}$ \\
GW200225\_060421 & GWTC-3 & $<1\times10^{-5}$ & $14.21_{-1.33}^{+1.47}$ & $0.73_{-0.28}^{+0.24}$ & $-0.11_{-0.28}^{+0.18}$ & $0.55_{-0.40}^{+0.41}$ \\
GW200302\_015811 & GWTC-3 & $1.12\times10^{-1}$ & $23.36_{-2.92}^{+4.67}$ & $0.53_{-0.20}^{+0.36}$ & $0.01_{-0.26}^{+0.25}$ & $0.38_{-0.29}^{+0.44}$ \\
GW200311\_115853 & GWTC-3 & $<1\times10^{-5}$ & $26.58_{-1.92}^{+2.35}$ & $0.81_{-0.27}^{+0.16}$ & $-0.02_{-0.19}^{+0.16}$ & $0.44_{-0.33}^{+0.45}$ \\
GW200316\_215756 & GWTC-3 & $<1\times10^{-5}$ & $8.75_{-0.55}^{+0.65}$ & $0.59_{-0.38}^{+0.34}$ & $0.13_{-0.10}^{+0.28}$ & $0.31_{-0.20}^{+0.38}$ \\
\hline
\caption{Binary BH mergers that enter our analysis. Each event passes the cut $\mathrm{FAR} < 1~\mathrm{yr}^{-1}$ in at least one of the searches. We exclude any events that potentially contain a neutron star. For each event, we list the catalog in which it was first reported and the minimum FAR. We list the medians and 90\% symmetric intervals for the chirp mass $\Mchirp$, mass ratio $q$, effective aligned spin $\chieff$, and effective precessing spin $\chip$. The reference frequency used to measure $\chip$ is 20~Hz for all events except GW190521\_030229, which is measured at 11~Hz.}
\end{longtable*}

\bibliography{draft}

\begin{thebibliography}{152}%
\makeatletter
\providecommand \@ifxundefined [1]{%
 \@ifx{#1\undefined}
}%
\providecommand \@ifnum [1]{%
 \ifnum #1\expandafter \@firstoftwo
 \else \expandafter \@secondoftwo
 \fi
}%
\providecommand \@ifx [1]{%
 \ifx #1\expandafter \@firstoftwo
 \else \expandafter \@secondoftwo
 \fi
}%
\providecommand \natexlab [1]{#1}%
\providecommand \enquote  [1]{``#1''}%
\providecommand \bibnamefont  [1]{#1}%
\providecommand \bibfnamefont [1]{#1}%
\providecommand \citenamefont [1]{#1}%
\providecommand \href@noop [0]{\@secondoftwo}%
\providecommand \href [0]{\begingroup \@sanitize@url \@href}%
\providecommand \@href[1]{\@@startlink{#1}\@@href}%
\providecommand \@@href[1]{\endgroup#1\@@endlink}%
\providecommand \@sanitize@url [0]{\catcode `\\12\catcode `\$12\catcode
  `\&12\catcode `\#12\catcode `\^12\catcode `\_12\catcode `\%12\relax}%
\providecommand \@@startlink[1]{}%
\providecommand \@@endlink[0]{}%
\providecommand \url  [0]{\begingroup\@sanitize@url \@url }%
\providecommand \@url [1]{\endgroup\@href {#1}{\urlprefix }}%
\providecommand \urlprefix  [0]{URL }%
\providecommand \Eprint [0]{\href }%
\providecommand \doibase [0]{https://doi.org/}%
\providecommand \selectlanguage [0]{\@gobble}%
\providecommand \bibinfo  [0]{\@secondoftwo}%
\providecommand \bibfield  [0]{\@secondoftwo}%
\providecommand \translation [1]{[#1]}%
\providecommand \BibitemOpen [0]{}%
\providecommand \bibitemStop [0]{}%
\providecommand \bibitemNoStop [0]{.\EOS\space}%
\providecommand \EOS [0]{\spacefactor3000\relax}%
\providecommand \BibitemShut  [1]{\csname bibitem#1\endcsname}%
\let\auto@bib@innerbib\@empty
\bibitem [{\citenamefont {{Aasi}}\ \emph {et~al.}(2015)\citenamefont {{Aasi}}
  \emph {et~al.}}]{2015CQGra..32g4001L}%
  \BibitemOpen
  \bibfield  {author} {\bibinfo {author} {\bibfnamefont {J.}~\bibnamefont
  {{Aasi}}} \emph {et~al.} (\bibinfo {collaboration} {LIGO Collaboration}),\
  }\href {https://doi.org/10.1088/0264-9381/32/7/074001} {\bibfield  {journal}
  {\bibinfo  {journal} {\cqg}\ }\textbf {\bibinfo {volume} {32}},\ \bibinfo
  {eid} {074001} (\bibinfo {year} {2015})},\ \Eprint
  {https://arxiv.org/abs/1411.4547} {arXiv:1411.4547 [gr-qc]} \BibitemShut
  {NoStop}%
\bibitem [{\citenamefont {{Acernese}}\ \emph {et~al.}(2015)\citenamefont
  {{Acernese}} \emph {et~al.}}]{2015CQGra..32b4001A}%
  \BibitemOpen
  \bibfield  {author} {\bibinfo {author} {\bibfnamefont {F.}~\bibnamefont
  {{Acernese}}} \emph {et~al.} (\bibinfo {collaboration} {Virgo
  Collaboration}),\ }\href {https://doi.org/10.1088/0264-9381/32/2/024001}
  {\bibfield  {journal} {\bibinfo  {journal} {\cqg}\ }\textbf {\bibinfo
  {volume} {32}},\ \bibinfo {eid} {024001} (\bibinfo {year} {2015})},\ \Eprint
  {https://arxiv.org/abs/1408.3978} {arXiv:1408.3978 [gr-qc]} \BibitemShut
  {NoStop}%
\bibitem [{\citenamefont {{Abbott}}\ \emph
  {et~al.}(2019{\natexlab{a}})\citenamefont {{Abbott}} \emph
  {et~al.}}]{2019PhRvX...9c1040A}%
  \BibitemOpen
  \bibfield  {author} {\bibinfo {author} {\bibfnamefont {B.~P.}\ \bibnamefont
  {{Abbott}}} \emph {et~al.} (\bibinfo {collaboration} {LIGO and Virgo
  Collaboration}),\ }\href {https://doi.org/10.1103/PhysRevX.9.031040}
  {\bibfield  {journal} {\bibinfo  {journal} {\prx}\ }\textbf {\bibinfo
  {volume} {9}},\ \bibinfo {eid} {031040} (\bibinfo {year}
  {2019}{\natexlab{a}})},\ \Eprint {https://arxiv.org/abs/1811.12907}
  {arXiv:1811.12907 [astro-ph.HE]} \BibitemShut {NoStop}%
\bibitem [{\citenamefont {{Abbott}}\ \emph
  {et~al.}(2021{\natexlab{a}})\citenamefont {{Abbott}} \emph
  {et~al.}}]{2021PhRvX..11b1053A}%
  \BibitemOpen
  \bibfield  {author} {\bibinfo {author} {\bibfnamefont {R.}~\bibnamefont
  {{Abbott}}} \emph {et~al.} (\bibinfo {collaboration} {LIGO and Virgo
  Collaboration}),\ }\href {https://doi.org/10.1103/PhysRevX.11.021053}
  {\bibfield  {journal} {\bibinfo  {journal} {\prx}\ }\textbf {\bibinfo
  {volume} {11}},\ \bibinfo {eid} {021053} (\bibinfo {year}
  {2021}{\natexlab{a}})},\ \Eprint {https://arxiv.org/abs/2010.14527}
  {arXiv:2010.14527 [gr-qc]} \BibitemShut {NoStop}%
\bibitem [{\citenamefont {{Abbott}}\ \emph
  {et~al.}(2021{\natexlab{b}})\citenamefont {{Abbott}} \emph
  {et~al.}}]{2021arXiv210801045T}%
  \BibitemOpen
  \bibfield  {author} {\bibinfo {author} {\bibfnamefont {R.}~\bibnamefont
  {{Abbott}}} \emph {et~al.} (\bibinfo {collaboration} {LIGO and Virgo
  Collaboration}),\ }\href@noop {} {\bibfield  {journal} {\bibinfo  {journal}
  {{}}\ } (\bibinfo {year} {2021}{\natexlab{b}})},\ \Eprint
  {https://arxiv.org/abs/2108.01045} {arXiv:2108.01045 [gr-qc]} \BibitemShut
  {NoStop}%
\bibitem [{\citenamefont {{Abbott}}\ \emph
  {et~al.}(2021{\natexlab{c}})\citenamefont {{Abbott}} \emph
  {et~al.}}]{2021arXiv211103606T}%
  \BibitemOpen
  \bibfield  {author} {\bibinfo {author} {\bibfnamefont {R.}~\bibnamefont
  {{Abbott}}} \emph {et~al.} (\bibinfo {collaboration} {LIGO, Virgo, and KAGRA
  Collaboration}),\ }\href@noop {} {\bibfield  {journal} {\bibinfo  {journal}
  {{}}\ } (\bibinfo {year} {2021}{\natexlab{c}})},\ \Eprint
  {https://arxiv.org/abs/2111.03606} {arXiv:2111.03606 [gr-qc]} \BibitemShut
  {NoStop}%
\bibitem [{\citenamefont {{Nitz}}\ \emph
  {et~al.}(2021{\natexlab{a}})\citenamefont {{Nitz}}, \citenamefont {{Capano}},
  \citenamefont {{Kumar}}, \citenamefont {{Wang}}, \citenamefont {{Kastha}},
  \citenamefont {{Sch{\"a}fer}}, \citenamefont {{Dhurkunde}},\ and\
  \citenamefont {{Cabero}}}]{2021ApJ...922...76N}%
  \BibitemOpen
  \bibfield  {author} {\bibinfo {author} {\bibfnamefont {A.~H.}\ \bibnamefont
  {{Nitz}}}, \bibinfo {author} {\bibfnamefont {C.~D.}\ \bibnamefont
  {{Capano}}}, \bibinfo {author} {\bibfnamefont {S.}~\bibnamefont {{Kumar}}},
  \bibinfo {author} {\bibfnamefont {Y.-F.}\ \bibnamefont {{Wang}}}, \bibinfo
  {author} {\bibfnamefont {S.}~\bibnamefont {{Kastha}}}, \bibinfo {author}
  {\bibfnamefont {M.}~\bibnamefont {{Sch{\"a}fer}}}, \bibinfo {author}
  {\bibfnamefont {R.}~\bibnamefont {{Dhurkunde}}},\ and\ \bibinfo {author}
  {\bibfnamefont {M.}~\bibnamefont {{Cabero}}},\ }\href
  {https://doi.org/10.3847/1538-4357/ac1c03} {\bibfield  {journal} {\bibinfo
  {journal} {\apj}\ }\textbf {\bibinfo {volume} {922}},\ \bibinfo {eid} {76}
  (\bibinfo {year} {2021}{\natexlab{a}})},\ \Eprint
  {https://arxiv.org/abs/2105.09151} {arXiv:2105.09151 [astro-ph.HE]}
  \BibitemShut {NoStop}%
\bibitem [{\citenamefont {{Nitz}}\ \emph
  {et~al.}(2021{\natexlab{b}})\citenamefont {{Nitz}}, \citenamefont {{Kumar}},
  \citenamefont {{Wang}}, \citenamefont {{Kastha}}, \citenamefont {{Wu}},
  \citenamefont {{Sch{\"a}fer}}, \citenamefont {{Dhurkunde}},\ and\
  \citenamefont {{Capano}}}]{2021arXiv211206878N}%
  \BibitemOpen
  \bibfield  {author} {\bibinfo {author} {\bibfnamefont {A.~H.}\ \bibnamefont
  {{Nitz}}}, \bibinfo {author} {\bibfnamefont {S.}~\bibnamefont {{Kumar}}},
  \bibinfo {author} {\bibfnamefont {Y.-F.}\ \bibnamefont {{Wang}}}, \bibinfo
  {author} {\bibfnamefont {S.}~\bibnamefont {{Kastha}}}, \bibinfo {author}
  {\bibfnamefont {S.}~\bibnamefont {{Wu}}}, \bibinfo {author} {\bibfnamefont
  {M.}~\bibnamefont {{Sch{\"a}fer}}}, \bibinfo {author} {\bibfnamefont
  {R.}~\bibnamefont {{Dhurkunde}}},\ and\ \bibinfo {author} {\bibfnamefont
  {C.~D.}\ \bibnamefont {{Capano}}},\ }\href@noop {} {\bibfield  {journal}
  {\bibinfo  {journal} {{}}\ } (\bibinfo {year} {2021}{\natexlab{b}})},\
  \Eprint {https://arxiv.org/abs/2112.06878} {arXiv:2112.06878 [astro-ph.HE]}
  \BibitemShut {NoStop}%
\bibitem [{\citenamefont {{Olsen}}\ \emph {et~al.}(2022)\citenamefont
  {{Olsen}}, \citenamefont {{Venumadhav}}, \citenamefont {{Mushkin}},
  \citenamefont {{Roulet}}, \citenamefont {{Zackay}},\ and\ \citenamefont
  {{Zaldarriaga}}}]{2022arXiv220102252O}%
  \BibitemOpen
  \bibfield  {author} {\bibinfo {author} {\bibfnamefont {S.}~\bibnamefont
  {{Olsen}}}, \bibinfo {author} {\bibfnamefont {T.}~\bibnamefont
  {{Venumadhav}}}, \bibinfo {author} {\bibfnamefont {J.}~\bibnamefont
  {{Mushkin}}}, \bibinfo {author} {\bibfnamefont {J.}~\bibnamefont {{Roulet}}},
  \bibinfo {author} {\bibfnamefont {B.}~\bibnamefont {{Zackay}}},\ and\
  \bibinfo {author} {\bibfnamefont {M.}~\bibnamefont {{Zaldarriaga}}},\ }\href
  {https://doi.org/10.1103/PhysRevD.106.043009} {\bibfield  {journal} {\bibinfo
   {journal} {\prd}\ }\textbf {\bibinfo {volume} {106}},\ \bibinfo {eid}
  {043009} (\bibinfo {year} {2022})},\ \Eprint
  {https://arxiv.org/abs/2201.02252} {arXiv:2201.02252 [astro-ph.HE]}
  \BibitemShut {NoStop}%
\bibitem [{\citenamefont {{Postnov}}\ and\ \citenamefont
  {{Yungelson}}(2014)}]{2014LRR....17....3P}%
  \BibitemOpen
  \bibfield  {author} {\bibinfo {author} {\bibfnamefont {K.~A.}\ \bibnamefont
  {{Postnov}}}\ and\ \bibinfo {author} {\bibfnamefont {L.~R.}\ \bibnamefont
  {{Yungelson}}},\ }\href {https://doi.org/10.12942/lrr-2014-3} {\bibfield
  {journal} {\bibinfo  {journal} {\lrr}\ }\textbf {\bibinfo {volume} {17}},\
  \bibinfo {eid} {3} (\bibinfo {year} {2014})},\ \Eprint
  {https://arxiv.org/abs/1403.4754} {arXiv:1403.4754 [astro-ph.HE]}
  \BibitemShut {NoStop}%
\bibitem [{\citenamefont {{Benacquista}}\ and\ \citenamefont
  {{Downing}}(2013)}]{2013LRR....16....4B}%
  \BibitemOpen
  \bibfield  {author} {\bibinfo {author} {\bibfnamefont {M.~J.}\ \bibnamefont
  {{Benacquista}}}\ and\ \bibinfo {author} {\bibfnamefont {J.~M.~B.}\
  \bibnamefont {{Downing}}},\ }\href {https://doi.org/10.12942/lrr-2013-4}
  {\bibfield  {journal} {\bibinfo  {journal} {\lrr}\ }\textbf {\bibinfo
  {volume} {16}},\ \bibinfo {eid} {4} (\bibinfo {year} {2013})},\ \Eprint
  {https://arxiv.org/abs/1110.4423} {arXiv:1110.4423 [astro-ph.SR]}
  \BibitemShut {NoStop}%
\bibitem [{\citenamefont {{Farmer}}\ \emph {et~al.}(2019)\citenamefont
  {{Farmer}}, \citenamefont {{Renzo}}, \citenamefont {{de Mink}}, \citenamefont
  {{Marchant}},\ and\ \citenamefont {{Justham}}}]{2019ApJ...887...53F}%
  \BibitemOpen
  \bibfield  {author} {\bibinfo {author} {\bibfnamefont {R.}~\bibnamefont
  {{Farmer}}}, \bibinfo {author} {\bibfnamefont {M.}~\bibnamefont {{Renzo}}},
  \bibinfo {author} {\bibfnamefont {S.~E.}\ \bibnamefont {{de Mink}}}, \bibinfo
  {author} {\bibfnamefont {P.}~\bibnamefont {{Marchant}}},\ and\ \bibinfo
  {author} {\bibfnamefont {S.}~\bibnamefont {{Justham}}},\ }\href
  {https://doi.org/10.3847/1538-4357/ab518b} {\bibfield  {journal} {\bibinfo
  {journal} {\apj}\ }\textbf {\bibinfo {volume} {887}},\ \bibinfo {eid} {53}
  (\bibinfo {year} {2019})},\ \Eprint {https://arxiv.org/abs/1910.12874}
  {arXiv:1910.12874 [astro-ph.SR]} \BibitemShut {NoStop}%
\bibitem [{\citenamefont {{Woosley}}\ and\ \citenamefont
  {{Heger}}(2021)}]{2021ApJ...912L..31W}%
  \BibitemOpen
  \bibfield  {author} {\bibinfo {author} {\bibfnamefont {S.~E.}\ \bibnamefont
  {{Woosley}}}\ and\ \bibinfo {author} {\bibfnamefont {A.}~\bibnamefont
  {{Heger}}},\ }\href {https://doi.org/10.3847/2041-8213/abf2c4} {\bibfield
  {journal} {\bibinfo  {journal} {\apjl}\ }\textbf {\bibinfo {volume} {912}},\
  \bibinfo {eid} {L31} (\bibinfo {year} {2021})},\ \Eprint
  {https://arxiv.org/abs/2103.07933} {arXiv:2103.07933 [astro-ph.SR]}
  \BibitemShut {NoStop}%
\bibitem [{\citenamefont {{Kalogera}}(2000)}]{2000ApJ...541..319K}%
  \BibitemOpen
  \bibfield  {author} {\bibinfo {author} {\bibfnamefont {V.}~\bibnamefont
  {{Kalogera}}},\ }\href {https://doi.org/10.1086/309400} {\bibfield  {journal}
  {\bibinfo  {journal} {\apj}\ }\textbf {\bibinfo {volume} {541}},\ \bibinfo
  {pages} {319} (\bibinfo {year} {2000})},\ \Eprint
  {https://arxiv.org/abs/astro-ph/9911417} {arXiv:astro-ph/9911417 [astro-ph]}
  \BibitemShut {NoStop}%
\bibitem [{\citenamefont {{Gerosa}}\ \emph
  {et~al.}(2018{\natexlab{a}})\citenamefont {{Gerosa}}, \citenamefont
  {{Berti}}, \citenamefont {{O'Shaughnessy}}, \citenamefont {{Belczynski}},
  \citenamefont {{Kesden}}, \citenamefont {{Wysocki}},\ and\ \citenamefont
  {{Gladysz}}}]{2018PhRvD..98h4036G}%
  \BibitemOpen
  \bibfield  {author} {\bibinfo {author} {\bibfnamefont {D.}~\bibnamefont
  {{Gerosa}}}, \bibinfo {author} {\bibfnamefont {E.}~\bibnamefont {{Berti}}},
  \bibinfo {author} {\bibfnamefont {R.}~\bibnamefont {{O'Shaughnessy}}},
  \bibinfo {author} {\bibfnamefont {K.}~\bibnamefont {{Belczynski}}}, \bibinfo
  {author} {\bibfnamefont {M.}~\bibnamefont {{Kesden}}}, \bibinfo {author}
  {\bibfnamefont {D.}~\bibnamefont {{Wysocki}}},\ and\ \bibinfo {author}
  {\bibfnamefont {W.}~\bibnamefont {{Gladysz}}},\ }\href
  {https://doi.org/10.1103/PhysRevD.98.084036} {\bibfield  {journal} {\bibinfo
  {journal} {\prd}\ }\textbf {\bibinfo {volume} {98}},\ \bibinfo {eid} {084036}
  (\bibinfo {year} {2018}{\natexlab{a}})},\ \Eprint
  {https://arxiv.org/abs/1808.02491} {arXiv:1808.02491 [astro-ph.HE]}
  \BibitemShut {NoStop}%
\bibitem [{\citenamefont {{Steinle}}\ and\ \citenamefont
  {{Kesden}}(2021)}]{2021PhRvD.103f3032S}%
  \BibitemOpen
  \bibfield  {author} {\bibinfo {author} {\bibfnamefont {N.}~\bibnamefont
  {{Steinle}}}\ and\ \bibinfo {author} {\bibfnamefont {M.}~\bibnamefont
  {{Kesden}}},\ }\href {https://doi.org/10.1103/PhysRevD.103.063032} {\bibfield
   {journal} {\bibinfo  {journal} {\prd}\ }\textbf {\bibinfo {volume} {103}},\
  \bibinfo {eid} {063032} (\bibinfo {year} {2021})},\ \Eprint
  {https://arxiv.org/abs/2010.00078} {arXiv:2010.00078 [astro-ph.HE]}
  \BibitemShut {NoStop}%
\bibitem [{\citenamefont {{Gerosa}}\ and\ \citenamefont
  {{Fishbach}}(2021)}]{2021NatAs...5..749G}%
  \BibitemOpen
  \bibfield  {author} {\bibinfo {author} {\bibfnamefont {D.}~\bibnamefont
  {{Gerosa}}}\ and\ \bibinfo {author} {\bibfnamefont {M.}~\bibnamefont
  {{Fishbach}}},\ }\href {https://doi.org/10.1038/s41550-021-01398-w}
  {\bibfield  {journal} {\bibinfo  {journal} {\natastro}\ }\textbf {\bibinfo
  {volume} {5}},\ \bibinfo {pages} {749} (\bibinfo {year} {2021})},\ \Eprint
  {https://arxiv.org/abs/2105.03439} {arXiv:2105.03439 [astro-ph.HE]}
  \BibitemShut {NoStop}%
\bibitem [{\citenamefont {{Bogdanovi{\'c}}}\ \emph {et~al.}(2007)\citenamefont
  {{Bogdanovi{\'c}}}, \citenamefont {{Reynolds}},\ and\ \citenamefont
  {{Miller}}}]{2007ApJ...661L.147B}%
  \BibitemOpen
  \bibfield  {author} {\bibinfo {author} {\bibfnamefont {T.}~\bibnamefont
  {{Bogdanovi{\'c}}}}, \bibinfo {author} {\bibfnamefont {C.~S.}\ \bibnamefont
  {{Reynolds}}},\ and\ \bibinfo {author} {\bibfnamefont {M.~C.}\ \bibnamefont
  {{Miller}}},\ }\href {https://doi.org/10.1086/518769} {\bibfield  {journal}
  {\bibinfo  {journal} {\apjl}\ }\textbf {\bibinfo {volume} {661}},\ \bibinfo
  {pages} {L147} (\bibinfo {year} {2007})},\ \Eprint
  {https://arxiv.org/abs/astro-ph/0703054} {arXiv:astro-ph/0703054 [astro-ph]}
  \BibitemShut {NoStop}%
\bibitem [{\citenamefont {{Gerosa}}\ \emph {et~al.}(2015)\citenamefont
  {{Gerosa}}, \citenamefont {{Kesden}}, \citenamefont {{Sperhake}},
  \citenamefont {{Berti}},\ and\ \citenamefont
  {{O'Shaughnessy}}}]{2015PhRvD..92f4016G}%
  \BibitemOpen
  \bibfield  {author} {\bibinfo {author} {\bibfnamefont {D.}~\bibnamefont
  {{Gerosa}}}, \bibinfo {author} {\bibfnamefont {M.}~\bibnamefont {{Kesden}}},
  \bibinfo {author} {\bibfnamefont {U.}~\bibnamefont {{Sperhake}}}, \bibinfo
  {author} {\bibfnamefont {E.}~\bibnamefont {{Berti}}},\ and\ \bibinfo {author}
  {\bibfnamefont {R.}~\bibnamefont {{O'Shaughnessy}}},\ }\href
  {https://doi.org/10.1103/PhysRevD.92.064016} {\bibfield  {journal} {\bibinfo
  {journal} {\prd}\ }\textbf {\bibinfo {volume} {92}},\ \bibinfo {eid} {064016}
  (\bibinfo {year} {2015})},\ \Eprint {https://arxiv.org/abs/1506.03492}
  {arXiv:1506.03492 [gr-qc]} \BibitemShut {NoStop}%
\bibitem [{\citenamefont {{Mandel}}\ and\ \citenamefont
  {{Farmer}}(2022)}]{2022PhR...955....1M}%
  \BibitemOpen
  \bibfield  {author} {\bibinfo {author} {\bibfnamefont {I.}~\bibnamefont
  {{Mandel}}}\ and\ \bibinfo {author} {\bibfnamefont {A.}~\bibnamefont
  {{Farmer}}},\ }\href {https://doi.org/10.1016/j.physrep.2022.01.003}
  {\bibfield  {journal} {\bibinfo  {journal} {\physrep}\ }\textbf {\bibinfo
  {volume} {955}},\ \bibinfo {pages} {1} (\bibinfo {year} {2022})},\ \Eprint
  {https://arxiv.org/abs/1806.05820} {arXiv:1806.05820 [astro-ph.HE]}
  \BibitemShut {NoStop}%
\bibitem [{\citenamefont {{Mandel}}\ and\ \citenamefont
  {{Broekgaarden}}(2022)}]{2022LRR....25....1M}%
  \BibitemOpen
  \bibfield  {author} {\bibinfo {author} {\bibfnamefont {I.}~\bibnamefont
  {{Mandel}}}\ and\ \bibinfo {author} {\bibfnamefont {F.~S.}\ \bibnamefont
  {{Broekgaarden}}},\ }\href {https://doi.org/10.1007/s41114-021-00034-3}
  {\bibfield  {journal} {\bibinfo  {journal} {\lrr}\ }\textbf {\bibinfo
  {volume} {25}},\ \bibinfo {eid} {1} (\bibinfo {year} {2022})},\ \Eprint
  {https://arxiv.org/abs/2107.14239} {arXiv:2107.14239 [astro-ph.HE]}
  \BibitemShut {NoStop}%
\bibitem [{\citenamefont {{Mapelli}}(2021)}]{2021hgwa.bookE...4M}%
  \BibitemOpen
  \bibfield  {author} {\bibinfo {author} {\bibfnamefont {M.}~\bibnamefont
  {{Mapelli}}},\ }in\ \href {https://doi.org/10.1007/978-981-15-4702-7\_16-1}
  {\emph {\bibinfo {booktitle} {Handbook of Gravitational Wave Astronomy}}}\
  (\bibinfo  {publisher} {Springer},\ \bibinfo {year} {2021})\ p.~\bibinfo
  {pages} {4}\BibitemShut {NoStop}%
\bibitem [{\citenamefont {{Barrett}}\ \emph {et~al.}(2018)\citenamefont
  {{Barrett}}, \citenamefont {{Gaebel}}, \citenamefont {{Neijssel}},
  \citenamefont {{Vigna-G{\'o}mez}}, \citenamefont {{Stevenson}}, \citenamefont
  {{Berry}}, \citenamefont {{Farr}},\ and\ \citenamefont
  {{Mandel}}}]{2018MNRAS.477.4685B}%
  \BibitemOpen
  \bibfield  {author} {\bibinfo {author} {\bibfnamefont {J.~W.}\ \bibnamefont
  {{Barrett}}}, \bibinfo {author} {\bibfnamefont {S.~M.}\ \bibnamefont
  {{Gaebel}}}, \bibinfo {author} {\bibfnamefont {C.~J.}\ \bibnamefont
  {{Neijssel}}}, \bibinfo {author} {\bibfnamefont {A.}~\bibnamefont
  {{Vigna-G{\'o}mez}}}, \bibinfo {author} {\bibfnamefont {S.}~\bibnamefont
  {{Stevenson}}}, \bibinfo {author} {\bibfnamefont {C.~P.~L.}\ \bibnamefont
  {{Berry}}}, \bibinfo {author} {\bibfnamefont {W.~M.}\ \bibnamefont
  {{Farr}}},\ and\ \bibinfo {author} {\bibfnamefont {I.}~\bibnamefont
  {{Mandel}}},\ }\href {https://doi.org/10.1093/mnras/sty908} {\bibfield
  {journal} {\bibinfo  {journal} {\mnras}\ }\textbf {\bibinfo {volume} {477}},\
  \bibinfo {pages} {4685} (\bibinfo {year} {2018})},\ \Eprint
  {https://arxiv.org/abs/1711.06287} {arXiv:1711.06287 [astro-ph.HE]}
  \BibitemShut {NoStop}%
\bibitem [{\citenamefont {{Belczynski}}\ \emph {et~al.}(2022)\citenamefont
  {{Belczynski}}, \citenamefont {{Romagnolo}}, \citenamefont {{Olejak}},
  \citenamefont {{Klencki}}, \citenamefont {{Chattopadhyay}}, \citenamefont
  {{Stevenson}}, \citenamefont {{Coleman Miller}}, \citenamefont {{Lasota}},\
  and\ \citenamefont {{Crowther}}}]{2022ApJ...925...69B}%
  \BibitemOpen
  \bibfield  {author} {\bibinfo {author} {\bibfnamefont {K.}~\bibnamefont
  {{Belczynski}}}, \bibinfo {author} {\bibfnamefont {A.}~\bibnamefont
  {{Romagnolo}}}, \bibinfo {author} {\bibfnamefont {A.}~\bibnamefont
  {{Olejak}}}, \bibinfo {author} {\bibfnamefont {J.}~\bibnamefont {{Klencki}}},
  \bibinfo {author} {\bibfnamefont {D.}~\bibnamefont {{Chattopadhyay}}},
  \bibinfo {author} {\bibfnamefont {S.}~\bibnamefont {{Stevenson}}}, \bibinfo
  {author} {\bibfnamefont {M.}~\bibnamefont {{Coleman Miller}}}, \bibinfo
  {author} {\bibfnamefont {J.~P.}\ \bibnamefont {{Lasota}}},\ and\ \bibinfo
  {author} {\bibfnamefont {P.~A.}\ \bibnamefont {{Crowther}}},\ }\href
  {https://doi.org/10.3847/1538-4357/ac375a} {\bibfield  {journal} {\bibinfo
  {journal} {\apj}\ }\textbf {\bibinfo {volume} {925}},\ \bibinfo {eid} {69}
  (\bibinfo {year} {2022})},\ \Eprint {https://arxiv.org/abs/2108.10885}
  {arXiv:2108.10885 [astro-ph.HE]} \BibitemShut {NoStop}%
\bibitem [{\citenamefont {{Mandel}}\ \emph {et~al.}(2019)\citenamefont
  {{Mandel}}, \citenamefont {{Farr}},\ and\ \citenamefont
  {{Gair}}}]{2019MNRAS.486.1086M}%
  \BibitemOpen
  \bibfield  {author} {\bibinfo {author} {\bibfnamefont {I.}~\bibnamefont
  {{Mandel}}}, \bibinfo {author} {\bibfnamefont {W.~M.}\ \bibnamefont
  {{Farr}}},\ and\ \bibinfo {author} {\bibfnamefont {J.~R.}\ \bibnamefont
  {{Gair}}},\ }\href {https://doi.org/10.1093/mnras/stz896} {\bibfield
  {journal} {\bibinfo  {journal} {\mnras}\ }\textbf {\bibinfo {volume} {486}},\
  \bibinfo {pages} {1086} (\bibinfo {year} {2019})},\ \Eprint
  {https://arxiv.org/abs/1809.02063} {arXiv:1809.02063 [physics.data-an]}
  \BibitemShut {NoStop}%
\bibitem [{\citenamefont {{Vitale}}\ \emph {et~al.}(2022)\citenamefont
  {{Vitale}}, \citenamefont {{Gerosa}}, \citenamefont {{Farr}},\ and\
  \citenamefont {{Taylor}}}]{2022hgwa.bookE..45V}%
  \BibitemOpen
  \bibfield  {author} {\bibinfo {author} {\bibfnamefont {S.}~\bibnamefont
  {{Vitale}}}, \bibinfo {author} {\bibfnamefont {D.}~\bibnamefont {{Gerosa}}},
  \bibinfo {author} {\bibfnamefont {W.~M.}\ \bibnamefont {{Farr}}},\ and\
  \bibinfo {author} {\bibfnamefont {S.~R.}\ \bibnamefont {{Taylor}}},\ }in\
  \href {https://doi.org/10.1007/978-981-15-4702-7_45-1} {\emph {\bibinfo
  {booktitle} {Handbook of Gravitational Wave Astronomy}}}\ (\bibinfo
  {publisher} {Springer},\ \bibinfo {year} {2022})\ p.~\bibinfo {pages}
  {45}\BibitemShut {NoStop}%
\bibitem [{\citenamefont {{Zevin}}\ \emph {et~al.}(2021)\citenamefont
  {{Zevin}}, \citenamefont {{Bavera}}, \citenamefont {{Berry}}, \citenamefont
  {{Kalogera}}, \citenamefont {{Fragos}}, \citenamefont {{Marchant}},
  \citenamefont {{Rodriguez}}, \citenamefont {{Antonini}}, \citenamefont
  {{Holz}},\ and\ \citenamefont {{Pankow}}}]{2021ApJ...910..152Z}%
  \BibitemOpen
  \bibfield  {author} {\bibinfo {author} {\bibfnamefont {M.}~\bibnamefont
  {{Zevin}}}, \bibinfo {author} {\bibfnamefont {S.~S.}\ \bibnamefont
  {{Bavera}}}, \bibinfo {author} {\bibfnamefont {C.~P.~L.}\ \bibnamefont
  {{Berry}}}, \bibinfo {author} {\bibfnamefont {V.}~\bibnamefont {{Kalogera}}},
  \bibinfo {author} {\bibfnamefont {T.}~\bibnamefont {{Fragos}}}, \bibinfo
  {author} {\bibfnamefont {P.}~\bibnamefont {{Marchant}}}, \bibinfo {author}
  {\bibfnamefont {C.~L.}\ \bibnamefont {{Rodriguez}}}, \bibinfo {author}
  {\bibfnamefont {F.}~\bibnamefont {{Antonini}}}, \bibinfo {author}
  {\bibfnamefont {D.~E.}\ \bibnamefont {{Holz}}},\ and\ \bibinfo {author}
  {\bibfnamefont {C.}~\bibnamefont {{Pankow}}},\ }\href
  {https://doi.org/10.3847/1538-4357/abe40e} {\bibfield  {journal} {\bibinfo
  {journal} {\apj}\ }\textbf {\bibinfo {volume} {910}},\ \bibinfo {eid} {152}
  (\bibinfo {year} {2021})},\ \Eprint {https://arxiv.org/abs/2011.10057}
  {arXiv:2011.10057 [astro-ph.HE]} \BibitemShut {NoStop}%
\bibitem [{\citenamefont {{Bouffanais}}\ \emph {et~al.}(2021)\citenamefont
  {{Bouffanais}}, \citenamefont {{Mapelli}}, \citenamefont {{Santoliquido}},
  \citenamefont {{Giacobbo}}, \citenamefont {{Di Carlo}}, \citenamefont
  {{Rastello}}, \citenamefont {{Artale}},\ and\ \citenamefont
  {{Iorio}}}]{2021MNRAS.507.5224B}%
  \BibitemOpen
  \bibfield  {author} {\bibinfo {author} {\bibfnamefont {Y.}~\bibnamefont
  {{Bouffanais}}}, \bibinfo {author} {\bibfnamefont {M.}~\bibnamefont
  {{Mapelli}}}, \bibinfo {author} {\bibfnamefont {F.}~\bibnamefont
  {{Santoliquido}}}, \bibinfo {author} {\bibfnamefont {N.}~\bibnamefont
  {{Giacobbo}}}, \bibinfo {author} {\bibfnamefont {U.~N.}\ \bibnamefont {{Di
  Carlo}}}, \bibinfo {author} {\bibfnamefont {S.}~\bibnamefont {{Rastello}}},
  \bibinfo {author} {\bibfnamefont {M.~C.}\ \bibnamefont {{Artale}}},\ and\
  \bibinfo {author} {\bibfnamefont {G.}~\bibnamefont {{Iorio}}},\ }\href
  {https://doi.org/10.1093/mnras/stab2438} {\bibfield  {journal} {\bibinfo
  {journal} {\mnras}\ }\textbf {\bibinfo {volume} {507}},\ \bibinfo {pages}
  {5224} (\bibinfo {year} {2021})},\ \Eprint {https://arxiv.org/abs/2102.12495}
  {arXiv:2102.12495 [astro-ph.HE]} \BibitemShut {NoStop}%
\bibitem [{\citenamefont {{Abbott}}\ \emph
  {et~al.}(2019{\natexlab{b}})\citenamefont {{Abbott}} \emph
  {et~al.}}]{2019ApJ...882L..24A}%
  \BibitemOpen
  \bibfield  {author} {\bibinfo {author} {\bibfnamefont {B.~P.}\ \bibnamefont
  {{Abbott}}} \emph {et~al.} (\bibinfo {collaboration} {LIGO and Virgo
  Collaboration}),\ }\href {https://doi.org/10.3847/2041-8213/ab3800}
  {\bibfield  {journal} {\bibinfo  {journal} {\apjl}\ }\textbf {\bibinfo
  {volume} {882}},\ \bibinfo {eid} {L24} (\bibinfo {year}
  {2019}{\natexlab{b}})},\ \Eprint {https://arxiv.org/abs/1811.12940}
  {arXiv:1811.12940 [astro-ph.HE]} \BibitemShut {NoStop}%
\bibitem [{\citenamefont {{Abbott}}\ \emph
  {et~al.}(2021{\natexlab{d}})\citenamefont {{Abbott}} \emph
  {et~al.}}]{2021ApJ...913L...7A}%
  \BibitemOpen
  \bibfield  {author} {\bibinfo {author} {\bibfnamefont {R.}~\bibnamefont
  {{Abbott}}} \emph {et~al.} (\bibinfo {collaboration} {LIGO and Virgo
  Collaboration}),\ }\href {https://doi.org/10.3847/2041-8213/abe949}
  {\bibfield  {journal} {\bibinfo  {journal} {\apjl}\ }\textbf {\bibinfo
  {volume} {913}},\ \bibinfo {eid} {L7} (\bibinfo {year}
  {2021}{\natexlab{d}})},\ \Eprint {https://arxiv.org/abs/2010.14533}
  {arXiv:2010.14533 [astro-ph.HE]} \BibitemShut {NoStop}%
\bibitem [{\citenamefont {{Abbott}}\ \emph
  {et~al.}(2021{\natexlab{e}})\citenamefont {{Abbott}} \emph
  {et~al.}}]{2021arXiv211103634T}%
  \BibitemOpen
  \bibfield  {author} {\bibinfo {author} {\bibfnamefont {R.}~\bibnamefont
  {{Abbott}}} \emph {et~al.} (\bibinfo {collaboration} {LIGO, Virgo, and KAGRA
  Collaboration}),\ }\href@noop {} {\bibfield  {journal} {\bibinfo  {journal}
  {{}}\ } (\bibinfo {year} {2021}{\natexlab{e}})},\ \Eprint
  {https://arxiv.org/abs/2111.03634} {arXiv:2111.03634 [astro-ph.HE]}
  \BibitemShut {NoStop}%
\bibitem [{\citenamefont {{Callister}}\ \emph {et~al.}(2021)\citenamefont
  {{Callister}}, \citenamefont {{Haster}}, \citenamefont {{Ng}}, \citenamefont
  {{Vitale}},\ and\ \citenamefont {{Farr}}}]{2021ApJ...922L...5C}%
  \BibitemOpen
  \bibfield  {author} {\bibinfo {author} {\bibfnamefont {T.~A.}\ \bibnamefont
  {{Callister}}}, \bibinfo {author} {\bibfnamefont {C.-J.}\ \bibnamefont
  {{Haster}}}, \bibinfo {author} {\bibfnamefont {K.~K.~Y.}\ \bibnamefont
  {{Ng}}}, \bibinfo {author} {\bibfnamefont {S.}~\bibnamefont {{Vitale}}},\
  and\ \bibinfo {author} {\bibfnamefont {W.~M.}\ \bibnamefont {{Farr}}},\
  }\href {https://doi.org/10.3847/2041-8213/ac2ccc} {\bibfield  {journal}
  {\bibinfo  {journal} {\apjl}\ }\textbf {\bibinfo {volume} {922}},\ \bibinfo
  {eid} {L5} (\bibinfo {year} {2021})},\ \Eprint
  {https://arxiv.org/abs/2106.00521} {arXiv:2106.00521 [astro-ph.HE]}
  \BibitemShut {NoStop}%
\bibitem [{\citenamefont {{Franciolini}}\ and\ \citenamefont
  {{Pani}}(2022)}]{2022PhRvD.105l3024F}%
  \BibitemOpen
  \bibfield  {author} {\bibinfo {author} {\bibfnamefont {G.}~\bibnamefont
  {{Franciolini}}}\ and\ \bibinfo {author} {\bibfnamefont {P.}~\bibnamefont
  {{Pani}}},\ }\href {https://doi.org/10.1103/PhysRevD.105.123024} {\bibfield
  {journal} {\bibinfo  {journal} {\prd}\ }\textbf {\bibinfo {volume} {105}},\
  \bibinfo {eid} {123024} (\bibinfo {year} {2022})},\ \Eprint
  {https://arxiv.org/abs/2201.13098} {arXiv:2201.13098 [astro-ph.HE]}
  \BibitemShut {NoStop}%
\bibitem [{\citenamefont {{Roulet}}\ \emph {et~al.}(2021)\citenamefont
  {{Roulet}}, \citenamefont {{Chia}}, \citenamefont {{Olsen}}, \citenamefont
  {{Dai}}, \citenamefont {{Venumadhav}}, \citenamefont {{Zackay}},\ and\
  \citenamefont {{Zaldarriaga}}}]{2021PhRvD.104h3010R}%
  \BibitemOpen
  \bibfield  {author} {\bibinfo {author} {\bibfnamefont {J.}~\bibnamefont
  {{Roulet}}}, \bibinfo {author} {\bibfnamefont {H.~S.}\ \bibnamefont
  {{Chia}}}, \bibinfo {author} {\bibfnamefont {S.}~\bibnamefont {{Olsen}}},
  \bibinfo {author} {\bibfnamefont {L.}~\bibnamefont {{Dai}}}, \bibinfo
  {author} {\bibfnamefont {T.}~\bibnamefont {{Venumadhav}}}, \bibinfo {author}
  {\bibfnamefont {B.}~\bibnamefont {{Zackay}}},\ and\ \bibinfo {author}
  {\bibfnamefont {M.}~\bibnamefont {{Zaldarriaga}}},\ }\href
  {https://doi.org/10.1103/PhysRevD.104.083010} {\bibfield  {journal} {\bibinfo
   {journal} {\prd}\ }\textbf {\bibinfo {volume} {104}},\ \bibinfo {eid}
  {083010} (\bibinfo {year} {2021})},\ \Eprint
  {https://arxiv.org/abs/2105.10580} {arXiv:2105.10580 [astro-ph.HE]}
  \BibitemShut {NoStop}%
\bibitem [{\citenamefont {{Galaudage}}\ \emph {et~al.}(2021)\citenamefont
  {{Galaudage}}, \citenamefont {{Talbot}}, \citenamefont {{Nagar}},
  \citenamefont {{Jain}}, \citenamefont {{Thrane}},\ and\ \citenamefont
  {{Mandel}}}]{2021ApJ...921L..15G}%
  \BibitemOpen
  \bibfield  {author} {\bibinfo {author} {\bibfnamefont {S.}~\bibnamefont
  {{Galaudage}}}, \bibinfo {author} {\bibfnamefont {C.}~\bibnamefont
  {{Talbot}}}, \bibinfo {author} {\bibfnamefont {T.}~\bibnamefont {{Nagar}}},
  \bibinfo {author} {\bibfnamefont {D.}~\bibnamefont {{Jain}}}, \bibinfo
  {author} {\bibfnamefont {E.}~\bibnamefont {{Thrane}}},\ and\ \bibinfo
  {author} {\bibfnamefont {I.}~\bibnamefont {{Mandel}}},\ }\href
  {https://doi.org/10.3847/2041-8213/ac2f3c} {\bibfield  {journal} {\bibinfo
  {journal} {\apjl}\ }\textbf {\bibinfo {volume} {921}},\ \bibinfo {eid} {L15}
  (\bibinfo {year} {2021})},\ \Eprint {https://arxiv.org/abs/2109.02424}
  {arXiv:2109.02424 [gr-qc]} \BibitemShut {NoStop}%
\bibitem [{\citenamefont {{Mould}}\ and\ \citenamefont
  {{Gerosa}}(2022)}]{2022PhRvD.105b4076M}%
  \BibitemOpen
  \bibfield  {author} {\bibinfo {author} {\bibfnamefont {M.}~\bibnamefont
  {{Mould}}}\ and\ \bibinfo {author} {\bibfnamefont {D.}~\bibnamefont
  {{Gerosa}}},\ }\href {https://doi.org/10.1103/PhysRevD.105.024076} {\bibfield
   {journal} {\bibinfo  {journal} {\prd}\ }\textbf {\bibinfo {volume} {105}},\
  \bibinfo {eid} {024076} (\bibinfo {year} {2022})},\ \Eprint
  {https://arxiv.org/abs/2110.05507} {arXiv:2110.05507 [astro-ph.HE]}
  \BibitemShut {NoStop}%
\bibitem [{\citenamefont {{Romero-Shaw}}\ \emph {et~al.}(2022)\citenamefont
  {{Romero-Shaw}}, \citenamefont {{Thrane}},\ and\ \citenamefont
  {{Lasky}}}]{2022PASA...39...25R}%
  \BibitemOpen
  \bibfield  {author} {\bibinfo {author} {\bibfnamefont {I.~M.}\ \bibnamefont
  {{Romero-Shaw}}}, \bibinfo {author} {\bibfnamefont {E.}~\bibnamefont
  {{Thrane}}},\ and\ \bibinfo {author} {\bibfnamefont {P.~D.}\ \bibnamefont
  {{Lasky}}},\ }\href {https://doi.org/10.1017/pasa.2022.24} {\bibfield
  {journal} {\bibinfo  {journal} {\pasa}\ }\textbf {\bibinfo {volume} {39}},\
  \bibinfo {eid} {e025} (\bibinfo {year} {2022})},\ \Eprint
  {https://arxiv.org/abs/2202.05479} {arXiv:2202.05479 [astro-ph.IM]}
  \BibitemShut {NoStop}%
\bibitem [{\citenamefont {{Edelman}}\ \emph {et~al.}(2022)\citenamefont
  {{Edelman}}, \citenamefont {{Doctor}}, \citenamefont {{Godfrey}},\ and\
  \citenamefont {{Farr}}}]{2022ApJ...924..101E}%
  \BibitemOpen
  \bibfield  {author} {\bibinfo {author} {\bibfnamefont {B.}~\bibnamefont
  {{Edelman}}}, \bibinfo {author} {\bibfnamefont {Z.}~\bibnamefont {{Doctor}}},
  \bibinfo {author} {\bibfnamefont {J.}~\bibnamefont {{Godfrey}}},\ and\
  \bibinfo {author} {\bibfnamefont {B.}~\bibnamefont {{Farr}}},\ }\href
  {https://doi.org/10.3847/1538-4357/ac3667} {\bibfield  {journal} {\bibinfo
  {journal} {\apj}\ }\textbf {\bibinfo {volume} {924}},\ \bibinfo {eid} {101}
  (\bibinfo {year} {2022})},\ \Eprint {https://arxiv.org/abs/2109.06137}
  {arXiv:2109.06137 [astro-ph.HE]} \BibitemShut {NoStop}%
\bibitem [{\citenamefont {{Tiwari}}(2021)}]{2021CQGra..38o5007T}%
  \BibitemOpen
  \bibfield  {author} {\bibinfo {author} {\bibfnamefont {V.}~\bibnamefont
  {{Tiwari}}},\ }\href {https://doi.org/10.1088/1361-6382/ac0b54} {\bibfield
  {journal} {\bibinfo  {journal} {\cqg}\ }\textbf {\bibinfo {volume} {38}},\
  \bibinfo {eid} {155007} (\bibinfo {year} {2021})},\ \Eprint
  {https://arxiv.org/abs/2006.15047} {arXiv:2006.15047 [astro-ph.HE]}
  \BibitemShut {NoStop}%
\bibitem [{\citenamefont {{Mandel}}\ \emph {et~al.}(2017)\citenamefont
  {{Mandel}}, \citenamefont {{Farr}}, \citenamefont {{Colonna}}, \citenamefont
  {{Stevenson}}, \citenamefont {{Ti{\v{n}}o}},\ and\ \citenamefont
  {{Veitch}}}]{2017MNRAS.465.3254M}%
  \BibitemOpen
  \bibfield  {author} {\bibinfo {author} {\bibfnamefont {I.}~\bibnamefont
  {{Mandel}}}, \bibinfo {author} {\bibfnamefont {W.~M.}\ \bibnamefont
  {{Farr}}}, \bibinfo {author} {\bibfnamefont {A.}~\bibnamefont {{Colonna}}},
  \bibinfo {author} {\bibfnamefont {S.}~\bibnamefont {{Stevenson}}}, \bibinfo
  {author} {\bibfnamefont {P.}~\bibnamefont {{Ti{\v{n}}o}}},\ and\ \bibinfo
  {author} {\bibfnamefont {J.}~\bibnamefont {{Veitch}}},\ }\href
  {https://doi.org/10.1093/mnras/stw2883} {\bibfield  {journal} {\bibinfo
  {journal} {\mnras}\ }\textbf {\bibinfo {volume} {465}},\ \bibinfo {pages}
  {3254} (\bibinfo {year} {2017})},\ \Eprint {https://arxiv.org/abs/1608.08223}
  {arXiv:1608.08223 [astro-ph.HE]} \BibitemShut {NoStop}%
\bibitem [{\citenamefont {{Sadiq}}\ \emph {et~al.}(2022)\citenamefont
  {{Sadiq}}, \citenamefont {{Dent}},\ and\ \citenamefont
  {{Wysocki}}}]{2022PhRvD.105l3014S}%
  \BibitemOpen
  \bibfield  {author} {\bibinfo {author} {\bibfnamefont {J.}~\bibnamefont
  {{Sadiq}}}, \bibinfo {author} {\bibfnamefont {T.}~\bibnamefont {{Dent}}},\
  and\ \bibinfo {author} {\bibfnamefont {D.}~\bibnamefont {{Wysocki}}},\ }\href
  {https://doi.org/10.1103/PhysRevD.105.123014} {\bibfield  {journal} {\bibinfo
   {journal} {\prd}\ }\textbf {\bibinfo {volume} {105}},\ \bibinfo {eid}
  {123014} (\bibinfo {year} {2022})},\ \Eprint
  {https://arxiv.org/abs/2112.12659} {arXiv:2112.12659 [gr-qc]} \BibitemShut
  {NoStop}%
\bibitem [{\citenamefont {{Rinaldi}}\ and\ \citenamefont {{Del
  Pozzo}}(2021)}]{2022MNRAS.509.5454R}%
  \BibitemOpen
  \bibfield  {author} {\bibinfo {author} {\bibfnamefont {S.}~\bibnamefont
  {{Rinaldi}}}\ and\ \bibinfo {author} {\bibfnamefont {W.}~\bibnamefont {{Del
  Pozzo}}},\ }\href {https://doi.org/10.1093/mnras/stab3224} {\bibfield
  {journal} {\bibinfo  {journal} {\mnras}\ }\textbf {\bibinfo {volume} {509}},\
  \bibinfo {pages} {5454} (\bibinfo {year} {2021})},\ \Eprint
  {https://arxiv.org/abs/2109.05960} {arXiv:2109.05960 [astro-ph.IM]}
  \BibitemShut {NoStop}%
\bibitem [{\citenamefont {{Taylor}}\ and\ \citenamefont
  {{Gerosa}}(2018)}]{2018PhRvD..98h3017T}%
  \BibitemOpen
  \bibfield  {author} {\bibinfo {author} {\bibfnamefont {S.~R.}\ \bibnamefont
  {{Taylor}}}\ and\ \bibinfo {author} {\bibfnamefont {D.}~\bibnamefont
  {{Gerosa}}},\ }\href {https://doi.org/10.1103/PhysRevD.98.083017} {\bibfield
  {journal} {\bibinfo  {journal} {\prd}\ }\textbf {\bibinfo {volume} {98}},\
  \bibinfo {eid} {083017} (\bibinfo {year} {2018})},\ \Eprint
  {https://arxiv.org/abs/1806.08365} {arXiv:1806.08365 [astro-ph.HE]}
  \BibitemShut {NoStop}%
\bibitem [{\citenamefont {{Wong}}\ and\ \citenamefont
  {{Gerosa}}(2019)}]{2019PhRvD.100h3015W}%
  \BibitemOpen
  \bibfield  {author} {\bibinfo {author} {\bibfnamefont {K.~W.~K.}\
  \bibnamefont {{Wong}}}\ and\ \bibinfo {author} {\bibfnamefont
  {D.}~\bibnamefont {{Gerosa}}},\ }\href
  {https://doi.org/10.1103/PhysRevD.100.083015} {\bibfield  {journal} {\bibinfo
   {journal} {\prd}\ }\textbf {\bibinfo {volume} {100}},\ \bibinfo {eid}
  {083015} (\bibinfo {year} {2019})},\ \Eprint
  {https://arxiv.org/abs/1909.06373} {arXiv:1909.06373 [astro-ph.HE]}
  \BibitemShut {NoStop}%
\bibitem [{\citenamefont {{Cheung}}\ \emph {et~al.}(2021)\citenamefont
  {{Cheung}}, \citenamefont {{Wong}}, \citenamefont {{Hannuksela}},
  \citenamefont {{Li}},\ and\ \citenamefont {{Ho}}}]{2021arXiv211206707C}%
  \BibitemOpen
  \bibfield  {author} {\bibinfo {author} {\bibfnamefont {D.~H.~T.}\
  \bibnamefont {{Cheung}}}, \bibinfo {author} {\bibfnamefont {K.~W.~K.}\
  \bibnamefont {{Wong}}}, \bibinfo {author} {\bibfnamefont {O.~A.}\
  \bibnamefont {{Hannuksela}}}, \bibinfo {author} {\bibfnamefont {T.~G.~F.}\
  \bibnamefont {{Li}}},\ and\ \bibinfo {author} {\bibfnamefont
  {S.}~\bibnamefont {{Ho}}},\ }\href@noop {} {\bibfield  {journal} {\bibinfo
  {journal} {{}}\ } (\bibinfo {year} {2021})},\ \Eprint
  {https://arxiv.org/abs/2112.06707} {arXiv:2112.06707 [astro-ph.IM]}
  \BibitemShut {NoStop}%
\bibitem [{\citenamefont {{Wong}}\ \emph
  {et~al.}(2020{\natexlab{a}})\citenamefont {{Wong}}, \citenamefont
  {{Contardo}},\ and\ \citenamefont {{Ho}}}]{2020PhRvD.101l3005W}%
  \BibitemOpen
  \bibfield  {author} {\bibinfo {author} {\bibfnamefont {K.~W.~K.}\
  \bibnamefont {{Wong}}}, \bibinfo {author} {\bibfnamefont {G.}~\bibnamefont
  {{Contardo}}},\ and\ \bibinfo {author} {\bibfnamefont {S.}~\bibnamefont
  {{Ho}}},\ }\href {https://doi.org/10.1103/PhysRevD.101.123005} {\bibfield
  {journal} {\bibinfo  {journal} {\prd}\ }\textbf {\bibinfo {volume} {101}},\
  \bibinfo {eid} {123005} (\bibinfo {year} {2020}{\natexlab{a}})},\ \Eprint
  {https://arxiv.org/abs/2002.09491} {arXiv:2002.09491 [astro-ph.IM]}
  \BibitemShut {NoStop}%
\bibitem [{\citenamefont {{Papamakarios}}\ \emph {et~al.}(2017)\citenamefont
  {{Papamakarios}}, \citenamefont {{Pavlakou}},\ and\ \citenamefont
  {{Murray}}}]{2017arXiv170507057P}%
  \BibitemOpen
  \bibfield  {author} {\bibinfo {author} {\bibfnamefont {G.}~\bibnamefont
  {{Papamakarios}}}, \bibinfo {author} {\bibfnamefont {T.}~\bibnamefont
  {{Pavlakou}}},\ and\ \bibinfo {author} {\bibfnamefont {I.}~\bibnamefont
  {{Murray}}},\ }\href@noop {} {\bibfield  {journal} {\bibinfo  {journal} {{}}\
  } (\bibinfo {year} {2017})},\ \Eprint {https://arxiv.org/abs/1705.07057}
  {arXiv:1705.07057 [stat.ML]} \BibitemShut {NoStop}%
\bibitem [{\citenamefont {{Wong}}\ \emph
  {et~al.}(2021{\natexlab{a}})\citenamefont {{Wong}}, \citenamefont
  {{Franciolini}}, \citenamefont {{De Luca}}, \citenamefont {{Baibhav}},
  \citenamefont {{Berti}}, \citenamefont {{Pani}},\ and\ \citenamefont
  {{Riotto}}}]{2021PhRvD.103b3026W}%
  \BibitemOpen
  \bibfield  {author} {\bibinfo {author} {\bibfnamefont {K.~W.~K.}\
  \bibnamefont {{Wong}}}, \bibinfo {author} {\bibfnamefont {G.}~\bibnamefont
  {{Franciolini}}}, \bibinfo {author} {\bibfnamefont {V.}~\bibnamefont {{De
  Luca}}}, \bibinfo {author} {\bibfnamefont {V.}~\bibnamefont {{Baibhav}}},
  \bibinfo {author} {\bibfnamefont {E.}~\bibnamefont {{Berti}}}, \bibinfo
  {author} {\bibfnamefont {P.}~\bibnamefont {{Pani}}},\ and\ \bibinfo {author}
  {\bibfnamefont {A.}~\bibnamefont {{Riotto}}},\ }\href
  {https://doi.org/10.1103/PhysRevD.103.023026} {\bibfield  {journal} {\bibinfo
   {journal} {\prd}\ }\textbf {\bibinfo {volume} {103}},\ \bibinfo {eid}
  {023026} (\bibinfo {year} {2021}{\natexlab{a}})},\ \Eprint
  {https://arxiv.org/abs/2011.01865} {arXiv:2011.01865 [gr-qc]} \BibitemShut
  {NoStop}%
\bibitem [{\citenamefont {{Wong}}\ \emph
  {et~al.}(2021{\natexlab{b}})\citenamefont {{Wong}}, \citenamefont
  {{Breivik}}, \citenamefont {{Kremer}},\ and\ \citenamefont
  {{Callister}}}]{2021PhRvD.103h3021W}%
  \BibitemOpen
  \bibfield  {author} {\bibinfo {author} {\bibfnamefont {K.~W.~K.}\
  \bibnamefont {{Wong}}}, \bibinfo {author} {\bibfnamefont {K.}~\bibnamefont
  {{Breivik}}}, \bibinfo {author} {\bibfnamefont {K.}~\bibnamefont
  {{Kremer}}},\ and\ \bibinfo {author} {\bibfnamefont {T.}~\bibnamefont
  {{Callister}}},\ }\href {https://doi.org/10.1103/PhysRevD.103.083021}
  {\bibfield  {journal} {\bibinfo  {journal} {\prd}\ }\textbf {\bibinfo
  {volume} {103}},\ \bibinfo {eid} {083021} (\bibinfo {year}
  {2021}{\natexlab{b}})},\ \Eprint {https://arxiv.org/abs/2011.03564}
  {arXiv:2011.03564 [astro-ph.HE]} \BibitemShut {NoStop}%
\bibitem [{\citenamefont {{Gerosa}}\ \emph
  {et~al.}(2020{\natexlab{a}})\citenamefont {{Gerosa}}, \citenamefont
  {{Pratten}},\ and\ \citenamefont {{Vecchio}}}]{2020PhRvD.102j3020G}%
  \BibitemOpen
  \bibfield  {author} {\bibinfo {author} {\bibfnamefont {D.}~\bibnamefont
  {{Gerosa}}}, \bibinfo {author} {\bibfnamefont {G.}~\bibnamefont
  {{Pratten}}},\ and\ \bibinfo {author} {\bibfnamefont {A.}~\bibnamefont
  {{Vecchio}}},\ }\href {https://doi.org/10.1103/PhysRevD.102.103020}
  {\bibfield  {journal} {\bibinfo  {journal} {\prd}\ }\textbf {\bibinfo
  {volume} {102}},\ \bibinfo {eid} {103020} (\bibinfo {year}
  {2020}{\natexlab{a}})},\ \Eprint {https://arxiv.org/abs/2007.06585}
  {arXiv:2007.06585 [astro-ph.HE]} \BibitemShut {NoStop}%
\bibitem [{\citenamefont {{Talbot}}\ and\ \citenamefont
  {{Thrane}}(2022)}]{2022ApJ...927...76T}%
  \BibitemOpen
  \bibfield  {author} {\bibinfo {author} {\bibfnamefont {C.}~\bibnamefont
  {{Talbot}}}\ and\ \bibinfo {author} {\bibfnamefont {E.}~\bibnamefont
  {{Thrane}}},\ }\href {https://doi.org/10.3847/1538-4357/ac4bc0} {\bibfield
  {journal} {\bibinfo  {journal} {\apj}\ }\textbf {\bibinfo {volume} {927}},\
  \bibinfo {eid} {76} (\bibinfo {year} {2022})}\BibitemShut {NoStop}%
\bibitem [{\citenamefont {{Wong}}\ \emph
  {et~al.}(2020{\natexlab{b}})\citenamefont {{Wong}}, \citenamefont {{Ng}},\
  and\ \citenamefont {{Berti}}}]{2020arXiv200710350W}%
  \BibitemOpen
  \bibfield  {author} {\bibinfo {author} {\bibfnamefont {K.~W.~K.}\
  \bibnamefont {{Wong}}}, \bibinfo {author} {\bibfnamefont {K.~K.~Y.}\
  \bibnamefont {{Ng}}},\ and\ \bibinfo {author} {\bibfnamefont
  {E.}~\bibnamefont {{Berti}}},\ }\href@noop {} {\bibfield  {journal} {\bibinfo
   {journal} {{}}\ } (\bibinfo {year} {2020}{\natexlab{b}})},\ \Eprint
  {https://arxiv.org/abs/2007.10350} {arXiv:2007.10350 [astro-ph.HE]}
  \BibitemShut {NoStop}%
\bibitem [{\citenamefont {{Abbott}}\ \emph
  {et~al.}(2020{\natexlab{a}})\citenamefont {{Abbott}} \emph
  {et~al.}}]{2020PhRvD.102d3015A}%
  \BibitemOpen
  \bibfield  {author} {\bibinfo {author} {\bibfnamefont {R.}~\bibnamefont
  {{Abbott}}} \emph {et~al.} (\bibinfo {collaboration} {LIGO and Virgo
  Collaboration}),\ }\href {https://doi.org/10.1103/PhysRevD.102.043015}
  {\bibfield  {journal} {\bibinfo  {journal} {\prd}\ }\textbf {\bibinfo
  {volume} {102}},\ \bibinfo {eid} {043015} (\bibinfo {year}
  {2020}{\natexlab{a}})},\ \Eprint {https://arxiv.org/abs/2004.08342}
  {arXiv:2004.08342 [astro-ph.HE]} \BibitemShut {NoStop}%
\bibitem [{\citenamefont {{Gerosa}}\ \emph
  {et~al.}(2020{\natexlab{b}})\citenamefont {{Gerosa}}, \citenamefont
  {{Vitale}},\ and\ \citenamefont {{Berti}}}]{2020PhRvL.125j1103G}%
  \BibitemOpen
  \bibfield  {author} {\bibinfo {author} {\bibfnamefont {D.}~\bibnamefont
  {{Gerosa}}}, \bibinfo {author} {\bibfnamefont {S.}~\bibnamefont {{Vitale}}},\
  and\ \bibinfo {author} {\bibfnamefont {E.}~\bibnamefont {{Berti}}},\ }\href
  {https://doi.org/10.1103/PhysRevLett.125.101103} {\bibfield  {journal}
  {\bibinfo  {journal} {\prl}\ }\textbf {\bibinfo {volume} {125}},\ \bibinfo
  {eid} {101103} (\bibinfo {year} {2020}{\natexlab{b}})},\ \Eprint
  {https://arxiv.org/abs/2005.04243} {arXiv:2005.04243 [astro-ph.HE]}
  \BibitemShut {NoStop}%
\bibitem [{\citenamefont {{Rodriguez}}\ \emph {et~al.}(2020)\citenamefont
  {{Rodriguez}}, \citenamefont {{Kremer}}, \citenamefont {{Grudi{\'c}}},
  \citenamefont {{Hafen}}, \citenamefont {{Chatterjee}}, \citenamefont
  {{Fragione}}, \citenamefont {{Lamberts}}, \citenamefont {{Martinez}},
  \citenamefont {{Rasio}}, \citenamefont {{Weatherford}},\ and\ \citenamefont
  {{Ye}}}]{2020ApJ...896L..10R}%
  \BibitemOpen
  \bibfield  {author} {\bibinfo {author} {\bibfnamefont {C.~L.}\ \bibnamefont
  {{Rodriguez}}}, \bibinfo {author} {\bibfnamefont {K.}~\bibnamefont
  {{Kremer}}}, \bibinfo {author} {\bibfnamefont {M.~Y.}\ \bibnamefont
  {{Grudi{\'c}}}}, \bibinfo {author} {\bibfnamefont {Z.}~\bibnamefont
  {{Hafen}}}, \bibinfo {author} {\bibfnamefont {S.}~\bibnamefont
  {{Chatterjee}}}, \bibinfo {author} {\bibfnamefont {G.}~\bibnamefont
  {{Fragione}}}, \bibinfo {author} {\bibfnamefont {A.}~\bibnamefont
  {{Lamberts}}}, \bibinfo {author} {\bibfnamefont {M.~A.~S.}\ \bibnamefont
  {{Martinez}}}, \bibinfo {author} {\bibfnamefont {F.~A.}\ \bibnamefont
  {{Rasio}}}, \bibinfo {author} {\bibfnamefont {N.}~\bibnamefont
  {{Weatherford}}},\ and\ \bibinfo {author} {\bibfnamefont {C.~S.}\
  \bibnamefont {{Ye}}},\ }\href {https://doi.org/10.3847/2041-8213/ab961d}
  {\bibfield  {journal} {\bibinfo  {journal} {\apjl}\ }\textbf {\bibinfo
  {volume} {896}},\ \bibinfo {eid} {L10} (\bibinfo {year} {2020})},\ \Eprint
  {https://arxiv.org/abs/2005.04239} {arXiv:2005.04239 [astro-ph.HE]}
  \BibitemShut {NoStop}%
\bibitem [{\citenamefont {{Hamers}}\ and\ \citenamefont
  {{Safarzadeh}}(2020)}]{2020ApJ...898...99H}%
  \BibitemOpen
  \bibfield  {author} {\bibinfo {author} {\bibfnamefont {A.~S.}\ \bibnamefont
  {{Hamers}}}\ and\ \bibinfo {author} {\bibfnamefont {M.}~\bibnamefont
  {{Safarzadeh}}},\ }\href {https://doi.org/10.3847/1538-4357/ab9b27}
  {\bibfield  {journal} {\bibinfo  {journal} {\apj}\ }\textbf {\bibinfo
  {volume} {898}},\ \bibinfo {eid} {99} (\bibinfo {year} {2020})},\ \Eprint
  {https://arxiv.org/abs/2005.03045} {arXiv:2005.03045 [astro-ph.HE]}
  \BibitemShut {NoStop}%
\bibitem [{\citenamefont {{Safarzadeh}}\ and\ \citenamefont
  {{Hotokezaka}}(2020)}]{2020ApJ...897L...7S}%
  \BibitemOpen
  \bibfield  {author} {\bibinfo {author} {\bibfnamefont {M.}~\bibnamefont
  {{Safarzadeh}}}\ and\ \bibinfo {author} {\bibfnamefont {K.}~\bibnamefont
  {{Hotokezaka}}},\ }\href {https://doi.org/10.3847/2041-8213/ab9b79}
  {\bibfield  {journal} {\bibinfo  {journal} {\apjl}\ }\textbf {\bibinfo
  {volume} {897}},\ \bibinfo {eid} {L7} (\bibinfo {year} {2020})},\ \Eprint
  {https://arxiv.org/abs/2005.06519} {arXiv:2005.06519 [astro-ph.HE]}
  \BibitemShut {NoStop}%
\bibitem [{\citenamefont {{Abbott}}\ \emph
  {et~al.}(2020{\natexlab{b}})\citenamefont {{Abbott}} \emph
  {et~al.}}]{2020PhRvL.125j1102A}%
  \BibitemOpen
  \bibfield  {author} {\bibinfo {author} {\bibfnamefont {R.}~\bibnamefont
  {{Abbott}}} \emph {et~al.} (\bibinfo {collaboration} {LIGO and Virgo
  Collaboration}),\ }\href {https://doi.org/10.1103/PhysRevLett.125.101102}
  {\bibfield  {journal} {\bibinfo  {journal} {\prl}\ }\textbf {\bibinfo
  {volume} {125}},\ \bibinfo {eid} {101102} (\bibinfo {year}
  {2020}{\natexlab{b}})},\ \Eprint {https://arxiv.org/abs/2009.01075}
  {arXiv:2009.01075 [gr-qc]} \BibitemShut {NoStop}%
\bibitem [{\citenamefont {{Abbott}}\ \emph
  {et~al.}(2020{\natexlab{c}})\citenamefont {{Abbott}} \emph
  {et~al.}}]{2020ApJ...900L..13A}%
  \BibitemOpen
  \bibfield  {author} {\bibinfo {author} {\bibfnamefont {R.}~\bibnamefont
  {{Abbott}}} \emph {et~al.} (\bibinfo {collaboration} {LIGO and Virgo
  Collaboration}),\ }\href {https://doi.org/10.3847/2041-8213/aba493}
  {\bibfield  {journal} {\bibinfo  {journal} {\apjl}\ }\textbf {\bibinfo
  {volume} {900}},\ \bibinfo {eid} {L13} (\bibinfo {year}
  {2020}{\natexlab{c}})},\ \Eprint {https://arxiv.org/abs/2009.01190}
  {arXiv:2009.01190 [astro-ph.HE]} \BibitemShut {NoStop}%
\bibitem [{\citenamefont {{Romero-Shaw}}\ \emph
  {et~al.}(2020{\natexlab{a}})\citenamefont {{Romero-Shaw}}, \citenamefont
  {{Lasky}}, \citenamefont {{Thrane}},\ and\ \citenamefont {{Calder{\'o}n
  Bustillo}}}]{2020ApJ...903L...5R}%
  \BibitemOpen
  \bibfield  {author} {\bibinfo {author} {\bibfnamefont {I.}~\bibnamefont
  {{Romero-Shaw}}}, \bibinfo {author} {\bibfnamefont {P.~D.}\ \bibnamefont
  {{Lasky}}}, \bibinfo {author} {\bibfnamefont {E.}~\bibnamefont {{Thrane}}},\
  and\ \bibinfo {author} {\bibfnamefont {J.}~\bibnamefont {{Calder{\'o}n
  Bustillo}}},\ }\href {https://doi.org/10.3847/2041-8213/abbe26} {\bibfield
  {journal} {\bibinfo  {journal} {\apjl}\ }\textbf {\bibinfo {volume} {903}},\
  \bibinfo {eid} {L5} (\bibinfo {year} {2020}{\natexlab{a}})},\ \Eprint
  {https://arxiv.org/abs/2009.04771} {arXiv:2009.04771 [astro-ph.HE]}
  \BibitemShut {NoStop}%
\bibitem [{\citenamefont {{Kimball}}\ \emph {et~al.}(2021)\citenamefont
  {{Kimball}}, \citenamefont {{Talbot}}, \citenamefont {{Berry}}, \citenamefont
  {{Zevin}}, \citenamefont {{Thrane}}, \citenamefont {{Kalogera}},
  \citenamefont {{Buscicchio}}, \citenamefont {{Carney}}, \citenamefont
  {{Dent}}, \citenamefont {{Middleton}}, \citenamefont {{Payne}}, \citenamefont
  {{Veitch}},\ and\ \citenamefont {{Williams}}}]{2021ApJ...915L..35K}%
  \BibitemOpen
  \bibfield  {author} {\bibinfo {author} {\bibfnamefont {C.}~\bibnamefont
  {{Kimball}}}, \bibinfo {author} {\bibfnamefont {C.}~\bibnamefont {{Talbot}}},
  \bibinfo {author} {\bibfnamefont {C.~P.~L.}\ \bibnamefont {{Berry}}},
  \bibinfo {author} {\bibfnamefont {M.}~\bibnamefont {{Zevin}}}, \bibinfo
  {author} {\bibfnamefont {E.}~\bibnamefont {{Thrane}}}, \bibinfo {author}
  {\bibfnamefont {V.}~\bibnamefont {{Kalogera}}}, \bibinfo {author}
  {\bibfnamefont {R.}~\bibnamefont {{Buscicchio}}}, \bibinfo {author}
  {\bibfnamefont {M.}~\bibnamefont {{Carney}}}, \bibinfo {author}
  {\bibfnamefont {T.}~\bibnamefont {{Dent}}}, \bibinfo {author} {\bibfnamefont
  {H.}~\bibnamefont {{Middleton}}}, \bibinfo {author} {\bibfnamefont
  {E.}~\bibnamefont {{Payne}}}, \bibinfo {author} {\bibfnamefont
  {J.}~\bibnamefont {{Veitch}}},\ and\ \bibinfo {author} {\bibfnamefont
  {D.}~\bibnamefont {{Williams}}},\ }\href
  {https://doi.org/10.3847/2041-8213/ac0aef} {\bibfield  {journal} {\bibinfo
  {journal} {\apjl}\ }\textbf {\bibinfo {volume} {915}},\ \bibinfo {eid} {L35}
  (\bibinfo {year} {2021})},\ \Eprint {https://arxiv.org/abs/2011.05332}
  {arXiv:2011.05332 [astro-ph.HE]} \BibitemShut {NoStop}%
\bibitem [{\citenamefont {{Fragione}}\ \emph {et~al.}(2020)\citenamefont
  {{Fragione}}, \citenamefont {{Loeb}},\ and\ \citenamefont
  {{Rasio}}}]{2020ApJ...902L..26F}%
  \BibitemOpen
  \bibfield  {author} {\bibinfo {author} {\bibfnamefont {G.}~\bibnamefont
  {{Fragione}}}, \bibinfo {author} {\bibfnamefont {A.}~\bibnamefont {{Loeb}}},\
  and\ \bibinfo {author} {\bibfnamefont {F.~A.}\ \bibnamefont {{Rasio}}},\
  }\href {https://doi.org/10.3847/2041-8213/abbc0a} {\bibfield  {journal}
  {\bibinfo  {journal} {\apjl}\ }\textbf {\bibinfo {volume} {902}},\ \bibinfo
  {eid} {L26} (\bibinfo {year} {2020})},\ \Eprint
  {https://arxiv.org/abs/2009.05065} {arXiv:2009.05065 [astro-ph.GA]}
  \BibitemShut {NoStop}%
\bibitem [{\citenamefont {{Mapelli}}\ \emph {et~al.}(2021)\citenamefont
  {{Mapelli}}, \citenamefont {{Dall'Amico}}, \citenamefont {{Bouffanais}},
  \citenamefont {{Giacobbo}}, \citenamefont {{Arca Sedda}}, \citenamefont
  {{Artale}}, \citenamefont {{Ballone}}, \citenamefont {{Di Carlo}},
  \citenamefont {{Iorio}}, \citenamefont {{Santoliquido}},\ and\ \citenamefont
  {{Torniamenti}}}]{2021MNRAS.505..339M}%
  \BibitemOpen
  \bibfield  {author} {\bibinfo {author} {\bibfnamefont {M.}~\bibnamefont
  {{Mapelli}}}, \bibinfo {author} {\bibfnamefont {M.}~\bibnamefont
  {{Dall'Amico}}}, \bibinfo {author} {\bibfnamefont {Y.}~\bibnamefont
  {{Bouffanais}}}, \bibinfo {author} {\bibfnamefont {N.}~\bibnamefont
  {{Giacobbo}}}, \bibinfo {author} {\bibfnamefont {M.}~\bibnamefont {{Arca
  Sedda}}}, \bibinfo {author} {\bibfnamefont {M.~C.}\ \bibnamefont {{Artale}}},
  \bibinfo {author} {\bibfnamefont {A.}~\bibnamefont {{Ballone}}}, \bibinfo
  {author} {\bibfnamefont {U.~N.}\ \bibnamefont {{Di Carlo}}}, \bibinfo
  {author} {\bibfnamefont {G.}~\bibnamefont {{Iorio}}}, \bibinfo {author}
  {\bibfnamefont {F.}~\bibnamefont {{Santoliquido}}},\ and\ \bibinfo {author}
  {\bibfnamefont {S.}~\bibnamefont {{Torniamenti}}},\ }\href
  {https://doi.org/10.1093/mnras/stab1334} {\bibfield  {journal} {\bibinfo
  {journal} {\mnras}\ }\textbf {\bibinfo {volume} {505}},\ \bibinfo {pages}
  {339} (\bibinfo {year} {2021})},\ \Eprint {https://arxiv.org/abs/2103.05016}
  {arXiv:2103.05016 [astro-ph.HE]} \BibitemShut {NoStop}%
\bibitem [{\citenamefont {{Samsing}}\ \emph {et~al.}(2020)\citenamefont
  {{Samsing}}, \citenamefont {{Bartos}}, \citenamefont {{D'Orazio}},
  \citenamefont {{Haiman}}, \citenamefont {{Kocsis}}, \citenamefont {{Leigh}},
  \citenamefont {{Liu}}, \citenamefont {{Pessah}},\ and\ \citenamefont
  {{Tagawa}}}]{2020arXiv201009765S}%
  \BibitemOpen
  \bibfield  {author} {\bibinfo {author} {\bibfnamefont {J.}~\bibnamefont
  {{Samsing}}}, \bibinfo {author} {\bibfnamefont {I.}~\bibnamefont {{Bartos}}},
  \bibinfo {author} {\bibfnamefont {D.~J.}\ \bibnamefont {{D'Orazio}}},
  \bibinfo {author} {\bibfnamefont {Z.}~\bibnamefont {{Haiman}}}, \bibinfo
  {author} {\bibfnamefont {B.}~\bibnamefont {{Kocsis}}}, \bibinfo {author}
  {\bibfnamefont {N.~W.~C.}\ \bibnamefont {{Leigh}}}, \bibinfo {author}
  {\bibfnamefont {B.}~\bibnamefont {{Liu}}}, \bibinfo {author} {\bibfnamefont
  {M.~E.}\ \bibnamefont {{Pessah}}},\ and\ \bibinfo {author} {\bibfnamefont
  {H.}~\bibnamefont {{Tagawa}}},\ }\href@noop {} {\bibfield  {journal}
  {\bibinfo  {journal} {{}}\ } (\bibinfo {year} {2020})},\ \Eprint
  {https://arxiv.org/abs/2010.09765} {arXiv:2010.09765 [astro-ph.HE]}
  \BibitemShut {NoStop}%
\bibitem [{\citenamefont {{Arca-Sedda}}\ \emph {et~al.}(2021)\citenamefont
  {{Arca-Sedda}}, \citenamefont {{Rizzuto}}, \citenamefont {{Naab}},
  \citenamefont {{Ostriker}}, \citenamefont {{Giersz}},\ and\ \citenamefont
  {{Spurzem}}}]{2021ApJ...920..128A}%
  \BibitemOpen
  \bibfield  {author} {\bibinfo {author} {\bibfnamefont {M.}~\bibnamefont
  {{Arca-Sedda}}}, \bibinfo {author} {\bibfnamefont {F.~P.}\ \bibnamefont
  {{Rizzuto}}}, \bibinfo {author} {\bibfnamefont {T.}~\bibnamefont {{Naab}}},
  \bibinfo {author} {\bibfnamefont {J.}~\bibnamefont {{Ostriker}}}, \bibinfo
  {author} {\bibfnamefont {M.}~\bibnamefont {{Giersz}}},\ and\ \bibinfo
  {author} {\bibfnamefont {R.}~\bibnamefont {{Spurzem}}},\ }\href
  {https://doi.org/10.3847/1538-4357/ac1419} {\bibfield  {journal} {\bibinfo
  {journal} {\apj}\ }\textbf {\bibinfo {volume} {920}},\ \bibinfo {eid} {128}
  (\bibinfo {year} {2021})},\ \Eprint {https://arxiv.org/abs/2105.07003}
  {arXiv:2105.07003 [astro-ph.GA]} \BibitemShut {NoStop}%
\bibitem [{\citenamefont {{Fishbach}}\ and\ \citenamefont
  {{Holz}}(2020)}]{2020ApJ...904L..26F}%
  \BibitemOpen
  \bibfield  {author} {\bibinfo {author} {\bibfnamefont {M.}~\bibnamefont
  {{Fishbach}}}\ and\ \bibinfo {author} {\bibfnamefont {D.~E.}\ \bibnamefont
  {{Holz}}},\ }\href {https://doi.org/10.3847/2041-8213/abc827} {\bibfield
  {journal} {\bibinfo  {journal} {\apjl}\ }\textbf {\bibinfo {volume} {904}},\
  \bibinfo {eid} {L26} (\bibinfo {year} {2020})},\ \Eprint
  {https://arxiv.org/abs/2009.05472} {arXiv:2009.05472 [astro-ph.HE]}
  \BibitemShut {NoStop}%
\bibitem [{\citenamefont {{Essick}}\ \emph {et~al.}(2022)\citenamefont
  {{Essick}}, \citenamefont {{Farah}}, \citenamefont {{Galaudage}},
  \citenamefont {{Talbot}}, \citenamefont {{Fishbach}}, \citenamefont
  {{Thrane}},\ and\ \citenamefont {{Holz}}}]{2022ApJ...926...34E}%
  \BibitemOpen
  \bibfield  {author} {\bibinfo {author} {\bibfnamefont {R.}~\bibnamefont
  {{Essick}}}, \bibinfo {author} {\bibfnamefont {A.}~\bibnamefont {{Farah}}},
  \bibinfo {author} {\bibfnamefont {S.}~\bibnamefont {{Galaudage}}}, \bibinfo
  {author} {\bibfnamefont {C.}~\bibnamefont {{Talbot}}}, \bibinfo {author}
  {\bibfnamefont {M.}~\bibnamefont {{Fishbach}}}, \bibinfo {author}
  {\bibfnamefont {E.}~\bibnamefont {{Thrane}}},\ and\ \bibinfo {author}
  {\bibfnamefont {D.~E.}\ \bibnamefont {{Holz}}},\ }\href
  {https://doi.org/10.3847/1538-4357/ac3978} {\bibfield  {journal} {\bibinfo
  {journal} {\apj}\ }\textbf {\bibinfo {volume} {926}},\ \bibinfo {eid} {34}
  (\bibinfo {year} {2022})},\ \Eprint {https://arxiv.org/abs/2109.00418}
  {arXiv:2109.00418 [astro-ph.HE]} \BibitemShut {NoStop}%
\bibitem [{\citenamefont {{Abbott}}\ \emph
  {et~al.}(2020{\natexlab{d}})\citenamefont {{Abbott}} \emph
  {et~al.}}]{2020ApJ...896L..44A}%
  \BibitemOpen
  \bibfield  {author} {\bibinfo {author} {\bibfnamefont {R.}~\bibnamefont
  {{Abbott}}} \emph {et~al.} (\bibinfo {collaboration} {LIGO and Virgo
  Collaboration}),\ }\href {https://doi.org/10.3847/2041-8213/ab960f}
  {\bibfield  {journal} {\bibinfo  {journal} {\apjl}\ }\textbf {\bibinfo
  {volume} {896}},\ \bibinfo {eid} {L44} (\bibinfo {year}
  {2020}{\natexlab{d}})},\ \Eprint {https://arxiv.org/abs/2006.12611}
  {arXiv:2006.12611 [astro-ph.HE]} \BibitemShut {NoStop}%
\bibitem [{\citenamefont {{Liu}}\ and\ \citenamefont
  {{Lai}}(2021)}]{2021MNRAS.502.2049L}%
  \BibitemOpen
  \bibfield  {author} {\bibinfo {author} {\bibfnamefont {B.}~\bibnamefont
  {{Liu}}}\ and\ \bibinfo {author} {\bibfnamefont {D.}~\bibnamefont {{Lai}}},\
  }\href {https://doi.org/10.1093/mnras/stab178} {\bibfield  {journal}
  {\bibinfo  {journal} {\mnras}\ }\textbf {\bibinfo {volume} {502}},\ \bibinfo
  {pages} {2049} (\bibinfo {year} {2021})},\ \Eprint
  {https://arxiv.org/abs/2009.10068} {arXiv:2009.10068 [astro-ph.HE]}
  \BibitemShut {NoStop}%
\bibitem [{\citenamefont {{Tagawa}}\ \emph {et~al.}(2021)\citenamefont
  {{Tagawa}}, \citenamefont {{Kocsis}}, \citenamefont {{Haiman}}, \citenamefont
  {{Bartos}}, \citenamefont {{Omukai}},\ and\ \citenamefont
  {{Samsing}}}]{2021ApJ...908..194T}%
  \BibitemOpen
  \bibfield  {author} {\bibinfo {author} {\bibfnamefont {H.}~\bibnamefont
  {{Tagawa}}}, \bibinfo {author} {\bibfnamefont {B.}~\bibnamefont {{Kocsis}}},
  \bibinfo {author} {\bibfnamefont {Z.}~\bibnamefont {{Haiman}}}, \bibinfo
  {author} {\bibfnamefont {I.}~\bibnamefont {{Bartos}}}, \bibinfo {author}
  {\bibfnamefont {K.}~\bibnamefont {{Omukai}}},\ and\ \bibinfo {author}
  {\bibfnamefont {J.}~\bibnamefont {{Samsing}}},\ }\href
  {https://doi.org/10.3847/1538-4357/abd555} {\bibfield  {journal} {\bibinfo
  {journal} {\apj}\ }\textbf {\bibinfo {volume} {908}},\ \bibinfo {eid} {194}
  (\bibinfo {year} {2021})},\ \Eprint {https://arxiv.org/abs/2012.00011}
  {arXiv:2012.00011 [astro-ph.HE]} \BibitemShut {NoStop}%
\bibitem [{\citenamefont {{Lu}}\ \emph {et~al.}(2020)\citenamefont {{Lu}},
  \citenamefont {{Beniamini}},\ and\ \citenamefont
  {{Bonnerot}}}]{2021MNRAS.500.1817L}%
  \BibitemOpen
  \bibfield  {author} {\bibinfo {author} {\bibfnamefont {W.}~\bibnamefont
  {{Lu}}}, \bibinfo {author} {\bibfnamefont {P.}~\bibnamefont {{Beniamini}}},\
  and\ \bibinfo {author} {\bibfnamefont {C.}~\bibnamefont {{Bonnerot}}},\
  }\href {https://doi.org/10.1093/mnras/staa3372} {\bibfield  {journal}
  {\bibinfo  {journal} {\mnras}\ }\textbf {\bibinfo {volume} {500}},\ \bibinfo
  {pages} {1817} (\bibinfo {year} {2020})},\ \Eprint
  {https://arxiv.org/abs/2009.10082} {arXiv:2009.10082 [astro-ph.HE]}
  \BibitemShut {NoStop}%
\bibitem [{\citenamefont {{Gonz{\'a}lez}}\ \emph {et~al.}(2007)\citenamefont
  {{Gonz{\'a}lez}}, \citenamefont {{Hannam}}, \citenamefont {{Sperhake}},
  \citenamefont {{Br{\"u}gmann}},\ and\ \citenamefont
  {{Husa}}}]{2007PhRvL..98w1101G}%
  \BibitemOpen
  \bibfield  {author} {\bibinfo {author} {\bibfnamefont {J.~A.}\ \bibnamefont
  {{Gonz{\'a}lez}}}, \bibinfo {author} {\bibfnamefont {M.}~\bibnamefont
  {{Hannam}}}, \bibinfo {author} {\bibfnamefont {U.}~\bibnamefont
  {{Sperhake}}}, \bibinfo {author} {\bibfnamefont {B.}~\bibnamefont
  {{Br{\"u}gmann}}},\ and\ \bibinfo {author} {\bibfnamefont {S.}~\bibnamefont
  {{Husa}}},\ }\href {https://doi.org/10.1103/PhysRevLett.98.231101} {\bibfield
   {journal} {\bibinfo  {journal} {\prl}\ }\textbf {\bibinfo {volume} {98}},\
  \bibinfo {eid} {231101} (\bibinfo {year} {2007})},\ \Eprint
  {https://arxiv.org/abs/gr-qc/0702052} {arXiv:gr-qc/0702052 [gr-qc]}
  \BibitemShut {NoStop}%
\bibitem [{\citenamefont {{Campanelli}}\ \emph {et~al.}(2007)\citenamefont
  {{Campanelli}}, \citenamefont {{Lousto}}, \citenamefont {{Zlochower}},\ and\
  \citenamefont {{Merritt}}}]{2007ApJ...659L...5C}%
  \BibitemOpen
  \bibfield  {author} {\bibinfo {author} {\bibfnamefont {M.}~\bibnamefont
  {{Campanelli}}}, \bibinfo {author} {\bibfnamefont {C.}~\bibnamefont
  {{Lousto}}}, \bibinfo {author} {\bibfnamefont {Y.}~\bibnamefont
  {{Zlochower}}},\ and\ \bibinfo {author} {\bibfnamefont {D.}~\bibnamefont
  {{Merritt}}},\ }\href {https://doi.org/10.1086/516712} {\bibfield  {journal}
  {\bibinfo  {journal} {\apjl}\ }\textbf {\bibinfo {volume} {659}},\ \bibinfo
  {pages} {L5} (\bibinfo {year} {2007})},\ \Eprint
  {https://arxiv.org/abs/gr-qc/0701164} {arXiv:gr-qc/0701164 [gr-qc]}
  \BibitemShut {NoStop}%
\bibitem [{\citenamefont {{Gerosa}}\ \emph
  {et~al.}(2018{\natexlab{b}})\citenamefont {{Gerosa}}, \citenamefont
  {{H{\'e}bert}},\ and\ \citenamefont {{Stein}}}]{2018PhRvD..97j4049G}%
  \BibitemOpen
  \bibfield  {author} {\bibinfo {author} {\bibfnamefont {D.}~\bibnamefont
  {{Gerosa}}}, \bibinfo {author} {\bibfnamefont {F.}~\bibnamefont
  {{H{\'e}bert}}},\ and\ \bibinfo {author} {\bibfnamefont {L.~C.}\ \bibnamefont
  {{Stein}}},\ }\href {https://doi.org/10.1103/PhysRevD.97.104049} {\bibfield
  {journal} {\bibinfo  {journal} {\prd}\ }\textbf {\bibinfo {volume} {97}},\
  \bibinfo {eid} {104049} (\bibinfo {year} {2018}{\natexlab{b}})},\ \Eprint
  {https://arxiv.org/abs/1802.04276} {arXiv:1802.04276 [gr-qc]} \BibitemShut
  {NoStop}%
\bibitem [{\citenamefont {{Gerosa}}\ and\ \citenamefont
  {{Berti}}(2019)}]{2019PhRvD.100d1301G}%
  \BibitemOpen
  \bibfield  {author} {\bibinfo {author} {\bibfnamefont {D.}~\bibnamefont
  {{Gerosa}}}\ and\ \bibinfo {author} {\bibfnamefont {E.}~\bibnamefont
  {{Berti}}},\ }\href {https://doi.org/10.1103/PhysRevD.100.041301} {\bibfield
  {journal} {\bibinfo  {journal} {\prd}\ }\textbf {\bibinfo {volume} {100}},\
  \bibinfo {eid} {041301} (\bibinfo {year} {2019})},\ \Eprint
  {https://arxiv.org/abs/1906.05295} {arXiv:1906.05295 [astro-ph.HE]}
  \BibitemShut {NoStop}%
\bibitem [{\citenamefont {{Baibhav}}\ \emph {et~al.}(2021)\citenamefont
  {{Baibhav}}, \citenamefont {{Berti}}, \citenamefont {{Gerosa}}, \citenamefont
  {{Mould}},\ and\ \citenamefont {{Wong}}}]{2021PhRvD.104h4002B}%
  \BibitemOpen
  \bibfield  {author} {\bibinfo {author} {\bibfnamefont {V.}~\bibnamefont
  {{Baibhav}}}, \bibinfo {author} {\bibfnamefont {E.}~\bibnamefont {{Berti}}},
  \bibinfo {author} {\bibfnamefont {D.}~\bibnamefont {{Gerosa}}}, \bibinfo
  {author} {\bibfnamefont {M.}~\bibnamefont {{Mould}}},\ and\ \bibinfo {author}
  {\bibfnamefont {K.~W.~K.}\ \bibnamefont {{Wong}}},\ }\href
  {https://doi.org/10.1103/PhysRevD.104.084002} {\bibfield  {journal} {\bibinfo
   {journal} {\prd}\ }\textbf {\bibinfo {volume} {104}},\ \bibinfo {eid}
  {084002} (\bibinfo {year} {2021})},\ \Eprint
  {https://arxiv.org/abs/2105.12140} {arXiv:2105.12140 [gr-qc]} \BibitemShut
  {NoStop}%
\bibitem [{\citenamefont {{Gerosa}}\ \emph
  {et~al.}(2021{\natexlab{a}})\citenamefont {{Gerosa}}, \citenamefont
  {{Giacobbo}},\ and\ \citenamefont {{Vecchio}}}]{2021ApJ...915...56G}%
  \BibitemOpen
  \bibfield  {author} {\bibinfo {author} {\bibfnamefont {D.}~\bibnamefont
  {{Gerosa}}}, \bibinfo {author} {\bibfnamefont {N.}~\bibnamefont
  {{Giacobbo}}},\ and\ \bibinfo {author} {\bibfnamefont {A.}~\bibnamefont
  {{Vecchio}}},\ }\href {https://doi.org/10.3847/1538-4357/ac00bb} {\bibfield
  {journal} {\bibinfo  {journal} {\apj}\ }\textbf {\bibinfo {volume} {915}},\
  \bibinfo {eid} {56} (\bibinfo {year} {2021}{\natexlab{a}})},\ \Eprint
  {https://arxiv.org/abs/2104.11247} {arXiv:2104.11247 [astro-ph.HE]}
  \BibitemShut {NoStop}%
\bibitem [{\citenamefont {McKay}\ \emph {et~al.}(1979)\citenamefont {McKay},
  \citenamefont {Beckman},\ and\ \citenamefont {Conover}}]{10.2307/1268522}%
  \BibitemOpen
  \bibfield  {author} {\bibinfo {author} {\bibfnamefont {M.~D.}\ \bibnamefont
  {McKay}}, \bibinfo {author} {\bibfnamefont {R.~J.}\ \bibnamefont {Beckman}},\
  and\ \bibinfo {author} {\bibfnamefont {W.~J.}\ \bibnamefont {Conover}},\
  }\href {http://www.jstor.org/stable/1268522} {\bibfield  {journal} {\bibinfo
  {journal} {Technometrics}\ }\textbf {\bibinfo {volume} {21}},\ \bibinfo
  {pages} {239} (\bibinfo {year} {1979})}\BibitemShut {NoStop}%
\bibitem [{\citenamefont {{Heger}}\ \emph {et~al.}(2003)\citenamefont
  {{Heger}}, \citenamefont {{Fryer}}, \citenamefont {{Woosley}}, \citenamefont
  {{Langer}},\ and\ \citenamefont {{Hartmann}}}]{2003ApJ...591..288H}%
  \BibitemOpen
  \bibfield  {author} {\bibinfo {author} {\bibfnamefont {A.}~\bibnamefont
  {{Heger}}}, \bibinfo {author} {\bibfnamefont {C.~L.}\ \bibnamefont
  {{Fryer}}}, \bibinfo {author} {\bibfnamefont {S.~E.}\ \bibnamefont
  {{Woosley}}}, \bibinfo {author} {\bibfnamefont {N.}~\bibnamefont
  {{Langer}}},\ and\ \bibinfo {author} {\bibfnamefont {D.~H.}\ \bibnamefont
  {{Hartmann}}},\ }\href {https://doi.org/10.1086/375341} {\bibfield  {journal}
  {\bibinfo  {journal} {\apj}\ }\textbf {\bibinfo {volume} {591}},\ \bibinfo
  {pages} {288} (\bibinfo {year} {2003})},\ \Eprint
  {https://arxiv.org/abs/astro-ph/0212469} {arXiv:astro-ph/0212469 [astro-ph]}
  \BibitemShut {NoStop}%
\bibitem [{\citenamefont {{Woosley}}\ \emph {et~al.}(2007)\citenamefont
  {{Woosley}}, \citenamefont {{Blinnikov}},\ and\ \citenamefont
  {{Heger}}}]{2007Natur.450..390W}%
  \BibitemOpen
  \bibfield  {author} {\bibinfo {author} {\bibfnamefont {S.~E.}\ \bibnamefont
  {{Woosley}}}, \bibinfo {author} {\bibfnamefont {S.}~\bibnamefont
  {{Blinnikov}}},\ and\ \bibinfo {author} {\bibfnamefont {A.}~\bibnamefont
  {{Heger}}},\ }\href {https://doi.org/10.1038/nature06333} {\bibfield
  {journal} {\bibinfo  {journal} {\nat}\ }\textbf {\bibinfo {volume} {450}},\
  \bibinfo {pages} {390} (\bibinfo {year} {2007})},\ \Eprint
  {https://arxiv.org/abs/0710.3314} {arXiv:0710.3314 [astro-ph]} \BibitemShut
  {NoStop}%
\bibitem [{\citenamefont {{Belczynski}}\ \emph {et~al.}(2016)\citenamefont
  {{Belczynski}}, \citenamefont {{Heger}}, \citenamefont {{Gladysz}},
  \citenamefont {{Ruiter}}, \citenamefont {{Woosley}}, \citenamefont
  {{Wiktorowicz}}, \citenamefont {{Chen}}, \citenamefont {{Bulik}},
  \citenamefont {{O'Shaughnessy}}, \citenamefont {{Holz}}, \citenamefont
  {{Fryer}},\ and\ \citenamefont {{Berti}}}]{2016A&A...594A..97B}%
  \BibitemOpen
  \bibfield  {author} {\bibinfo {author} {\bibfnamefont {K.}~\bibnamefont
  {{Belczynski}}}, \bibinfo {author} {\bibfnamefont {A.}~\bibnamefont
  {{Heger}}}, \bibinfo {author} {\bibfnamefont {W.}~\bibnamefont {{Gladysz}}},
  \bibinfo {author} {\bibfnamefont {A.~J.}\ \bibnamefont {{Ruiter}}}, \bibinfo
  {author} {\bibfnamefont {S.}~\bibnamefont {{Woosley}}}, \bibinfo {author}
  {\bibfnamefont {G.}~\bibnamefont {{Wiktorowicz}}}, \bibinfo {author}
  {\bibfnamefont {H.~Y.}\ \bibnamefont {{Chen}}}, \bibinfo {author}
  {\bibfnamefont {T.}~\bibnamefont {{Bulik}}}, \bibinfo {author} {\bibfnamefont
  {R.}~\bibnamefont {{O'Shaughnessy}}}, \bibinfo {author} {\bibfnamefont
  {D.~E.}\ \bibnamefont {{Holz}}}, \bibinfo {author} {\bibfnamefont {C.~L.}\
  \bibnamefont {{Fryer}}},\ and\ \bibinfo {author} {\bibfnamefont
  {E.}~\bibnamefont {{Berti}}},\ }\href
  {https://doi.org/10.1051/0004-6361/201628980} {\bibfield  {journal} {\bibinfo
   {journal} {\aap}\ }\textbf {\bibinfo {volume} {594}},\ \bibinfo {eid} {A97}
  (\bibinfo {year} {2016})},\ \Eprint {https://arxiv.org/abs/1607.03116}
  {arXiv:1607.03116 [astro-ph.HE]} \BibitemShut {NoStop}%
\bibitem [{\citenamefont {{Farmer}}\ \emph {et~al.}(2020)\citenamefont
  {{Farmer}}, \citenamefont {{Renzo}}, \citenamefont {{de Mink}}, \citenamefont
  {{Fishbach}},\ and\ \citenamefont {{Justham}}}]{2020ApJ...902L..36F}%
  \BibitemOpen
  \bibfield  {author} {\bibinfo {author} {\bibfnamefont {R.}~\bibnamefont
  {{Farmer}}}, \bibinfo {author} {\bibfnamefont {M.}~\bibnamefont {{Renzo}}},
  \bibinfo {author} {\bibfnamefont {S.~E.}\ \bibnamefont {{de Mink}}}, \bibinfo
  {author} {\bibfnamefont {M.}~\bibnamefont {{Fishbach}}},\ and\ \bibinfo
  {author} {\bibfnamefont {S.}~\bibnamefont {{Justham}}},\ }\href
  {https://doi.org/10.3847/2041-8213/abbadd} {\bibfield  {journal} {\bibinfo
  {journal} {\apjl}\ }\textbf {\bibinfo {volume} {902}},\ \bibinfo {eid} {L36}
  (\bibinfo {year} {2020})},\ \Eprint {https://arxiv.org/abs/2006.06678}
  {arXiv:2006.06678 [astro-ph.HE]} \BibitemShut {NoStop}%
\bibitem [{\citenamefont {{Marchant}}\ and\ \citenamefont
  {{Moriya}}(2020)}]{2020A&A...640L..18M}%
  \BibitemOpen
  \bibfield  {author} {\bibinfo {author} {\bibfnamefont {P.}~\bibnamefont
  {{Marchant}}}\ and\ \bibinfo {author} {\bibfnamefont {T.~J.}\ \bibnamefont
  {{Moriya}}},\ }\href {https://doi.org/10.1051/0004-6361/202038902} {\bibfield
   {journal} {\bibinfo  {journal} {\aap}\ }\textbf {\bibinfo {volume} {640}},\
  \bibinfo {eid} {L18} (\bibinfo {year} {2020})},\ \Eprint
  {https://arxiv.org/abs/2007.06220} {arXiv:2007.06220 [astro-ph.HE]}
  \BibitemShut {NoStop}%
\bibitem [{\citenamefont {{Rice}}\ and\ \citenamefont
  {{Zhang}}(2021)}]{2021ApJ...908...59R}%
  \BibitemOpen
  \bibfield  {author} {\bibinfo {author} {\bibfnamefont {J.~R.}\ \bibnamefont
  {{Rice}}}\ and\ \bibinfo {author} {\bibfnamefont {B.}~\bibnamefont
  {{Zhang}}},\ }\href {https://doi.org/10.3847/1538-4357/abd6ea} {\bibfield
  {journal} {\bibinfo  {journal} {\apj}\ }\textbf {\bibinfo {volume} {908}},\
  \bibinfo {eid} {59} (\bibinfo {year} {2021})},\ \Eprint
  {https://arxiv.org/abs/2009.11326} {arXiv:2009.11326 [astro-ph.HE]}
  \BibitemShut {NoStop}%
\bibitem [{\citenamefont {{Safarzadeh}}\ and\ \citenamefont
  {{Haiman}}(2020)}]{2020ApJ...903L..21S}%
  \BibitemOpen
  \bibfield  {author} {\bibinfo {author} {\bibfnamefont {M.}~\bibnamefont
  {{Safarzadeh}}}\ and\ \bibinfo {author} {\bibfnamefont {Z.}~\bibnamefont
  {{Haiman}}},\ }\href {https://doi.org/10.3847/2041-8213/abc253} {\bibfield
  {journal} {\bibinfo  {journal} {\apjl}\ }\textbf {\bibinfo {volume} {903}},\
  \bibinfo {eid} {L21} (\bibinfo {year} {2020})},\ \Eprint
  {https://arxiv.org/abs/2009.09320} {arXiv:2009.09320 [astro-ph.HE]}
  \BibitemShut {NoStop}%
\bibitem [{\citenamefont {{Roupas}}\ and\ \citenamefont
  {{Kazanas}}(2019)}]{2019A&A...632L...8R}%
  \BibitemOpen
  \bibfield  {author} {\bibinfo {author} {\bibfnamefont {Z.}~\bibnamefont
  {{Roupas}}}\ and\ \bibinfo {author} {\bibfnamefont {D.}~\bibnamefont
  {{Kazanas}}},\ }\href {https://doi.org/10.1051/0004-6361/201937002}
  {\bibfield  {journal} {\bibinfo  {journal} {\aap}\ }\textbf {\bibinfo
  {volume} {632}},\ \bibinfo {eid} {L8} (\bibinfo {year} {2019})},\ \Eprint
  {https://arxiv.org/abs/1911.03915} {arXiv:1911.03915 [astro-ph.GA]}
  \BibitemShut {NoStop}%
\bibitem [{\citenamefont {{Natarajan}}(2020)}]{2021MNRAS.501.1413N}%
  \BibitemOpen
  \bibfield  {author} {\bibinfo {author} {\bibfnamefont {P.}~\bibnamefont
  {{Natarajan}}},\ }\href {https://doi.org/10.1093/mnras/staa3724} {\bibfield
  {journal} {\bibinfo  {journal} {\mnras}\ }\textbf {\bibinfo {volume} {501}},\
  \bibinfo {pages} {1413} (\bibinfo {year} {2020})},\ \Eprint
  {https://arxiv.org/abs/2009.09156} {arXiv:2009.09156 [astro-ph.GA]}
  \BibitemShut {NoStop}%
\bibitem [{\citenamefont {{van Son}}\ \emph {et~al.}(2020)\citenamefont {{van
  Son}}, \citenamefont {{De Mink}}, \citenamefont {{Broekgaarden}},
  \citenamefont {{Renzo}}, \citenamefont {{Justham}}, \citenamefont
  {{Laplace}}, \citenamefont {{Mor{\'a}n-Fraile}}, \citenamefont {{Hendriks}},\
  and\ \citenamefont {{Farmer}}}]{2020ApJ...897..100V}%
  \BibitemOpen
  \bibfield  {author} {\bibinfo {author} {\bibfnamefont {L.~A.~C.}\
  \bibnamefont {{van Son}}}, \bibinfo {author} {\bibfnamefont {S.~E.}\
  \bibnamefont {{De Mink}}}, \bibinfo {author} {\bibfnamefont {F.~S.}\
  \bibnamefont {{Broekgaarden}}}, \bibinfo {author} {\bibfnamefont
  {M.}~\bibnamefont {{Renzo}}}, \bibinfo {author} {\bibfnamefont
  {S.}~\bibnamefont {{Justham}}}, \bibinfo {author} {\bibfnamefont
  {E.}~\bibnamefont {{Laplace}}}, \bibinfo {author} {\bibfnamefont
  {J.}~\bibnamefont {{Mor{\'a}n-Fraile}}}, \bibinfo {author} {\bibfnamefont
  {D.~D.}\ \bibnamefont {{Hendriks}}},\ and\ \bibinfo {author} {\bibfnamefont
  {R.}~\bibnamefont {{Farmer}}},\ }\href
  {https://doi.org/10.3847/1538-4357/ab9809} {\bibfield  {journal} {\bibinfo
  {journal} {\apj}\ }\textbf {\bibinfo {volume} {897}},\ \bibinfo {eid} {100}
  (\bibinfo {year} {2020})},\ \Eprint {https://arxiv.org/abs/2004.05187}
  {arXiv:2004.05187 [astro-ph.HE]} \BibitemShut {NoStop}%
\bibitem [{\citenamefont {{Belczynski}}\ \emph
  {et~al.}(2020{\natexlab{a}})\citenamefont {{Belczynski}}, \citenamefont
  {{Hirschi}}, \citenamefont {{Kaiser}}, \citenamefont {{Liu}}, \citenamefont
  {{Casares}}, \citenamefont {{Lu}}, \citenamefont {{O'Shaughnessy}},
  \citenamefont {{Heger}}, \citenamefont {{Justham}},\ and\ \citenamefont
  {{Soria}}}]{2020ApJ...890..113B}%
  \BibitemOpen
  \bibfield  {author} {\bibinfo {author} {\bibfnamefont {K.}~\bibnamefont
  {{Belczynski}}}, \bibinfo {author} {\bibfnamefont {R.}~\bibnamefont
  {{Hirschi}}}, \bibinfo {author} {\bibfnamefont {E.~A.}\ \bibnamefont
  {{Kaiser}}}, \bibinfo {author} {\bibfnamefont {J.}~\bibnamefont {{Liu}}},
  \bibinfo {author} {\bibfnamefont {J.}~\bibnamefont {{Casares}}}, \bibinfo
  {author} {\bibfnamefont {Y.}~\bibnamefont {{Lu}}}, \bibinfo {author}
  {\bibfnamefont {R.}~\bibnamefont {{O'Shaughnessy}}}, \bibinfo {author}
  {\bibfnamefont {A.}~\bibnamefont {{Heger}}}, \bibinfo {author} {\bibfnamefont
  {S.}~\bibnamefont {{Justham}}},\ and\ \bibinfo {author} {\bibfnamefont
  {R.}~\bibnamefont {{Soria}}},\ }\href
  {https://doi.org/10.3847/1538-4357/ab6d77} {\bibfield  {journal} {\bibinfo
  {journal} {\apj}\ }\textbf {\bibinfo {volume} {890}},\ \bibinfo {eid} {113}
  (\bibinfo {year} {2020}{\natexlab{a}})},\ \Eprint
  {https://arxiv.org/abs/1911.12357} {arXiv:1911.12357 [astro-ph.HE]}
  \BibitemShut {NoStop}%
\bibitem [{\citenamefont {{Vink}}\ \emph {et~al.}(2021)\citenamefont {{Vink}},
  \citenamefont {{Higgins}}, \citenamefont {{Sander}},\ and\ \citenamefont
  {{Sabhahit}}}]{2021MNRAS.504..146V}%
  \BibitemOpen
  \bibfield  {author} {\bibinfo {author} {\bibfnamefont {J.~S.}\ \bibnamefont
  {{Vink}}}, \bibinfo {author} {\bibfnamefont {E.~R.}\ \bibnamefont
  {{Higgins}}}, \bibinfo {author} {\bibfnamefont {A.~A.~C.}\ \bibnamefont
  {{Sander}}},\ and\ \bibinfo {author} {\bibfnamefont {G.~N.}\ \bibnamefont
  {{Sabhahit}}},\ }\href {https://doi.org/10.1093/mnras/stab842} {\bibfield
  {journal} {\bibinfo  {journal} {\mnras}\ }\textbf {\bibinfo {volume} {504}},\
  \bibinfo {pages} {146} (\bibinfo {year} {2021})},\ \Eprint
  {https://arxiv.org/abs/2010.11730} {arXiv:2010.11730 [astro-ph.HE]}
  \BibitemShut {NoStop}%
\bibitem [{\citenamefont {{Tanikawa}}\ \emph {et~al.}(2021)\citenamefont
  {{Tanikawa}}, \citenamefont {{Susa}}, \citenamefont {{Yoshida}},
  \citenamefont {{Trani}},\ and\ \citenamefont
  {{Kinugawa}}}]{2021ApJ...910...30T}%
  \BibitemOpen
  \bibfield  {author} {\bibinfo {author} {\bibfnamefont {A.}~\bibnamefont
  {{Tanikawa}}}, \bibinfo {author} {\bibfnamefont {H.}~\bibnamefont {{Susa}}},
  \bibinfo {author} {\bibfnamefont {T.}~\bibnamefont {{Yoshida}}}, \bibinfo
  {author} {\bibfnamefont {A.~A.}\ \bibnamefont {{Trani}}},\ and\ \bibinfo
  {author} {\bibfnamefont {T.}~\bibnamefont {{Kinugawa}}},\ }\href
  {https://doi.org/10.3847/1538-4357/abe40d} {\bibfield  {journal} {\bibinfo
  {journal} {\apj}\ }\textbf {\bibinfo {volume} {910}},\ \bibinfo {eid} {30}
  (\bibinfo {year} {2021})},\ \Eprint {https://arxiv.org/abs/2008.01890}
  {arXiv:2008.01890 [astro-ph.HE]} \BibitemShut {NoStop}%
\bibitem [{\citenamefont {{Farrell}}\ \emph {et~al.}(2021)\citenamefont
  {{Farrell}}, \citenamefont {{Groh}}, \citenamefont {{Hirschi}}, \citenamefont
  {{Murphy}}, \citenamefont {{Kaiser}}, \citenamefont {{Ekstr{\"o}m}},
  \citenamefont {{Georgy}},\ and\ \citenamefont
  {{Meynet}}}]{2021MNRAS.502L..40F}%
  \BibitemOpen
  \bibfield  {author} {\bibinfo {author} {\bibfnamefont {E.}~\bibnamefont
  {{Farrell}}}, \bibinfo {author} {\bibfnamefont {J.~H.}\ \bibnamefont
  {{Groh}}}, \bibinfo {author} {\bibfnamefont {R.}~\bibnamefont {{Hirschi}}},
  \bibinfo {author} {\bibfnamefont {L.}~\bibnamefont {{Murphy}}}, \bibinfo
  {author} {\bibfnamefont {E.}~\bibnamefont {{Kaiser}}}, \bibinfo {author}
  {\bibfnamefont {S.}~\bibnamefont {{Ekstr{\"o}m}}}, \bibinfo {author}
  {\bibfnamefont {C.}~\bibnamefont {{Georgy}}},\ and\ \bibinfo {author}
  {\bibfnamefont {G.}~\bibnamefont {{Meynet}}},\ }\href
  {https://doi.org/10.1093/mnrasl/slaa196} {\bibfield  {journal} {\bibinfo
  {journal} {\mnras}\ }\textbf {\bibinfo {volume} {502}},\ \bibinfo {pages}
  {L40} (\bibinfo {year} {2021})},\ \Eprint {https://arxiv.org/abs/2009.06585}
  {arXiv:2009.06585 [astro-ph.SR]} \BibitemShut {NoStop}%
\bibitem [{\citenamefont {{Kinugawa}}\ \emph {et~al.}(2020)\citenamefont
  {{Kinugawa}}, \citenamefont {{Nakamura}},\ and\ \citenamefont
  {{Nakano}}}]{2021MNRAS.501L..49K}%
  \BibitemOpen
  \bibfield  {author} {\bibinfo {author} {\bibfnamefont {T.}~\bibnamefont
  {{Kinugawa}}}, \bibinfo {author} {\bibfnamefont {T.}~\bibnamefont
  {{Nakamura}}},\ and\ \bibinfo {author} {\bibfnamefont {H.}~\bibnamefont
  {{Nakano}}},\ }\href {https://doi.org/10.1093/mnrasl/slaa191} {\bibfield
  {journal} {\bibinfo  {journal} {\mnras}\ }\textbf {\bibinfo {volume} {501}},\
  \bibinfo {pages} {L49} (\bibinfo {year} {2020})},\ \Eprint
  {https://arxiv.org/abs/2009.06922} {arXiv:2009.06922 [astro-ph.HE]}
  \BibitemShut {NoStop}%
\bibitem [{\citenamefont {{Costa}}\ \emph {et~al.}(2021)\citenamefont
  {{Costa}}, \citenamefont {{Bressan}}, \citenamefont {{Mapelli}},
  \citenamefont {{Marigo}}, \citenamefont {{Iorio}},\ and\ \citenamefont
  {{Spera}}}]{2021MNRAS.501.4514C}%
  \BibitemOpen
  \bibfield  {author} {\bibinfo {author} {\bibfnamefont {G.}~\bibnamefont
  {{Costa}}}, \bibinfo {author} {\bibfnamefont {A.}~\bibnamefont {{Bressan}}},
  \bibinfo {author} {\bibfnamefont {M.}~\bibnamefont {{Mapelli}}}, \bibinfo
  {author} {\bibfnamefont {P.}~\bibnamefont {{Marigo}}}, \bibinfo {author}
  {\bibfnamefont {G.}~\bibnamefont {{Iorio}}},\ and\ \bibinfo {author}
  {\bibfnamefont {M.}~\bibnamefont {{Spera}}},\ }\href
  {https://doi.org/10.1093/mnras/staa3916} {\bibfield  {journal} {\bibinfo
  {journal} {\mnras}\ }\textbf {\bibinfo {volume} {501}},\ \bibinfo {pages}
  {4514} (\bibinfo {year} {2021})},\ \Eprint {https://arxiv.org/abs/2010.02242}
  {arXiv:2010.02242 [astro-ph.SR]} \BibitemShut {NoStop}%
\bibitem [{\citenamefont {{Fuller}}\ and\ \citenamefont
  {{Ma}}(2019)}]{2019ApJ...881L...1F}%
  \BibitemOpen
  \bibfield  {author} {\bibinfo {author} {\bibfnamefont {J.}~\bibnamefont
  {{Fuller}}}\ and\ \bibinfo {author} {\bibfnamefont {L.}~\bibnamefont
  {{Ma}}},\ }\href {https://doi.org/10.3847/2041-8213/ab339b} {\bibfield
  {journal} {\bibinfo  {journal} {\apjl}\ }\textbf {\bibinfo {volume} {881}},\
  \bibinfo {eid} {L1} (\bibinfo {year} {2019})},\ \Eprint
  {https://arxiv.org/abs/1907.03714} {arXiv:1907.03714 [astro-ph.SR]}
  \BibitemShut {NoStop}%
\bibitem [{\citenamefont {{Belczynski}}\ \emph
  {et~al.}(2020{\natexlab{b}})\citenamefont {{Belczynski}}, \citenamefont
  {{Klencki}}, \citenamefont {{Fields}}, \citenamefont {{Olejak}},
  \citenamefont {{Berti}}, \citenamefont {{Meynet}}, \citenamefont {{Fryer}},
  \citenamefont {{Holz}}, \citenamefont {{O'Shaughnessy}}, \citenamefont
  {{Brown}}, \citenamefont {{Bulik}}, \citenamefont {{Leung}}, \citenamefont
  {{Nomoto}}, \citenamefont {{Madau}}, \citenamefont {{Hirschi}}, \citenamefont
  {{Kaiser}}, \citenamefont {{Jones}}, \citenamefont {{Mondal}}, \citenamefont
  {{Chruslinska}}, \citenamefont {{Drozda}}, \citenamefont {{Gerosa}},
  \citenamefont {{Doctor}}, \citenamefont {{Giersz}}, \citenamefont
  {{Ekstrom}}, \citenamefont {{Georgy}}, \citenamefont {{Askar}}, \citenamefont
  {{Baibhav}}, \citenamefont {{Wysocki}}, \citenamefont {{Natan}},
  \citenamefont {{Farr}}, \citenamefont {{Wiktorowicz}}, \citenamefont
  {{Coleman Miller}}, \citenamefont {{Farr}},\ and\ \citenamefont
  {{Lasota}}}]{2020A&A...636A.104B}%
  \BibitemOpen
  \bibfield  {author} {\bibinfo {author} {\bibfnamefont {K.}~\bibnamefont
  {{Belczynski}}}, \bibinfo {author} {\bibfnamefont {J.}~\bibnamefont
  {{Klencki}}}, \bibinfo {author} {\bibfnamefont {C.~E.}\ \bibnamefont
  {{Fields}}}, \bibinfo {author} {\bibfnamefont {A.}~\bibnamefont {{Olejak}}},
  \bibinfo {author} {\bibfnamefont {E.}~\bibnamefont {{Berti}}}, \bibinfo
  {author} {\bibfnamefont {G.}~\bibnamefont {{Meynet}}}, \bibinfo {author}
  {\bibfnamefont {C.~L.}\ \bibnamefont {{Fryer}}}, \bibinfo {author}
  {\bibfnamefont {D.~E.}\ \bibnamefont {{Holz}}}, \bibinfo {author}
  {\bibfnamefont {R.}~\bibnamefont {{O'Shaughnessy}}}, \bibinfo {author}
  {\bibfnamefont {D.~A.}\ \bibnamefont {{Brown}}}, \bibinfo {author}
  {\bibfnamefont {T.}~\bibnamefont {{Bulik}}}, \bibinfo {author} {\bibfnamefont
  {S.~C.}\ \bibnamefont {{Leung}}}, \bibinfo {author} {\bibfnamefont
  {K.}~\bibnamefont {{Nomoto}}}, \bibinfo {author} {\bibfnamefont
  {P.}~\bibnamefont {{Madau}}}, \bibinfo {author} {\bibfnamefont
  {R.}~\bibnamefont {{Hirschi}}}, \bibinfo {author} {\bibfnamefont
  {E.}~\bibnamefont {{Kaiser}}}, \bibinfo {author} {\bibfnamefont
  {S.}~\bibnamefont {{Jones}}}, \bibinfo {author} {\bibfnamefont
  {S.}~\bibnamefont {{Mondal}}}, \bibinfo {author} {\bibfnamefont
  {M.}~\bibnamefont {{Chruslinska}}}, \bibinfo {author} {\bibfnamefont
  {P.}~\bibnamefont {{Drozda}}}, \bibinfo {author} {\bibfnamefont
  {D.}~\bibnamefont {{Gerosa}}}, \bibinfo {author} {\bibfnamefont
  {Z.}~\bibnamefont {{Doctor}}}, \bibinfo {author} {\bibfnamefont
  {M.}~\bibnamefont {{Giersz}}}, \bibinfo {author} {\bibfnamefont
  {S.}~\bibnamefont {{Ekstrom}}}, \bibinfo {author} {\bibfnamefont
  {C.}~\bibnamefont {{Georgy}}}, \bibinfo {author} {\bibfnamefont
  {A.}~\bibnamefont {{Askar}}}, \bibinfo {author} {\bibfnamefont
  {V.}~\bibnamefont {{Baibhav}}}, \bibinfo {author} {\bibfnamefont
  {D.}~\bibnamefont {{Wysocki}}}, \bibinfo {author} {\bibfnamefont
  {T.}~\bibnamefont {{Natan}}}, \bibinfo {author} {\bibfnamefont {W.~M.}\
  \bibnamefont {{Farr}}}, \bibinfo {author} {\bibfnamefont {G.}~\bibnamefont
  {{Wiktorowicz}}}, \bibinfo {author} {\bibfnamefont {M.}~\bibnamefont
  {{Coleman Miller}}}, \bibinfo {author} {\bibfnamefont {B.}~\bibnamefont
  {{Farr}}},\ and\ \bibinfo {author} {\bibfnamefont {J.~P.}\ \bibnamefont
  {{Lasota}}},\ }\href {https://doi.org/10.1051/0004-6361/201936528} {\bibfield
   {journal} {\bibinfo  {journal} {\aap}\ }\textbf {\bibinfo {volume} {636}},\
  \bibinfo {eid} {A104} (\bibinfo {year} {2020}{\natexlab{b}})},\ \Eprint
  {https://arxiv.org/abs/1706.07053} {arXiv:1706.07053 [astro-ph.HE]}
  \BibitemShut {NoStop}%
\bibitem [{\citenamefont {{Gnedin}}\ \emph {et~al.}(2002)\citenamefont
  {{Gnedin}}, \citenamefont {{Zhao}}, \citenamefont {{Pringle}}, \citenamefont
  {{Fall}}, \citenamefont {{Livio}},\ and\ \citenamefont
  {{Meylan}}}]{2002ApJ...568L..23G}%
  \BibitemOpen
  \bibfield  {author} {\bibinfo {author} {\bibfnamefont {O.~Y.}\ \bibnamefont
  {{Gnedin}}}, \bibinfo {author} {\bibfnamefont {H.}~\bibnamefont {{Zhao}}},
  \bibinfo {author} {\bibfnamefont {J.~E.}\ \bibnamefont {{Pringle}}}, \bibinfo
  {author} {\bibfnamefont {S.~M.}\ \bibnamefont {{Fall}}}, \bibinfo {author}
  {\bibfnamefont {M.}~\bibnamefont {{Livio}}},\ and\ \bibinfo {author}
  {\bibfnamefont {G.}~\bibnamefont {{Meylan}}},\ }\href
  {https://doi.org/10.1086/340319} {\bibfield  {journal} {\bibinfo  {journal}
  {\apjl}\ }\textbf {\bibinfo {volume} {568}},\ \bibinfo {pages} {L23}
  (\bibinfo {year} {2002})},\ \Eprint {https://arxiv.org/abs/astro-ph/0202045}
  {arXiv:astro-ph/0202045 [astro-ph]} \BibitemShut {NoStop}%
\bibitem [{\citenamefont {{Merritt}}\ \emph {et~al.}(2004)\citenamefont
  {{Merritt}}, \citenamefont {{Milosavljevi{\'c}}}, \citenamefont {{Favata}},
  \citenamefont {{Hughes}},\ and\ \citenamefont
  {{Holz}}}]{2004ApJ...607L...9M}%
  \BibitemOpen
  \bibfield  {author} {\bibinfo {author} {\bibfnamefont {D.}~\bibnamefont
  {{Merritt}}}, \bibinfo {author} {\bibfnamefont {M.}~\bibnamefont
  {{Milosavljevi{\'c}}}}, \bibinfo {author} {\bibfnamefont {M.}~\bibnamefont
  {{Favata}}}, \bibinfo {author} {\bibfnamefont {S.~A.}\ \bibnamefont
  {{Hughes}}},\ and\ \bibinfo {author} {\bibfnamefont {D.~E.}\ \bibnamefont
  {{Holz}}},\ }\href {https://doi.org/10.1086/421551} {\bibfield  {journal}
  {\bibinfo  {journal} {\apjl}\ }\textbf {\bibinfo {volume} {607}},\ \bibinfo
  {pages} {L9} (\bibinfo {year} {2004})},\ \Eprint
  {https://arxiv.org/abs/astro-ph/0402057} {arXiv:astro-ph/0402057 [astro-ph]}
  \BibitemShut {NoStop}%
\bibitem [{\citenamefont {{Antonini}}\ and\ \citenamefont
  {{Rasio}}(2016)}]{2016ApJ...831..187A}%
  \BibitemOpen
  \bibfield  {author} {\bibinfo {author} {\bibfnamefont {F.}~\bibnamefont
  {{Antonini}}}\ and\ \bibinfo {author} {\bibfnamefont {F.~A.}\ \bibnamefont
  {{Rasio}}},\ }\href {https://doi.org/10.3847/0004-637X/831/2/187} {\bibfield
  {journal} {\bibinfo  {journal} {\apj}\ }\textbf {\bibinfo {volume} {831}},\
  \bibinfo {eid} {187} (\bibinfo {year} {2016})},\ \Eprint
  {https://arxiv.org/abs/1606.04889} {arXiv:1606.04889 [astro-ph.HE]}
  \BibitemShut {NoStop}%
\bibitem [{\citenamefont {{Gerosa}}\ and\ \citenamefont
  {{Kesden}}(2016)}]{2016PhRvD..93l4066G}%
  \BibitemOpen
  \bibfield  {author} {\bibinfo {author} {\bibfnamefont {D.}~\bibnamefont
  {{Gerosa}}}\ and\ \bibinfo {author} {\bibfnamefont {M.}~\bibnamefont
  {{Kesden}}},\ }\href {https://doi.org/10.1103/PhysRevD.93.124066} {\bibfield
  {journal} {\bibinfo  {journal} {\prd}\ }\textbf {\bibinfo {volume} {93}},\
  \bibinfo {eid} {124066} (\bibinfo {year} {2016})},\ \Eprint
  {https://arxiv.org/abs/1605.01067} {arXiv:1605.01067 [astro-ph.HE]}
  \BibitemShut {NoStop}%
\bibitem [{\citenamefont {{Barausse}}\ \emph {et~al.}(2012)\citenamefont
  {{Barausse}}, \citenamefont {{Morozova}},\ and\ \citenamefont
  {{Rezzolla}}}]{2012ApJ...758...63B}%
  \BibitemOpen
  \bibfield  {author} {\bibinfo {author} {\bibfnamefont {E.}~\bibnamefont
  {{Barausse}}}, \bibinfo {author} {\bibfnamefont {V.}~\bibnamefont
  {{Morozova}}},\ and\ \bibinfo {author} {\bibfnamefont {L.}~\bibnamefont
  {{Rezzolla}}},\ }\href {https://doi.org/10.1088/0004-637X/758/1/63}
  {\bibfield  {journal} {\bibinfo  {journal} {\apj}\ }\textbf {\bibinfo
  {volume} {758}},\ \bibinfo {eid} {63} (\bibinfo {year} {2012})},\ \Eprint
  {https://arxiv.org/abs/1206.3803} {arXiv:1206.3803 [gr-qc]} \BibitemShut
  {NoStop}%
\bibitem [{\citenamefont {{Barausse}}\ and\ \citenamefont
  {{Rezzolla}}(2009)}]{2009ApJ...704L..40B}%
  \BibitemOpen
  \bibfield  {author} {\bibinfo {author} {\bibfnamefont {E.}~\bibnamefont
  {{Barausse}}}\ and\ \bibinfo {author} {\bibfnamefont {L.}~\bibnamefont
  {{Rezzolla}}},\ }\href {https://doi.org/10.1088/0004-637X/704/1/L40}
  {\bibfield  {journal} {\bibinfo  {journal} {\apjl}\ }\textbf {\bibinfo
  {volume} {704}},\ \bibinfo {pages} {L40} (\bibinfo {year} {2009})},\ \Eprint
  {https://arxiv.org/abs/0904.2577} {arXiv:0904.2577 [gr-qc]} \BibitemShut
  {NoStop}%
\bibitem [{\citenamefont {{Hofmann}}\ \emph {et~al.}(2016)\citenamefont
  {{Hofmann}}, \citenamefont {{Barausse}},\ and\ \citenamefont
  {{Rezzolla}}}]{2016ApJ...825L..19H}%
  \BibitemOpen
  \bibfield  {author} {\bibinfo {author} {\bibfnamefont {F.}~\bibnamefont
  {{Hofmann}}}, \bibinfo {author} {\bibfnamefont {E.}~\bibnamefont
  {{Barausse}}},\ and\ \bibinfo {author} {\bibfnamefont {L.}~\bibnamefont
  {{Rezzolla}}},\ }\href {https://doi.org/10.3847/2041-8205/825/2/L19}
  {\bibfield  {journal} {\bibinfo  {journal} {\apjl}\ }\textbf {\bibinfo
  {volume} {825}},\ \bibinfo {eid} {L19} (\bibinfo {year} {2016})},\ \Eprint
  {https://arxiv.org/abs/1605.01938} {arXiv:1605.01938 [gr-qc]} \BibitemShut
  {NoStop}%
\bibitem [{\citenamefont {{Lousto}}\ and\ \citenamefont
  {{Zlochower}}(2008)}]{2008PhRvD..77d4028L}%
  \BibitemOpen
  \bibfield  {author} {\bibinfo {author} {\bibfnamefont {C.~O.}\ \bibnamefont
  {{Lousto}}}\ and\ \bibinfo {author} {\bibfnamefont {Y.}~\bibnamefont
  {{Zlochower}}},\ }\href {https://doi.org/10.1103/PhysRevD.77.044028}
  {\bibfield  {journal} {\bibinfo  {journal} {\prd}\ }\textbf {\bibinfo
  {volume} {77}},\ \bibinfo {eid} {044028} (\bibinfo {year} {2008})},\ \Eprint
  {https://arxiv.org/abs/0708.4048} {arXiv:0708.4048 [gr-qc]} \BibitemShut
  {NoStop}%
\bibitem [{\citenamefont {{Lousto}}\ \emph {et~al.}(2012)\citenamefont
  {{Lousto}}, \citenamefont {{Zlochower}}, \citenamefont {{Dotti}},\ and\
  \citenamefont {{Volonteri}}}]{2012PhRvD..85h4015L}%
  \BibitemOpen
  \bibfield  {author} {\bibinfo {author} {\bibfnamefont {C.~O.}\ \bibnamefont
  {{Lousto}}}, \bibinfo {author} {\bibfnamefont {Y.}~\bibnamefont
  {{Zlochower}}}, \bibinfo {author} {\bibfnamefont {M.}~\bibnamefont
  {{Dotti}}},\ and\ \bibinfo {author} {\bibfnamefont {M.}~\bibnamefont
  {{Volonteri}}},\ }\href {https://doi.org/10.1103/PhysRevD.85.084015}
  {\bibfield  {journal} {\bibinfo  {journal} {\prd}\ }\textbf {\bibinfo
  {volume} {85}},\ \bibinfo {eid} {084015} (\bibinfo {year} {2012})},\ \Eprint
  {https://arxiv.org/abs/1201.1923} {arXiv:1201.1923 [gr-qc]} \BibitemShut
  {NoStop}%
\bibitem [{\citenamefont {{Lousto}}\ and\ \citenamefont
  {{Zlochower}}(2013)}]{2013PhRvD..87h4027L}%
  \BibitemOpen
  \bibfield  {author} {\bibinfo {author} {\bibfnamefont {C.~O.}\ \bibnamefont
  {{Lousto}}}\ and\ \bibinfo {author} {\bibfnamefont {Y.}~\bibnamefont
  {{Zlochower}}},\ }\href {https://doi.org/10.1103/PhysRevD.87.084027}
  {\bibfield  {journal} {\bibinfo  {journal} {\prd}\ }\textbf {\bibinfo
  {volume} {87}},\ \bibinfo {eid} {084027} (\bibinfo {year} {2013})},\ \Eprint
  {https://arxiv.org/abs/1211.7099} {arXiv:1211.7099 [gr-qc]} \BibitemShut
  {NoStop}%
\bibitem [{\citenamefont {{Bekenstein}}(1973)}]{1973ApJ...183..657B}%
  \BibitemOpen
  \bibfield  {author} {\bibinfo {author} {\bibfnamefont {J.~D.}\ \bibnamefont
  {{Bekenstein}}},\ }\href {https://doi.org/10.1086/152255} {\bibfield
  {journal} {\bibinfo  {journal} {\apj}\ }\textbf {\bibinfo {volume} {183}},\
  \bibinfo {pages} {657} (\bibinfo {year} {1973})}\BibitemShut {NoStop}%
\bibitem [{\citenamefont {{Fitchett}}(1983)}]{1983MNRAS.203.1049F}%
  \BibitemOpen
  \bibfield  {author} {\bibinfo {author} {\bibfnamefont {M.~J.}\ \bibnamefont
  {{Fitchett}}},\ }\href {https://doi.org/10.1093/mnras/203.4.1049} {\bibfield
  {journal} {\bibinfo  {journal} {\mnras}\ }\textbf {\bibinfo {volume} {203}},\
  \bibinfo {pages} {1049} (\bibinfo {year} {1983})}\BibitemShut {NoStop}%
\bibitem [{\citenamefont {{Racine}}(2008)}]{2008PhRvD..78d4021R}%
  \BibitemOpen
  \bibfield  {author} {\bibinfo {author} {\bibfnamefont {{\'E}.}~\bibnamefont
  {{Racine}}},\ }\href {https://doi.org/10.1103/PhysRevD.78.044021} {\bibfield
  {journal} {\bibinfo  {journal} {\prd}\ }\textbf {\bibinfo {volume} {78}},\
  \bibinfo {eid} {044021} (\bibinfo {year} {2008})},\ \Eprint
  {https://arxiv.org/abs/0803.1820} {arXiv:0803.1820 [gr-qc]} \BibitemShut
  {NoStop}%
\bibitem [{\citenamefont {{Gerosa}}\ \emph
  {et~al.}(2021{\natexlab{b}})\citenamefont {{Gerosa}}, \citenamefont
  {{Mould}}, \citenamefont {{Gangardt}}, \citenamefont {{Schmidt}},
  \citenamefont {{Pratten}},\ and\ \citenamefont
  {{Thomas}}}]{2021PhRvD.103f4067G}%
  \BibitemOpen
  \bibfield  {author} {\bibinfo {author} {\bibfnamefont {D.}~\bibnamefont
  {{Gerosa}}}, \bibinfo {author} {\bibfnamefont {M.}~\bibnamefont {{Mould}}},
  \bibinfo {author} {\bibfnamefont {D.}~\bibnamefont {{Gangardt}}}, \bibinfo
  {author} {\bibfnamefont {P.}~\bibnamefont {{Schmidt}}}, \bibinfo {author}
  {\bibfnamefont {G.}~\bibnamefont {{Pratten}}},\ and\ \bibinfo {author}
  {\bibfnamefont {L.~M.}\ \bibnamefont {{Thomas}}},\ }\href
  {https://doi.org/10.1103/PhysRevD.103.064067} {\bibfield  {journal} {\bibinfo
   {journal} {\prd}\ }\textbf {\bibinfo {volume} {103}},\ \bibinfo {eid}
  {064067} (\bibinfo {year} {2021}{\natexlab{b}})},\ \Eprint
  {https://arxiv.org/abs/2011.11948} {arXiv:2011.11948 [gr-qc]} \BibitemShut
  {NoStop}%
\bibitem [{\citenamefont {{Gerosa}}\ and\ \citenamefont
  {{Berti}}(2017)}]{2017PhRvD..95l4046G}%
  \BibitemOpen
  \bibfield  {author} {\bibinfo {author} {\bibfnamefont {D.}~\bibnamefont
  {{Gerosa}}}\ and\ \bibinfo {author} {\bibfnamefont {E.}~\bibnamefont
  {{Berti}}},\ }\href {https://doi.org/10.1103/PhysRevD.95.124046} {\bibfield
  {journal} {\bibinfo  {journal} {\prd}\ }\textbf {\bibinfo {volume} {95}},\
  \bibinfo {eid} {124046} (\bibinfo {year} {2017})},\ \Eprint
  {https://arxiv.org/abs/1703.06223} {arXiv:1703.06223 [gr-qc]} \BibitemShut
  {NoStop}%
\bibitem [{\citenamefont {{Pretorius}}(2005)}]{2005PhRvL..95l1101P}%
  \BibitemOpen
  \bibfield  {author} {\bibinfo {author} {\bibfnamefont {F.}~\bibnamefont
  {{Pretorius}}},\ }\href {https://doi.org/10.1103/PhysRevLett.95.121101}
  {\bibfield  {journal} {\bibinfo  {journal} {\prl}\ }\textbf {\bibinfo
  {volume} {95}},\ \bibinfo {eid} {121101} (\bibinfo {year} {2005})},\ \Eprint
  {https://arxiv.org/abs/gr-qc/0507014} {arXiv:gr-qc/0507014 [gr-qc]}
  \BibitemShut {NoStop}%
\bibitem [{\citenamefont {{Abbott}}\ \emph
  {et~al.}(2016{\natexlab{a}})\citenamefont {{Abbott}} \emph
  {et~al.}}]{2016PhRvX...6d1015A}%
  \BibitemOpen
  \bibfield  {author} {\bibinfo {author} {\bibfnamefont {B.~P.}\ \bibnamefont
  {{Abbott}}} \emph {et~al.} (\bibinfo {collaboration} {LIGO and Virgo
  Collaboration}),\ }\href {https://doi.org/10.1103/PhysRevX.6.041015}
  {\bibfield  {journal} {\bibinfo  {journal} {\prx}\ }\textbf {\bibinfo
  {volume} {6}},\ \bibinfo {eid} {041015} (\bibinfo {year}
  {2016}{\natexlab{a}})},\ \Eprint {https://arxiv.org/abs/1606.04856}
  {arXiv:1606.04856 [gr-qc]} \BibitemShut {NoStop}%
\bibitem [{\citenamefont {{Fishbach}}\ \emph {et~al.}(2018)\citenamefont
  {{Fishbach}}, \citenamefont {{Holz}},\ and\ \citenamefont
  {{Farr}}}]{2018ApJ...863L..41F}%
  \BibitemOpen
  \bibfield  {author} {\bibinfo {author} {\bibfnamefont {M.}~\bibnamefont
  {{Fishbach}}}, \bibinfo {author} {\bibfnamefont {D.~E.}\ \bibnamefont
  {{Holz}}},\ and\ \bibinfo {author} {\bibfnamefont {W.~M.}\ \bibnamefont
  {{Farr}}},\ }\href {https://doi.org/10.3847/2041-8213/aad800} {\bibfield
  {journal} {\bibinfo  {journal} {\apjl}\ }\textbf {\bibinfo {volume} {863}},\
  \bibinfo {eid} {L41} (\bibinfo {year} {2018})},\ \Eprint
  {https://arxiv.org/abs/1805.10270} {arXiv:1805.10270 [astro-ph.HE]}
  \BibitemShut {NoStop}%
\bibitem [{\citenamefont {{Virtanen}}\ \emph {et~al.}(2020)\citenamefont
  {{Virtanen}} \emph {et~al.}}]{2020NatMe..17..261V}%
  \BibitemOpen
  \bibfield  {author} {\bibinfo {author} {\bibfnamefont {P.}~\bibnamefont
  {{Virtanen}}} \emph {et~al.},\ }\href
  {https://doi.org/10.1038/s41592-019-0686-2} {\bibfield  {journal} {\bibinfo
  {journal} {Nat. Methods}\ }\textbf {\bibinfo {volume} {17}},\ \bibinfo
  {pages} {261} (\bibinfo {year} {2020})},\ \Eprint
  {https://arxiv.org/abs/1907.10121} {arXiv:1907.10121 [cs.MS]} \BibitemShut
  {NoStop}%
\bibitem [{\citenamefont {{Silverman}}(1986)}]{1986desd.book.....S}%
  \BibitemOpen
  \bibfield  {author} {\bibinfo {author} {\bibfnamefont {B.~W.}\ \bibnamefont
  {{Silverman}}},\ }\href@noop {} {\emph {\bibinfo {title} {{Density estimation
  for statistics and data analysis}}}}\ (\bibinfo  {publisher} {Chapman and
  Hall},\ \bibinfo {year} {1986})\BibitemShut {NoStop}%
\bibitem [{\citenamefont {{Speagle}}(2020)}]{2020MNRAS.493.3132S}%
  \BibitemOpen
  \bibfield  {author} {\bibinfo {author} {\bibfnamefont {J.~S.}\ \bibnamefont
  {{Speagle}}},\ }\href {https://doi.org/10.1093/mnras/staa278} {\bibfield
  {journal} {\bibinfo  {journal} {\mnras}\ }\textbf {\bibinfo {volume} {493}},\
  \bibinfo {pages} {3132} (\bibinfo {year} {2020})},\ \Eprint
  {https://arxiv.org/abs/1904.02180} {arXiv:1904.02180 [astro-ph.IM]}
  \BibitemShut {NoStop}%
\bibitem [{\citenamefont {{Ashton}}\ \emph {et~al.}(2019)\citenamefont
  {{Ashton}}, \citenamefont {{H{\"u}bner}}, \citenamefont {{Lasky}},
  \citenamefont {{Talbot}}, \citenamefont {{Ackley}}, \citenamefont
  {{Biscoveanu}}, \citenamefont {{Chu}}, \citenamefont {{Divakarla}},
  \citenamefont {{Easter}}, \citenamefont {{Goncharov}}, \citenamefont
  {{Hernandez Vivanco}}, \citenamefont {{Harms}}, \citenamefont {{Lower}},
  \citenamefont {{Meadors}}, \citenamefont {{Melchor}}, \citenamefont
  {{Payne}}, \citenamefont {{Pitkin}}, \citenamefont {{Powell}}, \citenamefont
  {{Sarin}}, \citenamefont {{Smith}},\ and\ \citenamefont
  {{Thrane}}}]{2019ApJS..241...27A}%
  \BibitemOpen
  \bibfield  {author} {\bibinfo {author} {\bibfnamefont {G.}~\bibnamefont
  {{Ashton}}}, \bibinfo {author} {\bibfnamefont {M.}~\bibnamefont
  {{H{\"u}bner}}}, \bibinfo {author} {\bibfnamefont {P.~D.}\ \bibnamefont
  {{Lasky}}}, \bibinfo {author} {\bibfnamefont {C.}~\bibnamefont {{Talbot}}},
  \bibinfo {author} {\bibfnamefont {K.}~\bibnamefont {{Ackley}}}, \bibinfo
  {author} {\bibfnamefont {S.}~\bibnamefont {{Biscoveanu}}}, \bibinfo {author}
  {\bibfnamefont {Q.}~\bibnamefont {{Chu}}}, \bibinfo {author} {\bibfnamefont
  {A.}~\bibnamefont {{Divakarla}}}, \bibinfo {author} {\bibfnamefont {P.~J.}\
  \bibnamefont {{Easter}}}, \bibinfo {author} {\bibfnamefont {B.}~\bibnamefont
  {{Goncharov}}}, \bibinfo {author} {\bibfnamefont {F.}~\bibnamefont
  {{Hernandez Vivanco}}}, \bibinfo {author} {\bibfnamefont {J.}~\bibnamefont
  {{Harms}}}, \bibinfo {author} {\bibfnamefont {M.~E.}\ \bibnamefont
  {{Lower}}}, \bibinfo {author} {\bibfnamefont {G.~D.}\ \bibnamefont
  {{Meadors}}}, \bibinfo {author} {\bibfnamefont {D.}~\bibnamefont
  {{Melchor}}}, \bibinfo {author} {\bibfnamefont {E.}~\bibnamefont {{Payne}}},
  \bibinfo {author} {\bibfnamefont {M.~D.}\ \bibnamefont {{Pitkin}}}, \bibinfo
  {author} {\bibfnamefont {J.}~\bibnamefont {{Powell}}}, \bibinfo {author}
  {\bibfnamefont {N.}~\bibnamefont {{Sarin}}}, \bibinfo {author} {\bibfnamefont
  {R.~J.~E.}\ \bibnamefont {{Smith}}},\ and\ \bibinfo {author} {\bibfnamefont
  {E.}~\bibnamefont {{Thrane}}},\ }\href
  {https://doi.org/10.3847/1538-4365/ab06fc} {\bibfield  {journal} {\bibinfo
  {journal} {\apjs}\ }\textbf {\bibinfo {volume} {241}},\ \bibinfo {eid} {27}
  (\bibinfo {year} {2019})},\ \Eprint {https://arxiv.org/abs/1811.02042}
  {arXiv:1811.02042 [astro-ph.IM]} \BibitemShut {NoStop}%
\bibitem [{\citenamefont {{Romero-Shaw}}\ \emph
  {et~al.}(2020{\natexlab{b}})\citenamefont {{Romero-Shaw}}, \citenamefont
  {{Talbot}}, \citenamefont {{Biscoveanu}}, \citenamefont {{D'Emilio}} \emph
  {et~al.}}]{2020MNRAS.499.3295R}%
  \BibitemOpen
  \bibfield  {author} {\bibinfo {author} {\bibfnamefont {I.~M.}\ \bibnamefont
  {{Romero-Shaw}}}, \bibinfo {author} {\bibfnamefont {C.}~\bibnamefont
  {{Talbot}}}, \bibinfo {author} {\bibfnamefont {S.}~\bibnamefont
  {{Biscoveanu}}}, \bibinfo {author} {\bibfnamefont {V.}~\bibnamefont
  {{D'Emilio}}}, \emph {et~al.},\ }\href
  {https://doi.org/10.1093/mnras/staa2850} {\bibfield  {journal} {\bibinfo
  {journal} {\mnras}\ }\textbf {\bibinfo {volume} {499}},\ \bibinfo {pages}
  {3295} (\bibinfo {year} {2020}{\natexlab{b}})},\ \Eprint
  {https://arxiv.org/abs/2006.00714} {arXiv:2006.00714 [astro-ph.IM]}
  \BibitemShut {NoStop}%
\bibitem [{\citenamefont {Odland}(2018)}]{tommy_odland_2018_2392268}%
  \BibitemOpen
  \bibfield  {author} {\bibinfo {author} {\bibfnamefont {T.}~\bibnamefont
  {Odland}},\ }\bibfield  {journal} {\bibinfo  {journal} {KDEpy}\ }\href
  {https://doi.org/10.5281/zenodo.2392268} {10.5281/zenodo.2392268} (\bibinfo
  {year} {2018})\BibitemShut {NoStop}%
\bibitem [{\citenamefont {{Botev}}\ \emph {et~al.}(2010)\citenamefont
  {{Botev}}, \citenamefont {{Grotowski}},\ and\ \citenamefont
  {{Kroese}}}]{2010arXiv1011.2602B}%
  \BibitemOpen
  \bibfield  {author} {\bibinfo {author} {\bibfnamefont {Z.~I.}\ \bibnamefont
  {{Botev}}}, \bibinfo {author} {\bibfnamefont {J.~F.}\ \bibnamefont
  {{Grotowski}}},\ and\ \bibinfo {author} {\bibfnamefont {D.~P.}\ \bibnamefont
  {{Kroese}}},\ }\href@noop {} {\bibfield  {journal} {\bibinfo  {journal} {{}}\
  } (\bibinfo {year} {2010})},\ \Eprint {https://arxiv.org/abs/1011.2602}
  {arXiv:1011.2602 [math.ST]} \BibitemShut {NoStop}%
\bibitem [{\citenamefont {Sheather}\ and\ \citenamefont
  {Jones}(1991)}]{10.2307/2345597}%
  \BibitemOpen
  \bibfield  {author} {\bibinfo {author} {\bibfnamefont {S.~J.}\ \bibnamefont
  {Sheather}}\ and\ \bibinfo {author} {\bibfnamefont {M.~C.}\ \bibnamefont
  {Jones}},\ }\href {http://www.jstor.org/stable/2345597} {\bibfield  {journal}
  {\bibinfo  {journal} {J. R. Stat. Soc. Series B Stat. Methodol.}\ }\textbf
  {\bibinfo {volume} {53}},\ \bibinfo {pages} {683} (\bibinfo {year}
  {1991})}\BibitemShut {NoStop}%
\bibitem [{\citenamefont {Abadi}\ \emph {et~al.}(2016)\citenamefont {Abadi},
  \citenamefont {Barham}, \citenamefont {Chen}, \citenamefont {Chen},
  \citenamefont {Davis}, \citenamefont {Dean}, \citenamefont {Devin},
  \citenamefont {Ghemawat}, \citenamefont {Irving}, \citenamefont {Isard},
  \citenamefont {Kudlur}, \citenamefont {Levenberg}, \citenamefont {Monga},
  \citenamefont {Moore}, \citenamefont {Murray}, \citenamefont {Steiner},
  \citenamefont {Tucker}, \citenamefont {Vasudevan}, \citenamefont {Warden},
  \citenamefont {Wicke}, \citenamefont {Yu},\ and\ \citenamefont
  {Zheng}}]{abadi2016tensorflow}%
  \BibitemOpen
  \bibfield  {author} {\bibinfo {author} {\bibfnamefont {M.}~\bibnamefont
  {Abadi}}, \bibinfo {author} {\bibfnamefont {P.}~\bibnamefont {Barham}},
  \bibinfo {author} {\bibfnamefont {J.}~\bibnamefont {Chen}}, \bibinfo {author}
  {\bibfnamefont {Z.}~\bibnamefont {Chen}}, \bibinfo {author} {\bibfnamefont
  {A.}~\bibnamefont {Davis}}, \bibinfo {author} {\bibfnamefont
  {J.}~\bibnamefont {Dean}}, \bibinfo {author} {\bibfnamefont {M.}~\bibnamefont
  {Devin}}, \bibinfo {author} {\bibfnamefont {S.}~\bibnamefont {Ghemawat}},
  \bibinfo {author} {\bibfnamefont {G.}~\bibnamefont {Irving}}, \bibinfo
  {author} {\bibfnamefont {M.}~\bibnamefont {Isard}}, \bibinfo {author}
  {\bibfnamefont {M.}~\bibnamefont {Kudlur}}, \bibinfo {author} {\bibfnamefont
  {J.}~\bibnamefont {Levenberg}}, \bibinfo {author} {\bibfnamefont
  {R.}~\bibnamefont {Monga}}, \bibinfo {author} {\bibfnamefont
  {S.}~\bibnamefont {Moore}}, \bibinfo {author} {\bibfnamefont {D.~G.}\
  \bibnamefont {Murray}}, \bibinfo {author} {\bibfnamefont {B.}~\bibnamefont
  {Steiner}}, \bibinfo {author} {\bibfnamefont {P.}~\bibnamefont {Tucker}},
  \bibinfo {author} {\bibfnamefont {V.}~\bibnamefont {Vasudevan}}, \bibinfo
  {author} {\bibfnamefont {P.}~\bibnamefont {Warden}}, \bibinfo {author}
  {\bibfnamefont {M.}~\bibnamefont {Wicke}}, \bibinfo {author} {\bibfnamefont
  {Y.}~\bibnamefont {Yu}},\ and\ \bibinfo {author} {\bibfnamefont
  {X.}~\bibnamefont {Zheng}},\ }in\ \href@noop {} {\emph {\bibinfo {booktitle}
  {Proceedings of the 12th USENIX Conference on Operating Systems Design and
  Implementation}}},\ \bibinfo {series and number} {OSDI'16}\ (\bibinfo
  {publisher} {USENIX Association},\ \bibinfo {address} {USA},\ \bibinfo {year}
  {2016})\ p.\ \bibinfo {pages} {265–283}\BibitemShut {NoStop}%
\bibitem [{\citenamefont {{Xu}}\ \emph {et~al.}(2015)\citenamefont {{Xu}},
  \citenamefont {{Wang}}, \citenamefont {{Chen}},\ and\ \citenamefont
  {{Li}}}]{2015arXiv150500853X}%
  \BibitemOpen
  \bibfield  {author} {\bibinfo {author} {\bibfnamefont {B.}~\bibnamefont
  {{Xu}}}, \bibinfo {author} {\bibfnamefont {N.}~\bibnamefont {{Wang}}},
  \bibinfo {author} {\bibfnamefont {T.}~\bibnamefont {{Chen}}},\ and\ \bibinfo
  {author} {\bibfnamefont {M.}~\bibnamefont {{Li}}},\ }\href@noop {} {\bibfield
   {journal} {\bibinfo  {journal} {{}}\ } (\bibinfo {year} {2015})},\ \Eprint
  {https://arxiv.org/abs/1505.00853} {arXiv:1505.00853 [cs.LG]} \BibitemShut
  {NoStop}%
\bibitem [{\citenamefont {{Kingma}}\ and\ \citenamefont
  {{Ba}}(2014)}]{2014arXiv1412.6980K}%
  \BibitemOpen
  \bibfield  {author} {\bibinfo {author} {\bibfnamefont {D.~P.}\ \bibnamefont
  {{Kingma}}}\ and\ \bibinfo {author} {\bibfnamefont {J.}~\bibnamefont
  {{Ba}}},\ }in\ \href@noop {} {\emph {\bibinfo {booktitle} {3rd International
  Conference for Learning Representations}}}\ (\bibinfo  {publisher}
  {Conference Track Proceedings},\ \bibinfo {year} {2014})\ \Eprint
  {https://arxiv.org/abs/1412.6980} {arXiv:1412.6980 [cs.LG]} \BibitemShut
  {NoStop}%
\bibitem [{\citenamefont {Hellinger}(1909)}]{Hellinger+1909+210+271}%
  \BibitemOpen
  \bibfield  {author} {\bibinfo {author} {\bibfnamefont {E.}~\bibnamefont
  {Hellinger}},\ }\href {https://doi.org/doi:10.1515/crll.1909.136.210}
  {\bibfield  {journal} {\bibinfo  {journal} {Journal f{\"u}r die reine und
  angewandte Mathematik}\ }\textbf {\bibinfo {volume} {1909}},\ \bibinfo
  {pages} {210} (\bibinfo {year} {1909})}\BibitemShut {NoStop}%
\bibitem [{\citenamefont {{Moore}}\ and\ \citenamefont
  {{Gerosa}}(2021)}]{2021PhRvD.104h3008M}%
  \BibitemOpen
  \bibfield  {author} {\bibinfo {author} {\bibfnamefont {C.~J.}\ \bibnamefont
  {{Moore}}}\ and\ \bibinfo {author} {\bibfnamefont {D.}~\bibnamefont
  {{Gerosa}}},\ }\href {https://doi.org/10.1103/PhysRevD.104.083008} {\bibfield
   {journal} {\bibinfo  {journal} {\prd}\ }\textbf {\bibinfo {volume} {104}},\
  \bibinfo {eid} {083008} (\bibinfo {year} {2021})},\ \Eprint
  {https://arxiv.org/abs/2108.02462} {arXiv:2108.02462 [gr-qc]} \BibitemShut
  {NoStop}%
\bibitem [{\citenamefont {{Finn}}\ and\ \citenamefont
  {{Chernoff}}(1993)}]{1993PhRvD..47.2198F}%
  \BibitemOpen
  \bibfield  {author} {\bibinfo {author} {\bibfnamefont {L.~S.}\ \bibnamefont
  {{Finn}}}\ and\ \bibinfo {author} {\bibfnamefont {D.~F.}\ \bibnamefont
  {{Chernoff}}},\ }\href {https://doi.org/10.1103/PhysRevD.47.2198} {\bibfield
  {journal} {\bibinfo  {journal} {\prd}\ }\textbf {\bibinfo {volume} {47}},\
  \bibinfo {pages} {2198} (\bibinfo {year} {1993})},\ \Eprint
  {https://arxiv.org/abs/gr-qc/9301003} {arXiv:gr-qc/9301003 [gr-qc]}
  \BibitemShut {NoStop}%
\bibitem [{\citenamefont {{Finn}}(1996)}]{1996PhRvD..53.2878F}%
  \BibitemOpen
  \bibfield  {author} {\bibinfo {author} {\bibfnamefont {L.~S.}\ \bibnamefont
  {{Finn}}},\ }\href {https://doi.org/10.1103/PhysRevD.53.2878} {\bibfield
  {journal} {\bibinfo  {journal} {\prd}\ }\textbf {\bibinfo {volume} {53}},\
  \bibinfo {pages} {2878} (\bibinfo {year} {1996})},\ \Eprint
  {https://arxiv.org/abs/gr-qc/9601048} {arXiv:gr-qc/9601048 [gr-qc]}
  \BibitemShut {NoStop}%
\bibitem [{\citenamefont {Gerosa}(2017)}]{davide_gerosa_2017_889966}%
  \BibitemOpen
  \bibfield  {author} {\bibinfo {author} {\bibfnamefont {D.}~\bibnamefont
  {Gerosa}},\ }\bibfield  {journal} {\bibinfo  {journal} {gwdet}\ }\href
  {https://doi.org/10.5281/zenodo.889966} {10.5281/zenodo.889966} (\bibinfo
  {year} {2017})\BibitemShut {NoStop}%
\bibitem [{\citenamefont {Nitz}\ \emph {et~al.}(2021)\citenamefont {Nitz},
  \citenamefont {Harry}, \citenamefont {Brown}, \citenamefont {Biwer} \emph
  {et~al.}}]{alex_nitz_2021_4556907}%
  \BibitemOpen
  \bibfield  {author} {\bibinfo {author} {\bibfnamefont {A.}~\bibnamefont
  {Nitz}}, \bibinfo {author} {\bibfnamefont {I.}~\bibnamefont {Harry}},
  \bibinfo {author} {\bibfnamefont {D.}~\bibnamefont {Brown}}, \bibinfo
  {author} {\bibfnamefont {C.~M.}\ \bibnamefont {Biwer}}, \emph {et~al.},\
  }\bibfield  {journal} {\bibinfo  {journal} {pycbc}\ }\href
  {https://doi.org/10.5281/zenodo.4556907} {10.5281/zenodo.4556907} (\bibinfo
  {year} {2021})\BibitemShut {NoStop}%
\bibitem [{\citenamefont {{Hannam}}\ \emph {et~al.}(2014)\citenamefont
  {{Hannam}}, \citenamefont {{Schmidt}}, \citenamefont {{Boh{\'e}}},
  \citenamefont {{Haegel}}, \citenamefont {{Husa}}, \citenamefont {{Ohme}},
  \citenamefont {{Pratten}},\ and\ \citenamefont
  {{P{\"u}rrer}}}]{2014PhRvL.113o1101H}%
  \BibitemOpen
  \bibfield  {author} {\bibinfo {author} {\bibfnamefont {M.}~\bibnamefont
  {{Hannam}}}, \bibinfo {author} {\bibfnamefont {P.}~\bibnamefont {{Schmidt}}},
  \bibinfo {author} {\bibfnamefont {A.}~\bibnamefont {{Boh{\'e}}}}, \bibinfo
  {author} {\bibfnamefont {L.}~\bibnamefont {{Haegel}}}, \bibinfo {author}
  {\bibfnamefont {S.}~\bibnamefont {{Husa}}}, \bibinfo {author} {\bibfnamefont
  {F.}~\bibnamefont {{Ohme}}}, \bibinfo {author} {\bibfnamefont
  {G.}~\bibnamefont {{Pratten}}},\ and\ \bibinfo {author} {\bibfnamefont
  {M.}~\bibnamefont {{P{\"u}rrer}}},\ }\href
  {https://doi.org/10.1103/PhysRevLett.113.151101} {\bibfield  {journal}
  {\bibinfo  {journal} {\prl}\ }\textbf {\bibinfo {volume} {113}},\ \bibinfo
  {eid} {151101} (\bibinfo {year} {2014})},\ \Eprint
  {https://arxiv.org/abs/1308.3271} {arXiv:1308.3271 [gr-qc]} \BibitemShut
  {NoStop}%
\bibitem [{\citenamefont {{Husa}}\ \emph {et~al.}(2016)\citenamefont {{Husa}},
  \citenamefont {{Khan}}, \citenamefont {{Hannam}}, \citenamefont
  {{P{\"u}rrer}}, \citenamefont {{Ohme}}, \citenamefont {{Forteza}},\ and\
  \citenamefont {{Boh{\'e}}}}]{2016PhRvD..93d4006H}%
  \BibitemOpen
  \bibfield  {author} {\bibinfo {author} {\bibfnamefont {S.}~\bibnamefont
  {{Husa}}}, \bibinfo {author} {\bibfnamefont {S.}~\bibnamefont {{Khan}}},
  \bibinfo {author} {\bibfnamefont {M.}~\bibnamefont {{Hannam}}}, \bibinfo
  {author} {\bibfnamefont {M.}~\bibnamefont {{P{\"u}rrer}}}, \bibinfo {author}
  {\bibfnamefont {F.}~\bibnamefont {{Ohme}}}, \bibinfo {author} {\bibfnamefont
  {X.~J.}\ \bibnamefont {{Forteza}}},\ and\ \bibinfo {author} {\bibfnamefont
  {A.}~\bibnamefont {{Boh{\'e}}}},\ }\href
  {https://doi.org/10.1103/PhysRevD.93.044006} {\bibfield  {journal} {\bibinfo
  {journal} {\prd}\ }\textbf {\bibinfo {volume} {93}},\ \bibinfo {eid} {044006}
  (\bibinfo {year} {2016})},\ \Eprint {https://arxiv.org/abs/1508.07250}
  {arXiv:1508.07250 [gr-qc]} \BibitemShut {NoStop}%
\bibitem [{\citenamefont {{Khan}}\ \emph {et~al.}(2016)\citenamefont {{Khan}},
  \citenamefont {{Husa}}, \citenamefont {{Hannam}}, \citenamefont {{Ohme}},
  \citenamefont {{P{\"u}rrer}}, \citenamefont {{Forteza}},\ and\ \citenamefont
  {{Boh{\'e}}}}]{2016PhRvD..93d4007K}%
  \BibitemOpen
  \bibfield  {author} {\bibinfo {author} {\bibfnamefont {S.}~\bibnamefont
  {{Khan}}}, \bibinfo {author} {\bibfnamefont {S.}~\bibnamefont {{Husa}}},
  \bibinfo {author} {\bibfnamefont {M.}~\bibnamefont {{Hannam}}}, \bibinfo
  {author} {\bibfnamefont {F.}~\bibnamefont {{Ohme}}}, \bibinfo {author}
  {\bibfnamefont {M.}~\bibnamefont {{P{\"u}rrer}}}, \bibinfo {author}
  {\bibfnamefont {X.~J.}\ \bibnamefont {{Forteza}}},\ and\ \bibinfo {author}
  {\bibfnamefont {A.}~\bibnamefont {{Boh{\'e}}}},\ }\href
  {https://doi.org/10.1103/PhysRevD.93.044007} {\bibfield  {journal} {\bibinfo
  {journal} {\prd}\ }\textbf {\bibinfo {volume} {93}},\ \bibinfo {eid} {044007}
  (\bibinfo {year} {2016})},\ \Eprint {https://arxiv.org/abs/1508.07253}
  {arXiv:1508.07253 [gr-qc]} \BibitemShut {NoStop}%
\bibitem [{\citenamefont {{Abbott}}\ \emph {et~al.}(2018)\citenamefont
  {{Abbott}} \emph {et~al.}}]{2018LRR....21....3A}%
  \BibitemOpen
  \bibfield  {author} {\bibinfo {author} {\bibfnamefont {B.~P.}\ \bibnamefont
  {{Abbott}}} \emph {et~al.} (\bibinfo {collaboration} {LIGO and Virgo
  Collaboration}),\ }\href {https://doi.org/10.1007/s41114-018-0012-9}
  {\bibfield  {journal} {\bibinfo  {journal} {\lrr}\ }\textbf {\bibinfo
  {volume} {21}},\ \bibinfo {eid} {3} (\bibinfo {year} {2018})},\ \Eprint
  {https://arxiv.org/abs/1304.0670} {arXiv:1304.0670 [gr-qc]} \BibitemShut
  {NoStop}%
\bibitem [{\citenamefont {{Abbott}}\ \emph
  {et~al.}(2016{\natexlab{b}})\citenamefont {{Abbott}} \emph
  {et~al.}}]{2016ApJS..227...14A}%
  \BibitemOpen
  \bibfield  {author} {\bibinfo {author} {\bibfnamefont {B.~P.}\ \bibnamefont
  {{Abbott}}} \emph {et~al.} (\bibinfo {collaboration} {LIGO and Virgo
  Collaboration}),\ }\href {https://doi.org/10.3847/0067-0049/227/2/14}
  {\bibfield  {journal} {\bibinfo  {journal} {\apjs}\ }\textbf {\bibinfo
  {volume} {227}},\ \bibinfo {eid} {14} (\bibinfo {year}
  {2016}{\natexlab{b}})},\ \Eprint {https://arxiv.org/abs/1606.03939}
  {arXiv:1606.03939 [astro-ph.HE]} \BibitemShut {NoStop}%
\bibitem [{\citenamefont {{Sesana}}\ \emph {et~al.}(2011)\citenamefont
  {{Sesana}}, \citenamefont {{Gair}}, \citenamefont {{Berti}},\ and\
  \citenamefont {{Volonteri}}}]{2011PhRvD..83d4036S}%
  \BibitemOpen
  \bibfield  {author} {\bibinfo {author} {\bibfnamefont {A.}~\bibnamefont
  {{Sesana}}}, \bibinfo {author} {\bibfnamefont {J.}~\bibnamefont {{Gair}}},
  \bibinfo {author} {\bibfnamefont {E.}~\bibnamefont {{Berti}}},\ and\ \bibinfo
  {author} {\bibfnamefont {M.}~\bibnamefont {{Volonteri}}},\ }\href
  {https://doi.org/10.1103/PhysRevD.83.044036} {\bibfield  {journal} {\bibinfo
  {journal} {\prd}\ }\textbf {\bibinfo {volume} {83}},\ \bibinfo {eid} {044036}
  (\bibinfo {year} {2011})},\ \Eprint {https://arxiv.org/abs/1011.5893}
  {arXiv:1011.5893 [astro-ph.CO]} \BibitemShut {NoStop}%
\bibitem [{\citenamefont {{Bouffanais}}\ \emph {et~al.}(2019)\citenamefont
  {{Bouffanais}}, \citenamefont {{Mapelli}}, \citenamefont {{Gerosa}},
  \citenamefont {{Di Carlo}}, \citenamefont {{Giacobbo}}, \citenamefont
  {{Berti}},\ and\ \citenamefont {{Baibhav}}}]{2019ApJ...886...25B}%
  \BibitemOpen
  \bibfield  {author} {\bibinfo {author} {\bibfnamefont {Y.}~\bibnamefont
  {{Bouffanais}}}, \bibinfo {author} {\bibfnamefont {M.}~\bibnamefont
  {{Mapelli}}}, \bibinfo {author} {\bibfnamefont {D.}~\bibnamefont {{Gerosa}}},
  \bibinfo {author} {\bibfnamefont {U.~N.}\ \bibnamefont {{Di Carlo}}},
  \bibinfo {author} {\bibfnamefont {N.}~\bibnamefont {{Giacobbo}}}, \bibinfo
  {author} {\bibfnamefont {E.}~\bibnamefont {{Berti}}},\ and\ \bibinfo {author}
  {\bibfnamefont {V.}~\bibnamefont {{Baibhav}}},\ }\href
  {https://doi.org/10.3847/1538-4357/ab4a79} {\bibfield  {journal} {\bibinfo
  {journal} {\apj}\ }\textbf {\bibinfo {volume} {886}},\ \bibinfo {eid} {25}
  (\bibinfo {year} {2019})},\ \Eprint {https://arxiv.org/abs/1905.11054}
  {arXiv:1905.11054 [astro-ph.HE]} \BibitemShut {NoStop}%
\bibitem [{\citenamefont {{Toubiana}}\ \emph {et~al.}(2021)\citenamefont
  {{Toubiana}}, \citenamefont {{Wong}}, \citenamefont {{Babak}}, \citenamefont
  {{Barausse}}, \citenamefont {{Berti}}, \citenamefont {{Gair}}, \citenamefont
  {{Marsat}},\ and\ \citenamefont {{Taylor}}}]{2021PhRvD.104h3027T}%
  \BibitemOpen
  \bibfield  {author} {\bibinfo {author} {\bibfnamefont {A.}~\bibnamefont
  {{Toubiana}}}, \bibinfo {author} {\bibfnamefont {K.~W.~K.}\ \bibnamefont
  {{Wong}}}, \bibinfo {author} {\bibfnamefont {S.}~\bibnamefont {{Babak}}},
  \bibinfo {author} {\bibfnamefont {E.}~\bibnamefont {{Barausse}}}, \bibinfo
  {author} {\bibfnamefont {E.}~\bibnamefont {{Berti}}}, \bibinfo {author}
  {\bibfnamefont {J.~R.}\ \bibnamefont {{Gair}}}, \bibinfo {author}
  {\bibfnamefont {S.}~\bibnamefont {{Marsat}}},\ and\ \bibinfo {author}
  {\bibfnamefont {S.~R.}\ \bibnamefont {{Taylor}}},\ }\href
  {https://doi.org/10.1103/PhysRevD.104.083027} {\bibfield  {journal} {\bibinfo
   {journal} {\prd}\ }\textbf {\bibinfo {volume} {104}},\ \bibinfo {eid}
  {083027} (\bibinfo {year} {2021})},\ \Eprint
  {https://arxiv.org/abs/2106.13819} {arXiv:2106.13819 [gr-qc]} \BibitemShut
  {NoStop}%
\bibitem [{\citenamefont {{Kroupa}}(2001)}]{2001MNRAS.322..231K}%
  \BibitemOpen
  \bibfield  {author} {\bibinfo {author} {\bibfnamefont {P.}~\bibnamefont
  {{Kroupa}}},\ }\href {https://doi.org/10.1046/j.1365-8711.2001.04022.x}
  {\bibfield  {journal} {\bibinfo  {journal} {\mnras}\ }\textbf {\bibinfo
  {volume} {322}},\ \bibinfo {pages} {231} (\bibinfo {year} {2001})},\ \Eprint
  {https://arxiv.org/abs/astro-ph/0009005} {arXiv:astro-ph/0009005 [astro-ph]}
  \BibitemShut {NoStop}%
\bibitem [{\citenamefont {{Belczynski}}(2020)}]{2020ApJ...905L..15B}%
  \BibitemOpen
  \bibfield  {author} {\bibinfo {author} {\bibfnamefont {K.}~\bibnamefont
  {{Belczynski}}},\ }\href {https://doi.org/10.3847/2041-8213/abcbf1}
  {\bibfield  {journal} {\bibinfo  {journal} {\apjl}\ }\textbf {\bibinfo
  {volume} {905}},\ \bibinfo {eid} {L15} (\bibinfo {year} {2020})},\ \Eprint
  {https://arxiv.org/abs/2009.13526} {arXiv:2009.13526 [astro-ph.HE]}
  \BibitemShut {NoStop}%
\bibitem [{\citenamefont {{Tiwari}}\ and\ \citenamefont
  {{Fairhurst}}(2021)}]{2021ApJ...913L..19T}%
  \BibitemOpen
  \bibfield  {author} {\bibinfo {author} {\bibfnamefont {V.}~\bibnamefont
  {{Tiwari}}}\ and\ \bibinfo {author} {\bibfnamefont {S.}~\bibnamefont
  {{Fairhurst}}},\ }\href {https://doi.org/10.3847/2041-8213/abfbe7} {\bibfield
   {journal} {\bibinfo  {journal} {\apjl}\ }\textbf {\bibinfo {volume} {913}},\
  \bibinfo {eid} {L19} (\bibinfo {year} {2021})},\ \Eprint
  {https://arxiv.org/abs/2011.04502} {arXiv:2011.04502 [astro-ph.HE]}
  \BibitemShut {NoStop}%
\bibitem [{\citenamefont {{Zaldarriaga}}\ \emph {et~al.}(2018)\citenamefont
  {{Zaldarriaga}}, \citenamefont {{Kushnir}},\ and\ \citenamefont
  {{Kollmeier}}}]{2018MNRAS.473.4174Z}%
  \BibitemOpen
  \bibfield  {author} {\bibinfo {author} {\bibfnamefont {M.}~\bibnamefont
  {{Zaldarriaga}}}, \bibinfo {author} {\bibfnamefont {D.}~\bibnamefont
  {{Kushnir}}},\ and\ \bibinfo {author} {\bibfnamefont {J.~A.}\ \bibnamefont
  {{Kollmeier}}},\ }\href {https://doi.org/10.1093/mnras/stx2577} {\bibfield
  {journal} {\bibinfo  {journal} {\mnras}\ }\textbf {\bibinfo {volume} {473}},\
  \bibinfo {pages} {4174} (\bibinfo {year} {2018})},\ \Eprint
  {https://arxiv.org/abs/1702.00885} {arXiv:1702.00885 [astro-ph.HE]}
  \BibitemShut {NoStop}%
\bibitem [{\citenamefont {{Bavera}}\ \emph {et~al.}(2020)\citenamefont
  {{Bavera}}, \citenamefont {{Fragos}}, \citenamefont {{Qin}}, \citenamefont
  {{Zapartas}}, \citenamefont {{Neijssel}}, \citenamefont {{Mandel}},
  \citenamefont {{Batta}}, \citenamefont {{Gaebel}}, \citenamefont
  {{Kimball}},\ and\ \citenamefont {{Stevenson}}}]{2020A&A...635A..97B}%
  \BibitemOpen
  \bibfield  {author} {\bibinfo {author} {\bibfnamefont {S.~S.}\ \bibnamefont
  {{Bavera}}}, \bibinfo {author} {\bibfnamefont {T.}~\bibnamefont {{Fragos}}},
  \bibinfo {author} {\bibfnamefont {Y.}~\bibnamefont {{Qin}}}, \bibinfo
  {author} {\bibfnamefont {E.}~\bibnamefont {{Zapartas}}}, \bibinfo {author}
  {\bibfnamefont {C.~J.}\ \bibnamefont {{Neijssel}}}, \bibinfo {author}
  {\bibfnamefont {I.}~\bibnamefont {{Mandel}}}, \bibinfo {author}
  {\bibfnamefont {A.}~\bibnamefont {{Batta}}}, \bibinfo {author} {\bibfnamefont
  {S.~M.}\ \bibnamefont {{Gaebel}}}, \bibinfo {author} {\bibfnamefont
  {C.}~\bibnamefont {{Kimball}}},\ and\ \bibinfo {author} {\bibfnamefont
  {S.}~\bibnamefont {{Stevenson}}},\ }\href
  {https://doi.org/10.1051/0004-6361/201936204} {\bibfield  {journal} {\bibinfo
   {journal} {\aap}\ }\textbf {\bibinfo {volume} {635}},\ \bibinfo {eid} {A97}
  (\bibinfo {year} {2020})},\ \Eprint {https://arxiv.org/abs/1906.12257}
  {arXiv:1906.12257 [astro-ph.HE]} \BibitemShut {NoStop}%
\bibitem [{\citenamefont {{Schmidt}}\ \emph {et~al.}(2015)\citenamefont
  {{Schmidt}}, \citenamefont {{Ohme}},\ and\ \citenamefont
  {{Hannam}}}]{2015PhRvD..91b4043S}%
  \BibitemOpen
  \bibfield  {author} {\bibinfo {author} {\bibfnamefont {P.}~\bibnamefont
  {{Schmidt}}}, \bibinfo {author} {\bibfnamefont {F.}~\bibnamefont {{Ohme}}},\
  and\ \bibinfo {author} {\bibfnamefont {M.}~\bibnamefont {{Hannam}}},\ }\href
  {https://doi.org/10.1103/PhysRevD.91.024043} {\bibfield  {journal} {\bibinfo
  {journal} {\prd}\ }\textbf {\bibinfo {volume} {91}},\ \bibinfo {eid} {024043}
  (\bibinfo {year} {2015})},\ \Eprint {https://arxiv.org/abs/1408.1810}
  {arXiv:1408.1810 [gr-qc]} \BibitemShut {NoStop}%
\bibitem [{\citenamefont {{Henshaw}}\ \emph {et~al.}(2022)\citenamefont
  {{Henshaw}}, \citenamefont {{O'Shaughnessy}},\ and\ \citenamefont
  {{Cadonati}}}]{2022CQGra..39l5003H}%
  \BibitemOpen
  \bibfield  {author} {\bibinfo {author} {\bibfnamefont {C.}~\bibnamefont
  {{Henshaw}}}, \bibinfo {author} {\bibfnamefont {R.}~\bibnamefont
  {{O'Shaughnessy}}},\ and\ \bibinfo {author} {\bibfnamefont {L.}~\bibnamefont
  {{Cadonati}}},\ }\href {https://doi.org/10.1088/1361-6382/ac6cc0} {\bibfield
  {journal} {\bibinfo  {journal} {\cqg}\ }\textbf {\bibinfo {volume} {39}},\
  \bibinfo {eid} {125003} (\bibinfo {year} {2022})},\ \Eprint
  {https://arxiv.org/abs/2201.05220} {arXiv:2201.05220 [gr-qc]} \BibitemShut
  {NoStop}%
\bibitem [{\citenamefont {{Doctor}}\ \emph {et~al.}(2020)\citenamefont
  {{Doctor}}, \citenamefont {{Wysocki}}, \citenamefont {{O'Shaughnessy}},
  \citenamefont {{Holz}},\ and\ \citenamefont {{Farr}}}]{2020ApJ...893...35D}%
  \BibitemOpen
  \bibfield  {author} {\bibinfo {author} {\bibfnamefont {Z.}~\bibnamefont
  {{Doctor}}}, \bibinfo {author} {\bibfnamefont {D.}~\bibnamefont {{Wysocki}}},
  \bibinfo {author} {\bibfnamefont {R.}~\bibnamefont {{O'Shaughnessy}}},
  \bibinfo {author} {\bibfnamefont {D.~E.}\ \bibnamefont {{Holz}}},\ and\
  \bibinfo {author} {\bibfnamefont {B.}~\bibnamefont {{Farr}}},\ }\href
  {https://doi.org/10.3847/1538-4357/ab7fac} {\bibfield  {journal} {\bibinfo
  {journal} {\apj}\ }\textbf {\bibinfo {volume} {893}},\ \bibinfo {eid} {35}
  (\bibinfo {year} {2020})},\ \Eprint {https://arxiv.org/abs/1911.04424}
  {arXiv:1911.04424 [astro-ph.HE]} \BibitemShut {NoStop}%
\bibitem [{\citenamefont {{Kimball}}\ \emph {et~al.}(2020)\citenamefont
  {{Kimball}}, \citenamefont {{Talbot}}, \citenamefont {{Berry}}, \citenamefont
  {{Carney}}, \citenamefont {{Zevin}}, \citenamefont {{Thrane}},\ and\
  \citenamefont {{Kalogera}}}]{2020ApJ...900..177K}%
  \BibitemOpen
  \bibfield  {author} {\bibinfo {author} {\bibfnamefont {C.}~\bibnamefont
  {{Kimball}}}, \bibinfo {author} {\bibfnamefont {C.}~\bibnamefont {{Talbot}}},
  \bibinfo {author} {\bibfnamefont {C.~P.~L.}\ \bibnamefont {{Berry}}},
  \bibinfo {author} {\bibfnamefont {M.}~\bibnamefont {{Carney}}}, \bibinfo
  {author} {\bibfnamefont {M.}~\bibnamefont {{Zevin}}}, \bibinfo {author}
  {\bibfnamefont {E.}~\bibnamefont {{Thrane}}},\ and\ \bibinfo {author}
  {\bibfnamefont {V.}~\bibnamefont {{Kalogera}}},\ }\href
  {https://doi.org/10.3847/1538-4357/aba518} {\bibfield  {journal} {\bibinfo
  {journal} {\apj}\ }\textbf {\bibinfo {volume} {900}},\ \bibinfo {eid} {177}
  (\bibinfo {year} {2020})},\ \Eprint {https://arxiv.org/abs/2005.00023}
  {arXiv:2005.00023 [astro-ph.HE]} \BibitemShut {NoStop}%
\bibitem [{\citenamefont {{McKernan}}\ \emph {et~al.}(2020)\citenamefont
  {{McKernan}}, \citenamefont {{Ford}}, \citenamefont {{O'Shaugnessy}},\ and\
  \citenamefont {{Wysocki}}}]{2020MNRAS.494.1203M}%
  \BibitemOpen
  \bibfield  {author} {\bibinfo {author} {\bibfnamefont {B.}~\bibnamefont
  {{McKernan}}}, \bibinfo {author} {\bibfnamefont {K.~E.~S.}\ \bibnamefont
  {{Ford}}}, \bibinfo {author} {\bibfnamefont {R.}~\bibnamefont
  {{O'Shaugnessy}}},\ and\ \bibinfo {author} {\bibfnamefont {D.}~\bibnamefont
  {{Wysocki}}},\ }\href {https://doi.org/10.1093/mnras/staa740} {\bibfield
  {journal} {\bibinfo  {journal} {\mnras}\ }\textbf {\bibinfo {volume} {494}},\
  \bibinfo {pages} {1203} (\bibinfo {year} {2020})},\ \Eprint
  {https://arxiv.org/abs/1907.04356} {arXiv:1907.04356 [astro-ph.HE]}
  \BibitemShut {NoStop}%
\bibitem [{\citenamefont {{Fishbach}}\ \emph {et~al.}(2022)\citenamefont
  {{Fishbach}}, \citenamefont {{Kimball}},\ and\ \citenamefont
  {{Kalogera}}}]{2022ApJ...935L..26F}%
  \BibitemOpen
  \bibfield  {author} {\bibinfo {author} {\bibfnamefont {M.}~\bibnamefont
  {{Fishbach}}}, \bibinfo {author} {\bibfnamefont {C.}~\bibnamefont
  {{Kimball}}},\ and\ \bibinfo {author} {\bibfnamefont {V.}~\bibnamefont
  {{Kalogera}}},\ }\href {https://doi.org/10.3847/2041-8213/ac86c4} {\bibfield
  {journal} {\bibinfo  {journal} {\apjl}\ }\textbf {\bibinfo {volume} {935}},\
  \bibinfo {eid} {L26} (\bibinfo {year} {2022})},\ \Eprint
  {https://arxiv.org/abs/2207.02924} {arXiv:2207.02924 [astro-ph.HE]}
  \BibitemShut {NoStop}%
\bibitem [{\citenamefont {{van den Oord}}\ \emph {et~al.}(2017)\citenamefont
  {{van den Oord}}, \citenamefont {{Li}}, \citenamefont {{Babuschkin}},
  \citenamefont {{Simonyan}}, \citenamefont {{Vinyals}}, \citenamefont
  {{Kavukcuoglu}}, \citenamefont {{van den Driessche}}, \citenamefont
  {{Lockhart}}, \citenamefont {{Cobo}}, \citenamefont {{Stimberg}},
  \citenamefont {{Casagrande}}, \citenamefont {{Grewe}}, \citenamefont
  {{Noury}}, \citenamefont {{Dieleman}}, \citenamefont {{Elsen}}, \citenamefont
  {{Kalchbrenner}}, \citenamefont {{Zen}}, \citenamefont {{Graves}},
  \citenamefont {{King}}, \citenamefont {{Walters}}, \citenamefont {{Belov}},\
  and\ \citenamefont {{Hassabis}}}]{2017arXiv171110433V}%
  \BibitemOpen
  \bibfield  {author} {\bibinfo {author} {\bibfnamefont {A.}~\bibnamefont {{van
  den Oord}}}, \bibinfo {author} {\bibfnamefont {Y.}~\bibnamefont {{Li}}},
  \bibinfo {author} {\bibfnamefont {I.}~\bibnamefont {{Babuschkin}}}, \bibinfo
  {author} {\bibfnamefont {K.}~\bibnamefont {{Simonyan}}}, \bibinfo {author}
  {\bibfnamefont {O.}~\bibnamefont {{Vinyals}}}, \bibinfo {author}
  {\bibfnamefont {K.}~\bibnamefont {{Kavukcuoglu}}}, \bibinfo {author}
  {\bibfnamefont {G.}~\bibnamefont {{van den Driessche}}}, \bibinfo {author}
  {\bibfnamefont {E.}~\bibnamefont {{Lockhart}}}, \bibinfo {author}
  {\bibfnamefont {L.~C.}\ \bibnamefont {{Cobo}}}, \bibinfo {author}
  {\bibfnamefont {F.}~\bibnamefont {{Stimberg}}}, \bibinfo {author}
  {\bibfnamefont {N.}~\bibnamefont {{Casagrande}}}, \bibinfo {author}
  {\bibfnamefont {D.}~\bibnamefont {{Grewe}}}, \bibinfo {author} {\bibfnamefont
  {S.}~\bibnamefont {{Noury}}}, \bibinfo {author} {\bibfnamefont
  {S.}~\bibnamefont {{Dieleman}}}, \bibinfo {author} {\bibfnamefont
  {E.}~\bibnamefont {{Elsen}}}, \bibinfo {author} {\bibfnamefont
  {N.}~\bibnamefont {{Kalchbrenner}}}, \bibinfo {author} {\bibfnamefont
  {H.}~\bibnamefont {{Zen}}}, \bibinfo {author} {\bibfnamefont
  {A.}~\bibnamefont {{Graves}}}, \bibinfo {author} {\bibfnamefont
  {H.}~\bibnamefont {{King}}}, \bibinfo {author} {\bibfnamefont
  {T.}~\bibnamefont {{Walters}}}, \bibinfo {author} {\bibfnamefont
  {D.}~\bibnamefont {{Belov}}},\ and\ \bibinfo {author} {\bibfnamefont
  {D.}~\bibnamefont {{Hassabis}}},\ }\href@noop {} {\bibfield  {journal}
  {\bibinfo  {journal} {{}}\ } (\bibinfo {year} {2017})},\ \Eprint
  {https://arxiv.org/abs/1711.10433} {arXiv:1711.10433 [cs.LG]} \BibitemShut
  {NoStop}%
\bibitem [{\citenamefont {{Valentin Jospin}}\ \emph {et~al.}(2020)\citenamefont
  {{Valentin Jospin}}, \citenamefont {{Buntine}}, \citenamefont {{Boussaid}},
  \citenamefont {{Laga}},\ and\ \citenamefont
  {{Bennamoun}}}]{2020arXiv200706823V}%
  \BibitemOpen
  \bibfield  {author} {\bibinfo {author} {\bibfnamefont {L.}~\bibnamefont
  {{Valentin Jospin}}}, \bibinfo {author} {\bibfnamefont {W.}~\bibnamefont
  {{Buntine}}}, \bibinfo {author} {\bibfnamefont {F.}~\bibnamefont
  {{Boussaid}}}, \bibinfo {author} {\bibfnamefont {H.}~\bibnamefont {{Laga}}},\
  and\ \bibinfo {author} {\bibfnamefont {M.}~\bibnamefont {{Bennamoun}}},\
  }\href@noop {} {\bibfield  {journal} {\bibinfo  {journal} {{}}\ } (\bibinfo
  {year} {2020})},\ \Eprint {https://arxiv.org/abs/2007.06823}
  {arXiv:2007.06823 [cs.LG]} \BibitemShut {NoStop}%
\end{thebibliography}%

\end{document}